\documentclass[twocolumn,tighten]{aastex62}
\hypersetup{linkcolor=red,citecolor=blue,filecolor=cyan,urlcolor=magenta}
\usepackage{amsmath}
\usepackage{natbib}
\usepackage{color}
\usepackage{hyperref}
\usepackage{url}

\newcommand{\Oo}{\textcolor{black}{17}}
\newcommand{\Oi}{\textcolor{black}{9}}
\newcommand{\Oii}{\textcolor{black}{27}}
\newcommand{\Oiii}{\textcolor{black}{10}}
\newcommand{\Oiiii}{\textcolor{black}{7}}
\newcommand{\Ov}{\textcolor{black}{4}}
\newcommand{\Ovi}{\textcolor{black}{5}}
\newcommand{\Ovii}{\textcolor{black}{18}}
\newcommand{\Oviii}{\textcolor{black}{13}}
\newcommand{\Oviiii}{\textcolor{black}{2}}
\newcommand{\Io}{\textcolor{black}{15}}
\newcommand{\Ii}{\textcolor{black}{23}}
\newcommand{\Iii}{\textcolor{black}{28}}
\newcommand{\Iiii}{\textcolor{black}{21}}
\newcommand{\Iiiii}{\textcolor{black}{1}}
\newcommand{\Iv}{\textcolor{black}{8}}
\newcommand{\Ivi}{\textcolor{black}{19}}
\newcommand{\Ivii}{\textcolor{black}{26}}
\newcommand{\Iviii}{\textcolor{black}{11}}
\newcommand{\Iviiii}{\textcolor{black}{12}}
\newcommand{\IIo}{\textcolor{black}{3}}
\newcommand{\IIi}{\textcolor{black}{16}}
\newcommand{\IIii}{\textcolor{black}{6}}
\newcommand{\IIiii}{\textcolor{black}{24}}
\newcommand{\IIiiii}{\textcolor{black}{25}}
\newcommand{\IIv}{\textcolor{black}{29}}
\newcommand{\IIvi}{\textcolor{black}{20}}
\newcommand{\IIvii}{\textcolor{black}{30}}
\newcommand{\IIviii}{\textcolor{black}{22}}
\newcommand{\IIviiii}{\textcolor{black}{14}}

\def\vector#1{\mbox{\boldmath $#1$}}

\newcommand{\kpc}{\ensuremath{\,\mathrm{kpc}}}

\newcommand{\Gyr}{\ensuremath{\,\mathrm{Gyr}}}
\newcommand{\kms}{\ensuremath{\,\mathrm{km\ s}^{-1}}}

\newcommand{\vlos}{v_{\ensuremath{\mathrm{los}}}}

\newcommand{\mualpha}{\mu_{\alpha*}}
\newcommand{\mudelta}{\mu_\delta}

\newcommand{\eq}[1]{\begin{align}#1\end{align}}

\newcommand{\new}[1]{{#1}}

\newcommand{\pen}{\mathrm{penalty}}

\begin{document}

\title{%
Finding $r$-II sibling stars in the Milky Way 
with the Greedy Optimistic Clustering algorithm 
}

\shorttitle{Clustering of $r$-II stars}
\shortauthors{Hattori et al.}

\author[0000-0001-6924-8862]{Kohei~Hattori}
\affiliation{National Astronomical Observatory of Japan, 2-21-1 Osawa, Mitaka, Tokyo 181-8588, Japan}
\affiliation{The Institute of Statistical Mathematics, 10-3 Midoricho, Tachikawa, Tokyo 190-8562, Japan}
\affiliation{Department of Astronomy, University of Michigan,
1085 S.\ University Avenue, Ann Arbor, MI 48109, USA}
\email{Email:\ khattori@ism.ac.jp}

\author[0000-0001-9621-8853]{Akifumi~Okuno}
\affiliation{The Institute of Statistical Mathematics, 10-3 Midoricho, Tachikawa, Tokyo 190-8562, Japan}
\affiliation{RIKEN Center for Advanced Intelligence Project, 1-4-1 Nihonbashi, Chuo-ku, Tokyo, 103-0027, Japan} 

\author[0000-0001-5107-8930]{Ian~U.~Roederer}
\affiliation{Department of Astronomy, University of Michigan,
1085 S.\ University Avenue, Ann Arbor, MI 48109, USA}
\affiliation{%
Joint Institute for Nuclear Astrophysics -- Center for the
Evolution of the Elements (JINA-CEE), USA}

\begin{abstract}

$R$-process enhanced stars with [Eu/Fe]$\geq+0.7$ (so-called $r$-II stars) 
are believed to have 
formed in an extremely neutron-rich environment in which 
a rare astrophysical event (e.g., a neutron star merger) occurred. 
This scenario is supported by the existence of an ultra-faint dwarf galaxy, Reticulum~II, where most of the stars are highly enhanced in $r$-process elements.
In this scenario, 
some small fraction of dwarf galaxies around the Milky Way were $r$ enhanced. 
When each $r$-enhanced dwarf galaxy accreted to the Milky Way, 
it deposited many $r$-II stars in the Galactic halo with similar orbital actions. 
To search for the remnants of the $r$-enhanced systems, 
we analyzed the distribution of the orbital actions of $N=161$ $r$-II stars in the Solar neighborhood by using the Gaia EDR3 data. 
Since the observational uncertainty is not negligible, 
we applied a newly-developed {\it greedy optimistic clustering method} 
to the orbital actions of our sample stars. 
We found six clusters of $r$-II stars that have similar orbits and chemistry, one of which is a new discovery. 
Given the apparent phase-mixed orbits of the member stars, 
we interpret that these clusters 
\new{are good candidates for }
remnants of completely disrupted $r$-enhanced dwarf galaxies that merged with the ancient Milky Way.

\end{abstract}

\section{Introduction}

\subsection{Reconstructing the Galactic merger history}

In the standard paradigm of galaxy formation, 
it is believed that the stellar halo of the Milky Way 
was formed through numerous mergers with smaller stellar systems, 
such as dwarf galaxies \citep{White1978MNRAS.183..341W, Blumenthal1984Natur.311..517B}. 
Identifying and recovering the past merger events in the Milky Way 
from the stellar position $\vector{x}$, velocity $\vector{v}$, 
chemistry, and age 
is one of the ultimate goals in Galactic astronomy.

Except for the surviving dwarf galaxies 
that are currently moving around the Milky Way at large Galactocentric radii, 
most of the dwarf galaxies that merged with the Milky Way 
are tidally disrupted. 
On the one hand, 
recently accreted dwarf galaxies are being disrupted 
to form stellar streams.  
These stellar streams have a spatially coherent structure, 
and they are extensively searched for 
from large photometric and astrometric surveys
\citep{Belokurov2006ApJ...642L.137B,Malhan2018MNRAS.481.3442M,Shipp2018ApJ...862..114S,Ibata2021ApJ...914..123I,Martin2022arXiv220101310M}. 
To date, nearly a hundred stellar streams have been found
\citep{Mateu2022arXiv220410326M}, 
and many of these streams are 
thought to be associated with large merger events \citep{Bonaca2021ApJ...909L..26B,Malhan2022ApJ...926..107M}. 
On the other hand, 
dwarf galaxies that merged with the ancient Milky Way 
have been completely disrupted \citep{Bullock2005ApJ...635..931B, Wu2022MNRAS.509.5882W, Brauer2022arXiv220607057B}. 
The stars that originated in these disrupted dwarf galaxies 
show a spatially smooth distribution 
due to the phase-mixing of the orbits. 
The lack of spatial coherence 
makes it hard to identify the remnants of the disrupted dwarf galaxies. 
To unravel the ancient merger history of the Milky Way, 
it is crucial to find the remnants of the completely disrupted dwarf galaxies 
from the `field' halo stars.

Because the stars stripped from the progenitor dwarf galaxy 
move on orbits that are similar to the progenitor's orbit, 
the remnants of the disrupted dwarf galaxies 
show a clumped distribution in the phase-space spanned by 
the conserved orbital properties, 
such as 
the orbital action $\vector{J}(\vector{x}, \vector{v})=(J_r,J_\phi,J_z)$, 
angular momentum $\vector{L}=\vector{x}\times\vector{v}$, 
or orbital energy, $E$ 
\citep{Helmi1999Natur.402...53H, Gomez2010MNRAS.408..935G}. 
Thus, the most promising way to find these remnants 
is to find substructure in the dynamical phase-space by using clustering methods 
\citep{Roederer2018, Myeong2018ApJ...856L..26M, Myeong2019MNRAS.488.1235M, Matsuno2019ApJ...874L..35M, Li2019ApJ...874...74L, Koppelman2019A&A...631L...9K, Yuan2020ApJ...891...39Y, Buder2022MNRAS.510.2407B, Shank2022ApJ...926...26S, Lovdal2022arXiv220102404S, Brauer2022arXiv220607057B}.

\subsection{$r$-II stars as the tracers of ancient merger history}

Spectroscopic observations of old stars in the Milky Way 
have revealed that a few percent of nearby field halo stars 
are highly enhanced in $r$-process elements, such as europium (Eu, $Z = 63$) 
\citep{Beers2005ARA&A..43..531B}. 
Currently, more than 160 stars are known to have [Eu/Fe]$>+0.7$ (and [Eu/Ba]$>0$) 
and are classified as $r$-II stars
\citep{Holmbeck2020}. 
Their extremely enhanced Eu abundances and their old ages imply that 
each $r$-II star was born in a system 
that was polluted by Eu-rich ejecta from one $r$-process event, such as a neutron star merger \citep{Hotokezaka2015NatPh..11.1042H}.

The birthplace of the field $r$-II stars has been a mystery for decades. 
Recently, \cite{Ji2016Natur.531..610J} and \cite{Roederer2016AJ....151...82R} 
discovered that 
the ultra-faint dwarf galaxy (UFD) named Reticulum~II 
contains seven $r$-II stars among nine spectroscopically observed stars in this galaxy. 
The discovery of this $r$-enhanced UFD 
favors a scenario wherein a majority of 
$r$-II stars in the halo were originally born in UFDs similar to Reticulum~II, 
in which very rare $r$-process events, such as a neutron-star merger, occurred 
\new{\citep{Tsujimoto2014A&A...565L...5T,Tsujimoto2014ApJ...795L..18T}}. 
In this scenario, $r$-II stars were later deposited to the Galactic halo when the progenitor system was disrupted 
\new{\citep{Brauer2022arXiv220607057B}}.

This view is supported by 
the distribution of the orbital action $\vector{J}$ of field $r$-II stars. 
\cite{Roederer2018}
performed a clustering analysis for 
$N=35$ $r$-II stars in the Solar neighborhood 
and found several clusters of stars 
with similar $\vector{J}$. 
Although metallicity information was not used in the clustering analysis, 
the discovered clusters turned out to have a tight distribution in [Fe/H]. 
These clusters of stars have similar orbits and chemistry, 
and at least some of the clusters 
may be the remnants of dwarf galaxies 
that were completely disrupted long ago.

\subsection{The scope of this paper}

One of the limitations in previous studies of the clustering analysis of halo stars 
is that the clustering of stars in the orbital action 
$\vector{J}(\vector{x}, \vector{v})$  
is blurred when the observational uncertainty in $(\vector{x}, \vector{v})$ 
is not negligible. 
For example, 
a modest distance uncertainty of $\Delta d / d = 0.2$ 
of Solar-neighbor halo stars 
would result in an uncertainty in action of $\Delta J \sim 400 \kpc\kms$. 
This blurring effect is a serious problem when we aim to 
find dynamically cold substructure of halo stars. 
Indeed, 
the remnants of disrupted low-mass dwarf galaxies (with stellar mass $M_{*} < 10^6 M_\odot$) -- including UFDs -- 
are expected to have a small internal dispersion of 
\textcolor{red}{$\sigma_J$} 
$\sim 100 \kpc \kms$ in galaxy formation simulations 
(e.g., \citealt{Bullock2005ApJ...635..931B}). 
Therefore, 
if we are to use conventional clustering methods to find dynamically cold substructure, 
we need to use a sample of halo stars with very good distance estimates. 
In fact, \cite{Roederer2018}
discarded $58\%$ of the stars from their original catalog of $r$-II stars 
because their stellar distances were associated with more than 12.5\% uncertainty. 
In general, 
the requirement of using very accurate data implies 
that we need to discard a large fraction of data available, 
which reduces the scientific impact of surveys. 
Therefore, 
it is crucial to invent a new clustering algorithm  
that allows data sets with large uncertainty.

The originality of this paper is to introduce a {\it greedy optimistic clustering} method, which is applicable for finding clusters from noisy data sets. 
This method simultaneously estimates both the centroid of each cluster 
and the denoised ({\it true}) value for each data point. 
A general mathematical formulation of this method is described in the accompanying paper \citep{OkunoHattori2022}. 
As demonstrated in \cite{OkunoHattori2022}, 
this new technique can successfully 
find the remnants of the completely disrupted dwarf galaxies 
from the realistic kinematical data of field halo stars with a Gaia EDR3-like astrometric error. 
In this paper, 
we apply one flavor of the greedy optimistic clustering method, 
a greedy optimistic clustering method using Gaussian mixture model (GMM)
(or {\it greedy optimistic GMM}), 
to $N=161$ $r$-II halo stars 
to find the remnants of completely disrupted dwarf galaxies.  
This sample is an extended version of the $r$-II star catalog presented in \cite{Roederer2018}, including most of the $r$-II stars published before the end of 2020. 
We emphasize that, with our new technique, 
we do not need to discard the sample stars 
even if the distance uncertainty is large.

In this paper, 
we first describe our catalog of $r$-II stars in Section \ref{sec:data}. 
In Section \ref{sec:why}, 
we explain that some of our sample stars 
have a large uncertainty in $\vector{J}$ because of the observational uncertainty. 
This is why we use the greedy optimistic clustering method 
instead of the conventional clustering methods. 
In Section \ref{sec:preprocess}, 
we explain the concept of the {\it uncertainty set} and 
describe how we generate it. 
This procedure is a preprocessing of the data 
that is required for our clustering method. 
The detailed implementation of our method is presented in Section \ref{sec:GMM}.
(An intuitive illustration of our method is presented in Appendix \ref{sec:idea}.)
Section \ref{sec:analysis} describes how we perform the clustering analysis 
and presents the fiducial results. 
We find that the results in \cite{Roederer2018} 
are recovered in our analysis. 
In Section \ref{sec:discussion}, 
we discuss the implications from our results. 
In particular, 
we validate our clustering results by using the chemistry data in Section \ref{subsec:percentile}. 
We find six clusters with a tight distribution in chemical abundances that played no role in the clustering process or the initial selection for inclusion in our sample, which we interpret as revealing the remnants of completely disrupted dwarf galaxies (Section \ref{subsubsec:tier1}). 
We further analyze the connection of our results 
to the Galactic merger history (Sections \ref{subsec:Malhan_groups} and \ref{subsec:new_merger_group}). 
In Section \ref{sec:conclusion}, we summarize our paper.

\section{Observational data} \label{sec:data}

\begin{figure}
\centering
\includegraphics[width=3.2in]{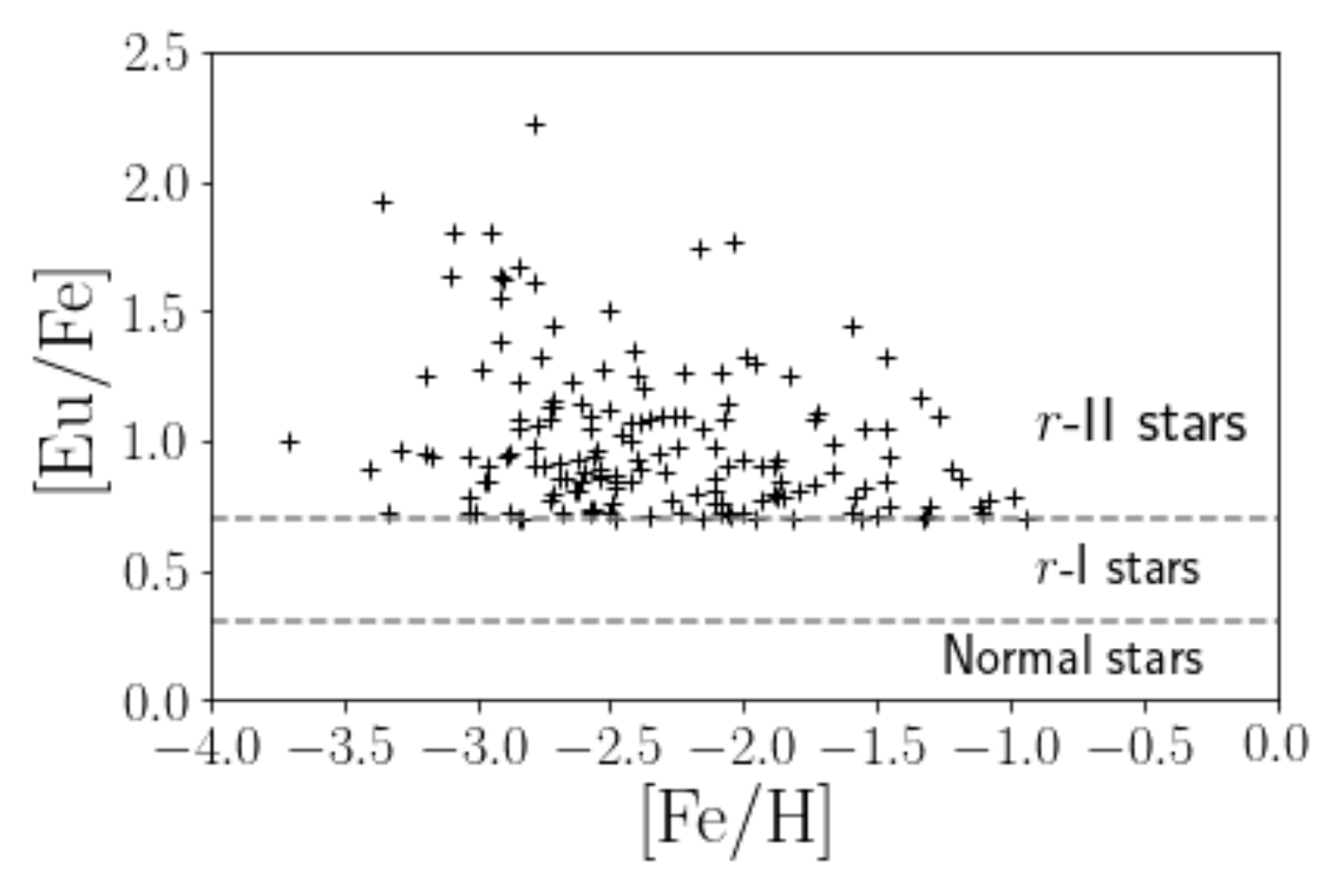}
\caption{
The chemistry of our sample of $N=161$ stars with [Eu/Fe]$\geq+0.7$, 
which we classify as $r$-II stars. 
Our sample is an extended version of the $r$-II star sample in \cite{Roederer2018}. 
}
\label{fig:FeH_EuFe}
\end{figure}

We extend the sample in \cite{Roederer2018} 
and construct a catalog of $r$-II stars 
for which reliable measurements 
of chemical abundances 
and the line-of-sight velocity $\vlos$ 
are available from high-resolution spectra. 
In constructing the catalog, 
we select $N=161$ stars from literature (published before the end of 2020) 
that satisfy the following criteria:\footnote{
Our criteria on [Eu/Fe] is stricter than 
that used in \cite{Gudin2021ApJ...908...79G} 
([Eu/Fe]$>+0.3$), 
because the formation sites of mildly $r$-enhanced stars 
(so-called $r$-I stars) may not be UFDs 
\citep{Wanajo2021MNRAS.505.5862W,Hirai2022MNRAS.517.4856H}.
}
\begin{itemize}
\item (i)   [Eu/Fe]$\geq +0.7$ and [Eu/Ba]$>0$;
\item (ii)  small $\vlos$ variability (no hint of binarity); and
\item (iii) not being a member of known dwarf galaxies or globular clusters. 
\end{itemize}
When observational uncertainties overlap with the boundaries defined by criterion (i), we apply our best judgment to assess inclusion in the catalog, based on spectral quality, other heavy-element abundance ratios, and---when available---confirmation by independent studies.
The distribution of [Fe/H] and [Eu/Fe] of our sample stars is shown in Fig.~\ref{fig:FeH_EuFe}. 
We combine the literature data of ([Fe/H], [Eu/Fe], [Eu/H], $\vlos$) 
and the astrometric data from Gaia EDR3 \citep{Gaia2021A&A...649A...1G} 
to construct the catalog to be analyzed in this paper.

For those stars that are reported in \cite{Roederer2018}, 
we adopt the uncertainty $\sigma_\mathrm{vlos}$ in the line-of-sight velocity 
as in \cite{Roederer2018}. 
For the rest of the stars, 
we use either the literature value or assign some reasonable value. 
In any case, the value of $\sigma_\mathrm{vlos}$ is small for all of our sample stars 
and thus the detailed value of $\sigma_\mathrm{vlos}$ does not affect our results.

For all of these stars, Gaia EDR3 provides 
reliable measurements of the five-dimensional astrometric quantities, 
$(\alpha, \delta, \varpi, \mualpha, \mudelta)$, 
as well as their associated uncertainties. 
In this paper, 
$\widehat{X}$ and $\sigma_X$ denote the 
point-estimate and the one-dimensional uncertainty for 
the quantity $X$, respectively. 
We assume that the uncertainties associated with 
the right ascension and declination 
$(\alpha, \delta)$ are negligible; 
and thus we assume $(\alpha, \delta)=(\widehat{\alpha}, \widehat{\delta})$. 
For parallax, we have the (zero-point corrected\footnote{
We corrected for the zero-point offset in parallax 
by using the prescription in \cite{Lindegren2021A&A...649A...4L}.
}) point estimate $\widehat{\varpi}$ 
and its uncertainty $\sigma_\varpi$. 
For proper motion, 
we have the point-estimate $(\widehat{\mualpha}, \widehat{\mudelta})$, 
their one-dimensional uncertainties $(\sigma_{\mu\alpha*}, \sigma_{\mu\delta})$, 
and the Pearson's correlation coefficient in these uncertainties $\rho_\mathrm{corr}$.

The name of $r$-II stars and their chemical abundances 
are listed in Appendix \ref{sec:list_of_stars}. 
We note that 
our catalog based on Gaia EDR3 
supersedes the catalog in \cite{Roederer2018}, 
which is based on Gaia DR2 
\citep{Gaia2016A&A...595A...1G, Gaia2018A&A...616A...1G}.

\begin{figure}
\centering
\includegraphics[width=3.2in]{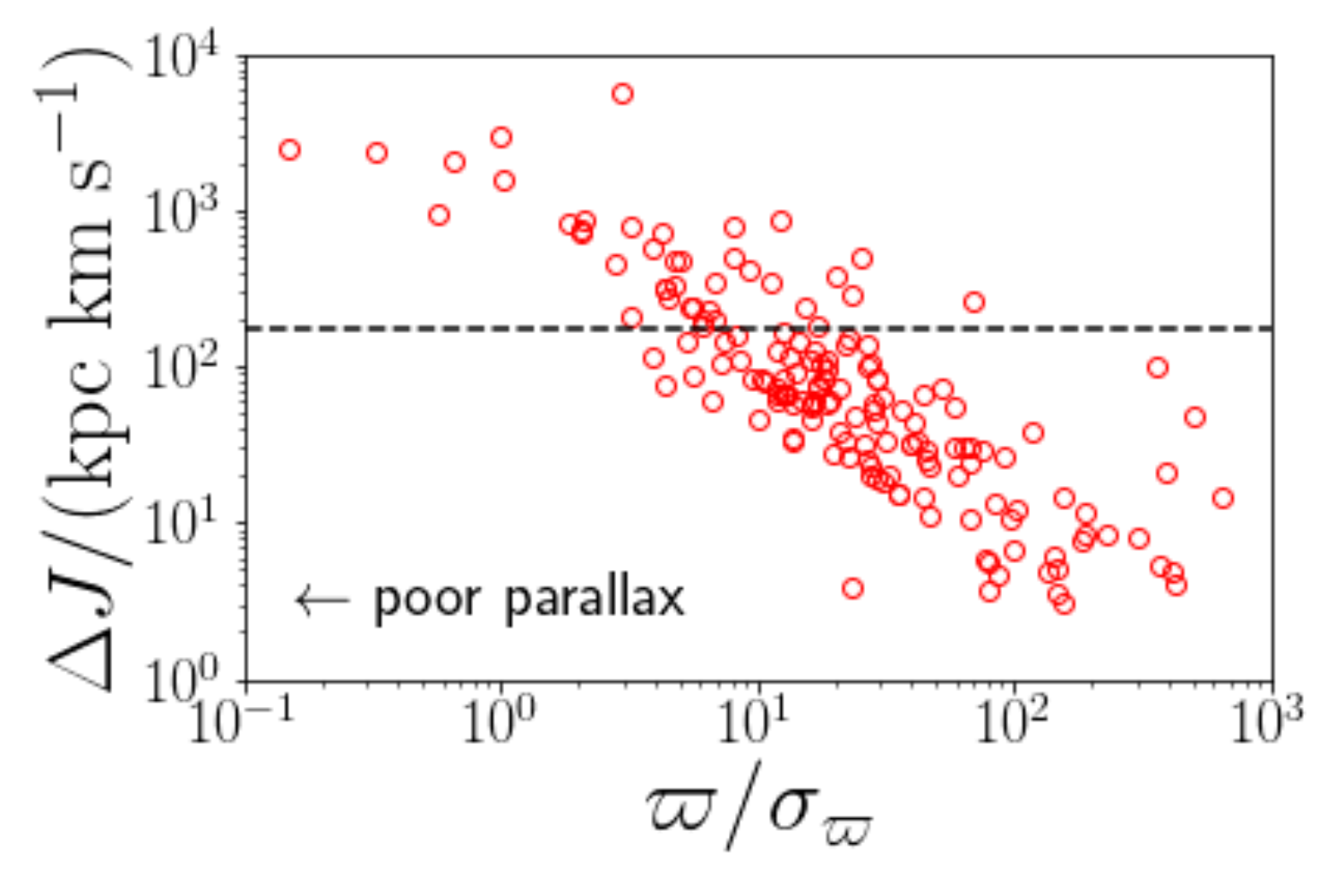}
\caption{
Observed uncertainty in parallax measured by $\varpi / \sigma_\varpi$ 
and the resultant uncertainty in action 
$\Delta J$ for our sample stars. 
We can see that stars with $\varpi / \sigma_\varpi \lesssim 10$ 
result in $\Delta J > \sqrt{3}$ \textcolor{red}{$\sigma_J$}
(horizontal dashed line), 
where we set \textcolor{red}{$\sigma_J$} $= 100 \kpc\kms$. 
Those stars with $\Delta J > \sqrt{3}$ \textcolor{red}{$\sigma_J$}
cannot be adequately handled with 
a conventional Gaussian Mixture Model 
comprising of $K$ Gaussian distributions with internal dispersion \textcolor{red}{$\sigma_J$}. 
To handle this noisy data set, 
we invent the greedy optimistic clustering method. 
}
\label{fig:delta_J}
\end{figure}

\section{Why do we need the greedy optimistic clustering?} \label{sec:why}

Before we perform a clustering analysis, 
we explain the difficulty of using the conventional clustering methods 
for our data set. 
We note that this Section is independent from the main analysis of this paper.

Let us compute the uncertainty in the orbital action 
given the error distribution of the observables in Section \ref{sec:data}. 
For each star in our $r$-II star catalog, 
we randomly draw 
$(\alpha, \delta, \varpi, \mualpha, \mudelta, \vlos)$ 
from the associated error distribution. 
After drawing a large enough number of instances (realizations), 
we discard instances with negative parallax 
and select $200$ instances with $\varpi>0$. 
By converting these observed quantities,\footnote{
The detailed procedure to compute the orbital action 
is the same as in Section \ref{sec:preprocess}. 
} 
we obtain $200$ instances of the orbital action, $\vector{J}=(J_r,J_z,J_\phi)$. 
For each star, 
we denote the standard deviation in each dimension as 
$(\Delta J_r,\Delta J_z,\Delta J_\phi)$. 
The uncertainty in the orbital action for each star is given by
$\Delta J = \sqrt{(\Delta J_r)^2 + (\Delta J_z)^2 + (\Delta J_\phi)^2}$.

The uncertainty $\Delta J$ is dominated by the uncertainty in parallax (or equivalently, distance). 
This is because the parallax uncertainty 
propagates to both the three-dimensional position and 
the two-dimensional tangential velocity. 
Fig.~\ref{fig:delta_J} shows $\Delta J$ 
as a function of the signal-to-noise ratio of the parallax, 
$\varpi/\sigma_\varpi$. 
On average,  
$\Delta J$ increases if the parallax measurement becomes poorer. 
This trend partially justifies previous studies 
in which sample stars are selected based on 
a threshold on $\varpi/\sigma_\varpi$ \citep{Roederer2018}. 
However, using such a threshold means that we end up discarding 
some fraction of the original data, which is not desired.

As we will describe in Section \ref{sec:GMM}, in this paper, 
we want to find clusters 
that typically have an intrinsic dispersion of 
\textcolor{red}{$\sigma_J$} 
$\sim 100 \kpc\kms$ 
in each dimension in the $\vector{J}$-space. 
Thus, if a star's orbital action is associated with 
an uncertainty $\Delta J \gtrsim \sqrt{3} \times 100 \kpc\kms$, 
conventional clustering methods 
may fail to assign the star to the correct cluster. 
In this regard, 
about a quarter of our sample have 
$\Delta J \gtrsim \sqrt{3} \times 100 \kpc\kms$ 
(see Fig.~\ref{fig:delta_J}) 
and therefore they are not suited for conventional clustering analyses.

\section{Preprocessing of the data: Generation of the uncertainty set}
\label{sec:preprocess}

An important difference between this paper and \cite{Roederer2018} 
is that we use 
the greedy optimistic clustering method, 
which allows a clustering analysis of noisy data 
\citep{OkunoHattori2022}. 
This method requires a special preprocessing of the data. 
Namely, for each star, 
we generate $M$ synthetic data points 
that represent the uncertainty in the data. 
We call these synthetic data points 
as an {\it uncertainty set}.\footnote{
In \cite{OkunoHattori2022} we used a term 
{\it empirical uncertainty set}, 
but we omit {\it empirical} for brevity in this paper. 
} \footnote{
The 
uncertainty set is not restricted to be a Monte-Carlo realization of the error distribution. Uncertainty set typically aims at approximating a confidence region, but not approximating the distribution itself. 
For example, 
in the main analysis of this paper, the parallax in the uncertainty set is drawn uniformly from a user-specified region.
}

In this Section, we describe how we generate the uncertainty set. 
In Section \ref{subsec:uncertainty_set_for_observables}, 
we describe the uncertainty set of the observed quantities, $D_i^\mathrm{obs}$, for $i$th star in the catalog.  
In Section \ref{subsec:uncertainty_set_for_J}, 
we convert $D_i^\mathrm{obs}$ to obtain the uncertainty set of the orbital action of $i$th star, $D_i^\mathrm{action}$. 
We note that $D_i^\mathrm{action}$ will be used in our analysis.

\subsection{Uncertain set of the observed quantities}
\label{subsec:uncertainty_set_for_observables}

As mentioned in Section \ref{sec:data}, 
the six-dimensional observables for $i$th star 
$(\alpha, \delta, \varpi, \mualpha, \mudelta, \vlos)_i$ 
are associated with observational uncertainties. 
To represent these uncertainties, 
we generate $M=101$ six-dimensional observable vectors, 
which we call the uncertainty set. 
To be specific, 
the uncertainty set of the observed quantities of $i$th star is expressed as 
\eq{
D^\text{obs}_i = \{ (\alpha, \delta, \varpi, \mualpha, \mudelta, \vlos)_{i,j} \mid -50 \leq j \leq 50 \} .
\label{eq:uncertainty_set_obs}
}
In the following, we describe the procedure to generate 
$j$th instance of the uncertainty set of $i$th star, 
$(\alpha, \delta, \varpi, \mualpha, \mudelta, \vlos)_{i,j}$. 

First, 
we set $j=0$th instance of the uncertainty set 
to be the same as the point-estimate of the observables:
\eq{
(\alpha, \delta, \varpi, \mualpha, \mudelta, \vlos)_{i,0} = (\widehat{\alpha}, \widehat{\delta}, \widehat{\varpi}, \widehat{\mualpha}, \widehat{\mudelta}, \widehat{\vlos})_{i}  .
}

Second, 
for $j \neq 0$, 
we neglect the tiny observational errors in the sky position 
$(\alpha, \delta)$, 
and we set 
\eq{
\alpha_{i,j} &=\widehat{\alpha}_{i}, \\
\delta_{i,j} &=\widehat{\delta}_{i} . 
}

Third, 
for $j \neq 0$,
we draw the line-of-sight velocity and proper motion 
from the corresponding error distribution:
\eq{
{\vlos}_{,i,j} &\sim \mathcal{N}(\widehat{{\vlos}}_{,i}; \sigma_{\mathrm{vlos},i}^2), \\
\begin{bmatrix}
{\mualpha} \\
{\mudelta} \\
\end{bmatrix}_{i,j}
&\sim 
\mathcal{N}
\left(
\begin{bmatrix}
\widehat{{\mualpha}} \\
\widehat{{\mudelta}} \\
\end{bmatrix}_i ; 
\begin{bmatrix}
\sigma^2_{\mu\alpha*}, \;
\rho_\mathrm{corr} \sigma_{\mu\alpha*} {\sigma_{\mu\delta}}  \\
\rho_\mathrm{corr} \sigma_{\mu\alpha*} {\sigma_{\mu\delta}}, \;
\sigma^2_{\mu\delta} \\
\end{bmatrix}_i 
\right).
}
Here, $A\sim B$ indicates that $A$ is drawn from a distribution $B$; 
and $\mathcal{N}(\vector{m}; S)$ corresponds to a Gaussian distribution 
with mean $\vector{m}$ and covariance matrix $S$.

Lastly, 
for $j \neq 0$,
we draw the parallax 
from a uniform distribution within two standard deviation range, 
$\widehat{\varpi}_{i} - 2\sigma_{\varpi,i} 
\leq \varpi_{i,j} \leq 
\widehat{\varpi}_{i} + 2\sigma_{\varpi,i}$. 
For computational simplicity, 
the parallax of $j$th instance of $D^\text{obs}_i$ is determined as
\eq{
\varpi_{i,j} = \widehat{\varpi}_{i} + \sigma_{\varpi,i} \times \frac{j}{25}.
}
With this setting, 
the parallaxes of $j=-50$th, $0$th, and $50$th 
instances of $D^\text{obs}_i$ is 
$\widehat{\varpi}_{i} - 2 \sigma_{\varpi,i}$, 
$\widehat{\varpi}_{i}$, and 
$\widehat{\varpi}_{i} + 2 \sigma_{\varpi,i}$,
respectively.\footnote{
\new{In generating the instances of the parallax, 
we do not use a Gaussian distribution. 
We note that using a Gaussian distribution (truncated within two standard deviation range) would not change the result significantly, as long as the instances of the parallax are densely distributed (e.g., $M \gtrsim 100$).}}
\footnote{
\new{We neglect the covariance between the proper motion and parallax. 
We have confirmed that the inclusion of the covariance hardly affects our results.}}

Up to this step, 
we have generated the uncertainty set $D^\text{obs}_i$ 
without taking care of their physical meanings. 
For example, 
if the parallax uncertainty $\sigma_{\varpi,i}$ is large for star $i$, 
the uncertainty set for this star may include an instance with negative parallax. 
To properly handle negative parallax (which is unphysical) 
and to compensate for the effect of observational uncertainty, 
we introduce a penalty function\footnote{
Suppose that a star has $\varpi_{i,j}>0$ for all $j$. 
Then, the penalty is 0 if $j=0$ 
(if $\varpi_{i,j} = \widehat{\varpi}_{i}$). 
Also, the penalty is $2\lambda$ if $j=\pm50$
($\varpi_{i,j}=\widehat{\varpi}_{i} \pm 2\sigma_{\varpi,i}$). 
} \footnote{
\new{
Among $N=161$ stars in our sample, there is no star whose observed parallax is negative; and there are only seven stars with \texttt{parallax\_over\_error} $< 2$. This means that only seven stars satisfy ($\varpi - 2 \sigma_\varpi < 0$). 
Thus the treatment of rejecting negative parallax is relevant for these seven stars only.}
} 
\eq{
\pen(i,j, \lambda) = 
\begin{cases}
\frac{1}{2} \lambda \left( \frac{j}{25} \right)^2 \;\; & \text{(if $\varpi_{i,j} > 0$)}\\
10^{10} (1 + \lambda)   \;\; & \text{(if $\varpi_{i,j} \leq 0$)} .\\
\end{cases}
} 
The hyper parameter $\lambda \geq 0$ is designed to tune the strength of the penalty (see also equation (\ref{eq:lnL_optimisticEM})). 
In the fiducial analysis of this paper, 
we set $\lambda=0$, 
and therefore 
the detailed implementation of the penalty function is not important in our fiducial analysis.

We repeat the same procedure for all $j$  
and generate the uncertainty set $D^\text{obs}_i$ 
consisting of $M=101$ instances of observable vectors. 
Then we repeat the same procedure for all the stars, 
$i=1,\cdots,N$.

\subsection{Uncertainty set of the orbital action}
\label{subsec:uncertainty_set_for_J}

To perform a clustering analysis in the orbital action space, 
we need the uncertainty set of the orbital action. 
Here we describe the procedure 
to map $D^\text{obs}_i$ 
(equation (\ref{eq:uncertainty_set_obs}))
to the uncertainty set of the orbital action of $i$th star 
\eq{
D^\text{action}_i = 
\{ \vector{J}_{i,j} \mid -50 \leq j \leq 50 \} . 
\label{eq:uncertainty_set_J}
} 

For each $(i,j)$, 
we first convert the observables 
$(\alpha, \delta, \varpi, \mualpha, \mudelta, \vlos)_{i,j}$ 
into the position and velocity $(\vector{x}, \vector{v})_{i,j}$ 
in the Galactocentric Cartesian coordinate. 
In this step, we assume the position and velocity of the Sun 
as described in \cite{Hattori2021MNRAS.508.5468H}. 
Then we map 
$(\vector{x}, \vector{v})_{i,j}$ 
to the orbital action 
$\vector{J}_{i,j} = \vector{J}(\vector{x}_{i,j}, \vector{v}_{i,j}) = (J_r,J_\phi,J_z)_{i,j}$. 
In computing the orbital action, 
we assume the gravitational potential of the Milky Way 
in \cite{McMillan2017} 
and 
we use the AGAMA package \citep{Vasiliev2019_AGAMA}.

In the top row in Fig.~\ref{fig:action_distribution}, 
the gray dots show 
the distribution of $D_i^\text{action}$ 
for our entire sample. 
We see that the uncertainty set for each star 
typically shows a banana-like shape, 
reflecting the fact that the parallax uncertainty dominates 
the uncertainty in $\vector{J}$.

\section{Greedy optimistic GMM} \label{sec:GMM}

\subsection{Gaussian Mixture Model in action space}

In this paper, 
we assume the Gaussian Mixture Model (GMM) 
to describe the intrinsic distribution of $\vector{J}$, 
which consists of $K$ 
isotropic Gaussian distributions 
with identical one-dimensional dispersion \textcolor{red}{$\sigma^2_J$}.
Under this assumption,\footnote{
\new{
We discuss the implication of this assumption 
in Section \ref{sec:caveat_action_distribution}.}
} 
the probability distribution of a randomly chosen star is expressed as 
\eq{\label{eq:GMM}
P(\vector{J} \mid \theta) = \sum_{k=1}^{K} 
\pi_k \mathcal{N}(\vector{J} \mid \langle \vector{J} \rangle_k, \sigma_J^2 \vector{I}).
}
Here, $\vector{I}$ denotes identity matrix, $k=1,\cdots,K$ is the index for the normal distributions representing clusters, with the model parameters $\theta=\{(\pi_k,\langle \vector{J} \rangle_k) \mid k=1,\cdots,K; \pi_k \ge 0; \sum_{k=1}^{K} \pi_k=1\}$ to be estimated. 
In GMM clustering shown below, $\langle \vector{J} \rangle_k \equiv (\langle J_r \rangle_k, \langle J_\phi \rangle_k, \langle J_z \rangle_k )$ corresponds to the centroid of $k$th cluster.

\subsection{Conventional GMM}

Let us first consider a case 
where we use the point-estimate of action $\widehat{\vector{J}}_i$
for each star $i$ to perform a clustering analysis. 
This case corresponds to an ideal case 
where we have no observational uncertainties. 

In this case, 
the logarithmic likelihood of the data $D = \{ \widehat{\vector{J}}_i \mid i=1,\cdots,N \}$ 
given 
the model parameters $\theta =\{(\pi_k, \langle \vector{J} \rangle_k ) \mid k=1,\cdots,K \}$ 
is given by 
\eq{
\ln L (D | \theta ) 
= 
\sum_{i=1}^{N} \ln 
\left\{
\sum_{k=1}^{K} \pi_k \mathcal{N} (\widehat{\vector{J}}_i \mid \langle \vector{J} \rangle_k, \sigma_J^2 \vector{I}) 
\right\} . \label{eq:lnL_standardEM}
}

The expectation-maximization (EM) algorithm tries to find the 
best parameters by iteratively improving them with the following steps \citep{Dempster10.2307/2984875,Neal1998,Bishop2006PRML}. 
\begin{itemize}
\item (E step) 
For a given set of $(\{ \pi_k, \langle \vector{J} \rangle_k \} \mid k=1,\cdots,K )$, 
compute $\gamma_{ik}$ and $N_k$ given by 
\eq{
\begin{cases}
\gamma_{ik} = 
{\displaystyle
\frac
{                 \pi_k \mathcal{N} (\widehat{\vector{J}}_i \mid \langle \vector{J} \rangle_k, \sigma_J^2 \vector{I}) }
{\sum_{\ell=1}^{K} \pi_\ell \mathcal{N} (\widehat{\vector{J}}_i \mid \langle \vector{J} \rangle_\ell, \sigma_J^2 \vector{I}) }
}\\
N_k = {\displaystyle \sum_{i=1}^{N} \gamma_{ik} }.
\end{cases}
}
\item (M step) 
Update $(\{ \pi_k, \langle \vector{J} \rangle_k \} \mid k=1,\cdots,K )$ with the following expressions:
\eq{
\begin{cases}
{\displaystyle
\langle \vector{J} \rangle_k \leftarrow 
\frac{1}{N_k} 
\sum_{i=1}^{N} \gamma_{ik} \widehat{\vector{J}}_i 
}, \\
{\displaystyle
\pi_k \leftarrow {\displaystyle \frac{N_k}{N} }
}.
\end{cases}
}
\end{itemize}
We note that $\gamma_{ik}$ is called responsibility, 
which quantifies the contribution of $i$th star 
to $k$th cluster. 
The quantity $N_k$ represents 
the effective size of $k$th cluster. 
The EM algorithm iteratively updates $\gamma_{ik}$ 
and the parameters such that the likelihood function 
is monotonically improved until convergence.

\subsection{Greedy optimistic GMM}

The above-mentioned conventional GMM  
does not take into account the uncertainty in the data. 
Therefore, the conventional GMM does not work when $\vector{J}$ is associated with large uncertainties. 
This is why we introduce a new flavor of clustering method, 
the greedy optimistic clustering method \citep{OkunoHattori2022}. 
An intuitive illustration of this method is given 
in Appendix \ref{sec:idea}.

The key ideas of our method are as follows: 
\begin{itemize}
\item[1.]
We make an {\it optimistic} assumption that the true orbital actions of $N$ stars 
$\{ \vector{J}_{i}^\mathrm{true} \mid i=1,\cdots,N \}$
are highly clustered in the $\vector{J}$-space. 

\item[2.]
Given the uncertainty set $D^\text{action}_i$ 
consisting of $M$ instances of the orbital action of $i$th star $\{ \vector{J}_{i,j} \}$, 
we assume that one of the instances  $\vector{J}_{i, \beta_i}$ is {\it very close} to the true orbital action $\vector{J}_{i}^\mathrm{true}$. 
Here, $\beta_i$ is the index of the `best' instance (among $M$ instances) in the uncertainty set $D^\text{action}_i$. 

\item[3.]
Given the two assumptions above, 
we {\it greedily} find the best configuration $\{ \vector{J}_{i, \beta_i} \mid i=1,\cdots,N \}$ of $N$ stars 
such that their distribution is most highly clustered in the $\vector{J}$-space. 
We call the best instance $\vector{J}_{i, \beta_i}$ 
the {\it greedy optimistic estimate} of the orbital action for star $i$. 

\end{itemize}

In the greedy optimistic clustering algorithm, 
we simultaneously estimate the cluster parameters 
$\{ (\pi_k, \langle \vector{J} \rangle_k ) \mid k=1,\cdots,K \}$ 
as well as the best indices $\{ \beta_i \mid i=1,\cdots,N \}$ 
such that the clustering of $N$ points $\{ \vector{J}_{i, \beta_i} \mid i=1,\cdots,N \}$ is most enhanced. 
For this purpose, we define an object function 
to be maximized 
\eq{
f (D^\text{action} | \theta ) 
&= 
\sum_{i=1}^{N} \ln 
\left\{
\sum_{k=1}^{K} \pi_k \mathcal{N} (\vector{J}_{i, \beta_i} \mid \langle \vector{J} \rangle_k, \sigma_J^2 \vector{I}) 
\right\} \nonumber \\ 
&\hspace{3em}- \sum_{i=1}^{N} \pen(i,\beta_i, \lambda) .
\label{eq:lnL_optimisticEM}
}
Here, $\lambda \geq 0$ is a hyper parameter that controls the 
strength of the penalty term. 
Also, \eq{
D^\text{action} = \{D^\text{action}_i \mid i=1,\cdots,N \}
}
is the uncertainty set of the entire sample, 
and 
\eq{
\theta = \{ 
\;&\{ (\pi_k, \langle \vector{J} \rangle_k ) \mid k=1,\cdots,K \}, \nonumber\\
\;&\{ \beta_i \mid i=1,\cdots,N \} \; \}
}
is the model parameters. 
We note that the model parameters now include $\{ \beta_i \}$.

To find the solution, we use an iteration algorithm, 
which we refer to as the GOEM algorithm.
In the GOEM algorithm, 
we introduce a `GO step' (greedy optimistic step) in addition to 
the E and M steps: 
\begin{itemize}
\item (GO step) 
For a given set of $\{ ( \pi_k, \langle \vector{J} \rangle_k ) \mid k=1,\cdots,K \}$, 
update $\{ \beta_i \mid i=1,\cdots,N \}$ so that the object function $f(D^\text{action}|\theta)$ is maximized.  
\item (E step) 
For a given set of $\{ ( \pi_k, \langle \vector{J} \rangle_k ) \mid k=1,\cdots,K \}$ 
and $\{ \beta_i \mid i=1,\cdots,N \}$, 
compute $\gamma_{ik}$ and $N_k$ given by 
\eq{
\begin{cases}
\gamma_{ik} = 
{\displaystyle
\frac
{               \pi_k \mathcal{N} (\vector{J}_{i, \beta_i} \mid \langle \vector{J} \rangle_k, \sigma_J^2 \vector{I}) }
{\sum_{\ell=1}^{K} \pi_\ell \mathcal{N} (\vector{J}_{i, \beta_i} \mid \langle \vector{J} \rangle_\ell, \sigma_J^2 \vector{I}) } 
}\\
{\displaystyle
N_k = \sum_{i=1}^{N} \gamma_{ik}.
}
\end{cases}
}
\item (M step) 
Update $\{ ( \pi_k, \langle \vector{J} \rangle_k ) \mid k=1,\cdots,K \}$ with the following expressions:
\eq{
\begin{cases}
{\displaystyle
\langle \vector{J} \rangle_k \leftarrow 
\frac{1}{N_k} 
\sum_{i=1}^{N} \gamma_{ik} \vector{J}_{i, \beta_i} ,
}\\
{\displaystyle
\pi_k \leftarrow \frac{N_k}{N} .
}
\end{cases}
}
\end{itemize}
We note that, in the GO step, 
we can optimize $\beta_i$ separately for each $i$. 

The GO, E, and M steps always improve the object function 
$f(D^\text{action}|\theta)$. 
Just like the conventional EM algorithm, 
the GOEM algorithm would converge to a local maximum, 
and there is no guarantee that the derived solution is the global maximum. 
To find a better solution, 
we try many initial conditions of the parameters and 
use the split-and-merge algorithm \citep{Ueda1998_NIPS1998_253f7b5d}.

\subsection{Difference between the conventional GMM and the greedy optimistic GMM}

In the greedy optimistic GMM, 
for each star $i$, 
we have a freedom to choose an instance from the uncertainty set $D_i^\text{action}$. 
In a sense, 
we can {\it move} the locations of the data points according to their uncertainties 
and can search for the best configuration 
such that the $N$ data points are most highly clustered. 
This characteristics is in contrast to the conventional GMM, 
where the data point is stuck to its point estimate. 
Note that the greedy optimistic GMM is distinct from extreme deconvolution~\citep{Bovy2011AOAS}, that further assumes that the centroid $\langle \vector{J} \rangle_k$ of $k$th cluster also follows a distribution, i.e., cluster center has uncertainty.

In the greedy optimistic GMM method, 
we introduce a penalty term in the object function (second term in equation (\ref{eq:lnL_optimisticEM})). 
This penalty term is designed to favor the instances 
whose parallax is closer to the point-estimate of the parallax. 
The strength of the penalty term is governed by the hyper parameter $\lambda$. 
On the one hand, 
in the limiting case of $\lambda \to \infty$, 
the optimal solution satisfies $\beta_i = \mathrm{argmin}_{\beta} \{ \pen(i,\beta,\lambda) \}$. 
In this limit, the greedy optimistic GMM reduces to the conventional GMM, 
because we no longer have the freedom to choose among $M$ instances. 
Thus, the greedy optimistic GMM is a generalization of the conventional GMM. 
On the other hand, 
in the limiting case of $\lambda = 0$, 
greedy optimistic GMM 
treats all the instances equally. 
Interestingly, 
the mock data analysis in \cite{OkunoHattori2022} 
suggests that the performance of the clustering 
is reasonably good when we set $\lambda = 0$.
Indeed, we adopt $\lambda = 0$ in the fiducial analysis of this paper.

\begin{figure*}
\centering
\includegraphics[width=6.3in]
{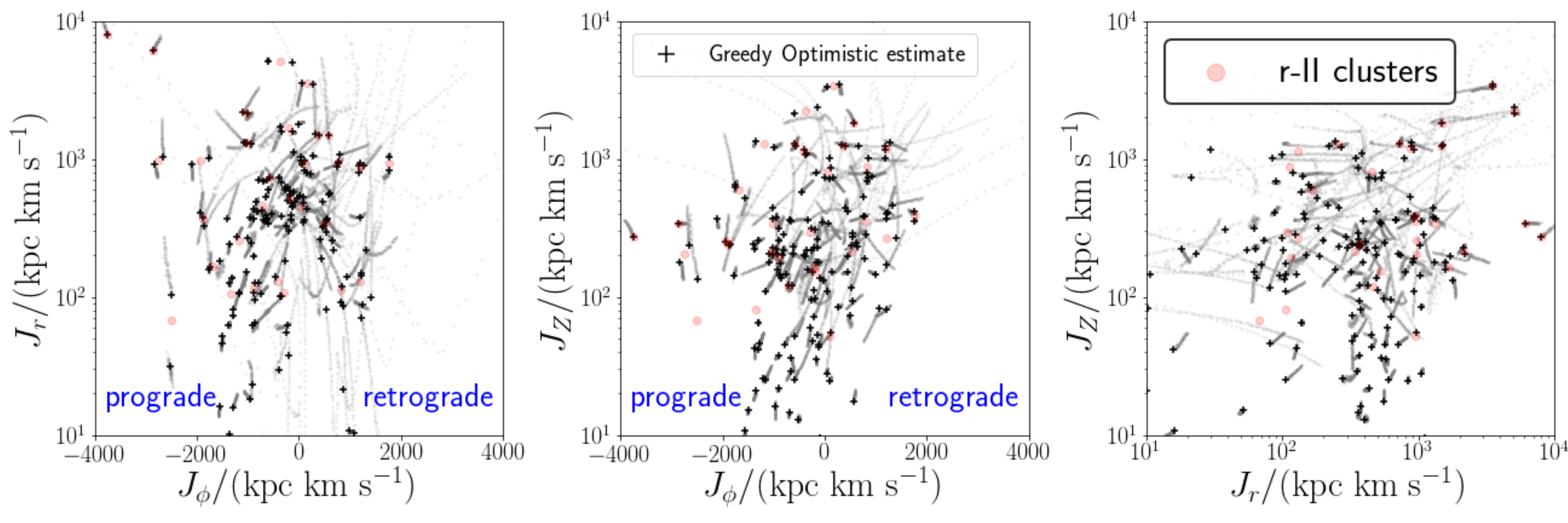}\\
\includegraphics[width=6.3in]
{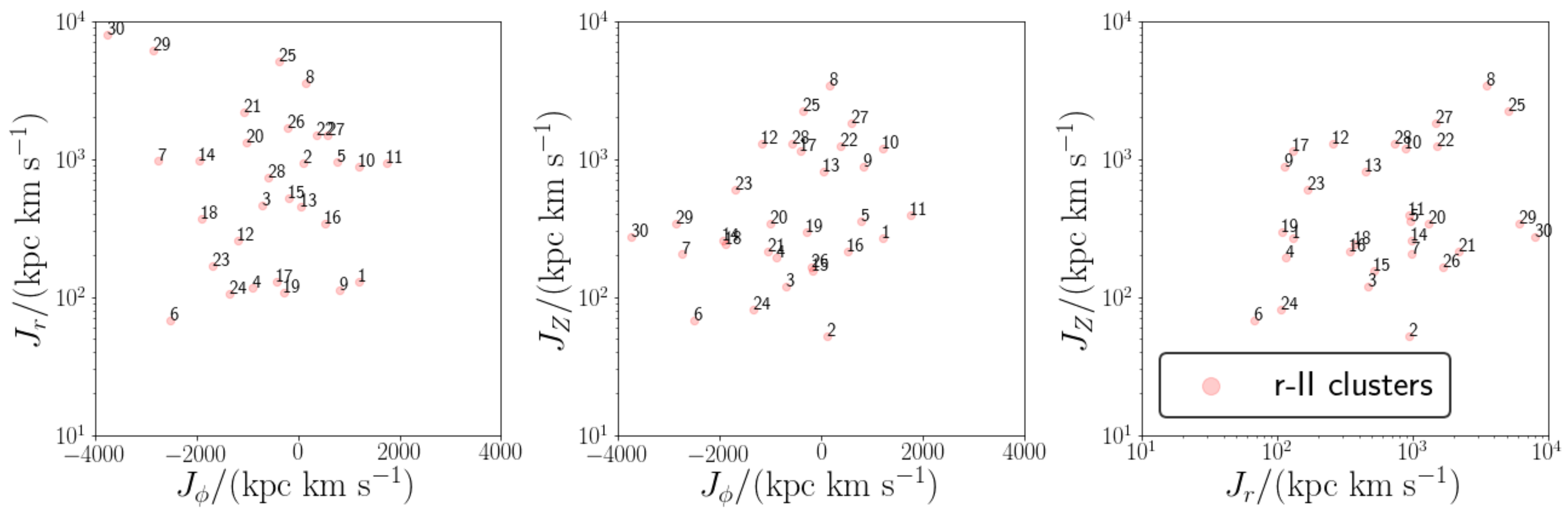}\\
\includegraphics[width=6.3in]
{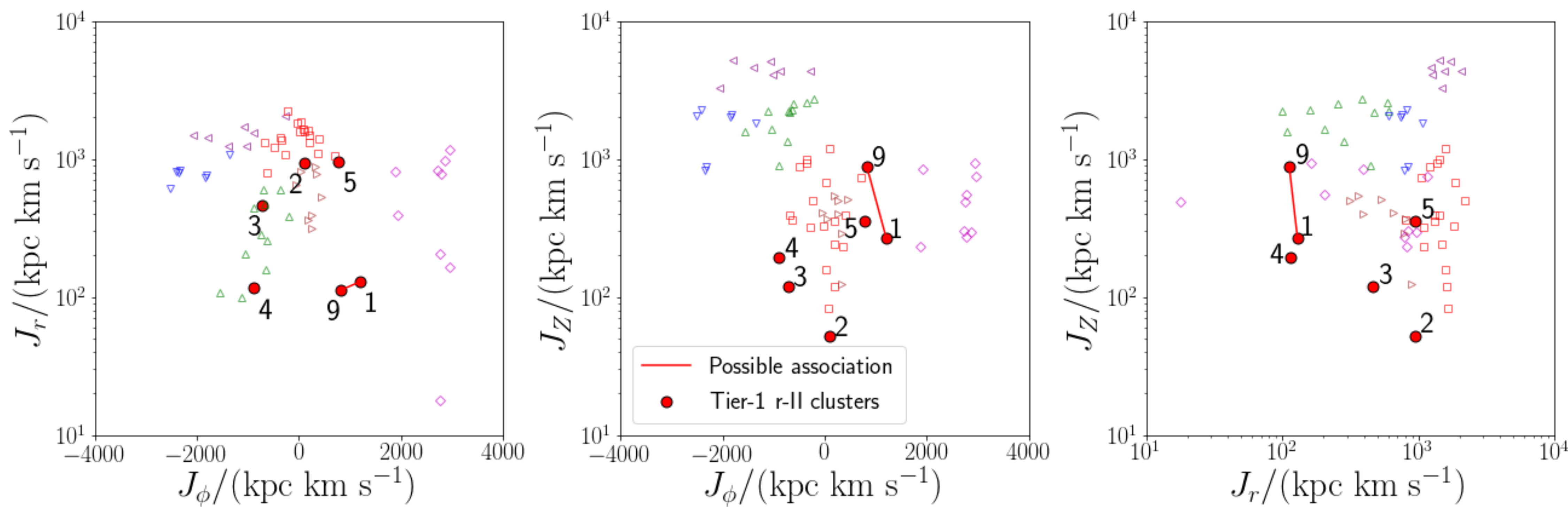}\\
\includegraphics[width=6.3in]
{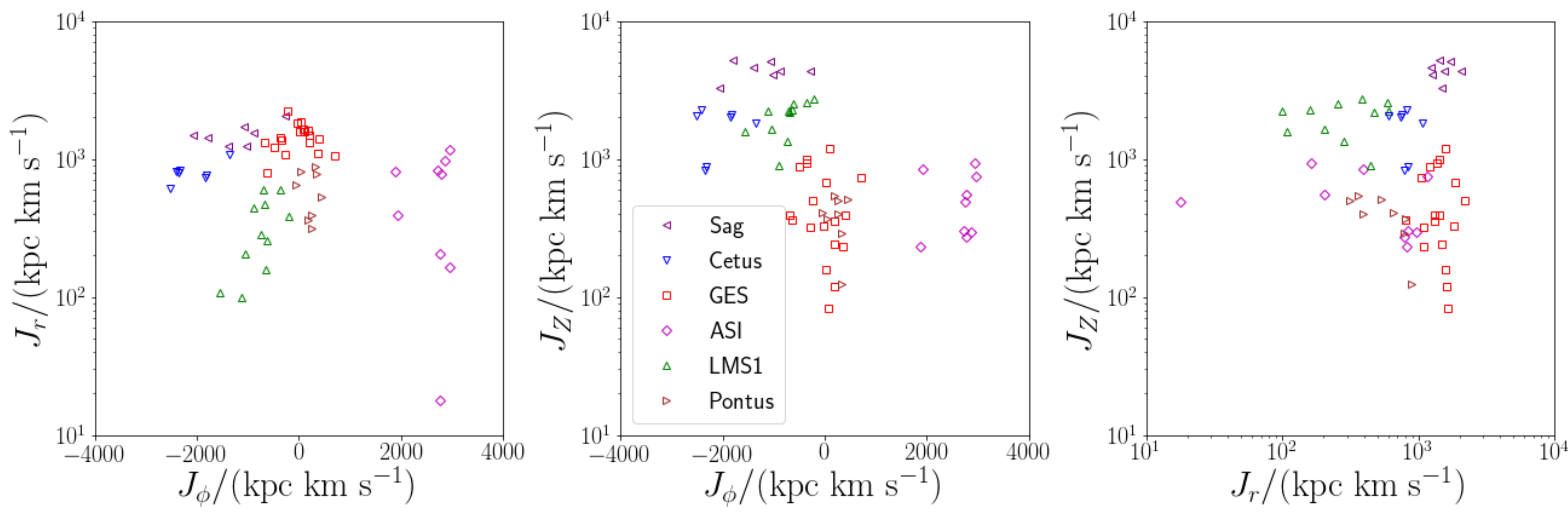}
\caption{
(Top row) 
Distribution of $N=161$ $r$-II stars in the orbital action space. 
The gray dots represent the uncertainty set, 
which typically has a banana-like shape. 
The black cross represents the greedy optimistic estimate of the orbital action in our fiducial analysis. 
(2nd row) 
Distribution of our $r$-II clusters in our fiducial analysis. 
The number indicates $k$ of the $r$-II cluster H22:DTC-$k$ ($1 \leq k \leq 30$). 
(3rd row) 
Distribution of Tier-1 clusters (i.e., most confident clusters). 
Two $r$-II clusters 
H22:DTC-$1$ and $9$ 
are possibly associated (see Section \ref{subsec:new_merger_group}). 
(Bottom row) 
Orbital actions of the stellar streams and globular clusters  associated with the six merger groups 
(the data are based on \citealt{Malhan2022ApJ...926..107M} 
and \citealt{Vasiliev2021MNRAS.505.5978V}). 
}
\label{fig:action_distribution}
\end{figure*}

\section{Analysis and results} \label{sec:analysis}

After preparing for the uncertainty set $\{ D_i^\text{action} \}$, 
we perform a greedy optimistic clustering analysis for our sample stars.

\subsection{List of hyper parameters} \label{sec:hyper_parameters}

In the greedy optimistic GMM, 
we need to fix three hyper parameters $(\sigma_J, \lambda, K)$, 
which will be described in the following.

\subsubsection{Internal dispersion of the cluster: $\sigma_J$}

The hyper parameter \textcolor{red}{$\sigma_J$} 
determines the internal dispersion of a cluster in $\vector{J}$-space. 
In realistic galaxy formation simulations, 
completely disrupted low-mass stellar systems 
typically have a dispersion of $\sim (100$-$200) \kpc \kms$ 
in $\vector{J}$-space \citep{Bullock2005ApJ...635..931B}.  
Motivated by this result, we fix 
\eq{
\sigma_J = 100 \kpc \kms .
}

This value of \textcolor{red}{$\sigma_J$} is also supported by the observed properties 
of UFDs hosting $r$-II stars,
such as Reticulum~II,
which has an internal velocity dispersion of $\sigma_\mathrm{v} \sim 3 \kms$
\citep{Koposov2015,Walker2015,Minor2019}.
If a stellar system with a similar velocity dispersion 
is disrupted in the inner part of the Milky Way 
($r_\text{disruption}\sim 10-30 \kpc$), 
we expect that the stars stripped from the system 
typically have a spread in action of $\sigma_\mathrm{v} \times r_\text{disruption} \sim (30-90) \kpc \kms$.

In principle, 
we can make \textcolor{red}{$\sigma_J$} 
as a free parameter to be fitted in the analysis. 
However, we fix \textcolor{red}{$\sigma_J$} to stabilize our analysis, 
following the mock data analysis in \cite{OkunoHattori2022}.

\subsubsection{Number of clusters: $K$}

The number of clusters $K$ is hard to choose, 
because fixing $K$ is equivalent to fixing the typical size of a cluster, $N/K$, 
which is unknown. 
After some experiments, 
we adopt six values, $K=10,20,30,40,50$, and $60$.

\subsubsection{Strength of the penalty term: $\lambda$}

The hyper parameter $\lambda$ governs 
the penalty term in our greedy optimistic GMM method. 
To see how our result is affected by $\lambda$, 
we adopt six values, 
$\lambda = 0, 10^{-4}, 10^{-3}, 10^{-2}, 1, 10^2$. 
On the one hand, $\lambda = 0$ 
corresponds to the most optimistic case 
where we pay the least respect to the observed parallax. 
On the other hand, 
the case with $\lambda = 10^2$ 
is almost equivalent to the conventional GMM.

\subsection{Fiducial hyper parameters}

As mentioned above, 
we try 36 sets of hyper parameters $(K, \lambda)$ 
while fixing \textcolor{red}{$\sigma_J$}$= 100 \kpc\kms$. 
Among 36 combinations, 
we choose the fiducial set of hyper parameters to be 
$(K, \lambda, \sigma_J) = (30, 0, 100 \kpc\kms)$ 
with the following procedure.

First, for each set of $(K, \lambda)$, 
we find the best solution 
$\{ (\pi_k, \langle \vector{J} \rangle_k) \mid k=1,\cdots,K \}$ and 
$\{ \beta_i                               \mid i=1,\cdots,N \}$  
that maximize the object function (equation (\ref{eq:lnL_optimisticEM})). 
Here, we use 400 initial conditions to efficiently 
explore the parameter space; 
and use the split-and-merge method \citep{Ueda1998_NIPS1998_253f7b5d}
to avoid local optimums.

After finding the best solution, 
we assign each star to the nearest centroid. 
Namely, $i$th star is assigned to $c_i$th cluster such that 
\eq{
c_i = \mathrm{argmin}_k || \vector{J}_{i, \beta_i} - \langle \vector{J} \rangle_k ||.
}
We count the number of member stars 
$N_{\mathrm{member},k}$ for $k$th cluster.\footnote{
We note that $N_{\mathrm{member},k} \simeq N_k (=\pi_k N)$. 
$N_{\mathrm{member},k}$ is an integer resulting from the hard assignment, 
while $N_k$ is a real number resulting from a soft assignment. 
} 
We note that this hard-assigment sometimes results in 
clusters with no members assigned, 
especially when $\lambda \leq 10^{-2}$ and $K \geq 50$. 
This is one of the reasons why we do not explore $K>60$ in our analysis.

If $k$th cluster has $N_{\mathrm{member},k}\geq2$ members, 
we compute the internal dispersion in $(J_r, J_z, J_\phi)$: 
\eq{
(\sigma_{Jr, k})^2 &= 
\frac{1}{N_{\mathrm{member},k}} \sum_{i \in \text{$k$th cluster}} || (\vector{J}_{i, \beta_i} - \langle \vector{J} \rangle_k)_r ||^2  \\
(\sigma_{Jz, k})^2 &= 
\frac{1}{N_{\mathrm{member},k}} \sum_{i \in \text{$k$th cluster}} || (\vector{J}_{i, \beta_i} - \langle \vector{J} \rangle_k)_z ||^2  \\
(\sigma_{J\phi, k})^2 &= 
\frac{1}{N_{\mathrm{member},k}} \sum_{i \in \text{$k$th cluster}} || (\vector{J}_{i, \beta_i} - \langle \vector{J} \rangle_k)_\phi ||^2 ,
}
and the total internal dispersion:
\eq{
\sigma_k = \sqrt{ \frac{1}{3} \left[ 
(\sigma_{Jr, k})^2 + (\sigma_{Jz, k})^2 + (\sigma_{J\phi, k})^2 
\right] } .  
} 
Here, $(\vector{J}_{i, \beta_i} - \langle \vector{J} \rangle_k)_q$ corresponds to the $q$-component of the vector 
$(\vector{J}_{i, \beta_i} - \langle \vector{J} \rangle_k)$, where $q=r,z,\phi$.

For each set of hyper parameters $(K, \lambda)$,  
we compute the median of $\{ \sigma_k \}$ for clusters with $N_{\mathrm{member},k}\geq2$. 
When $\lambda$ is fixed, 
the median value of $\{ \sigma_k \}$ is a decreasing function of $K$, 
so we choose the optimal $K$ such that this median 
is closest to \textcolor{red}{$\sigma_J$}$(=100 \kpc \kms)$, 
which is the fixed hyper parameter. 
As a result, we found that $K=30$ is optimal for $\lambda \leq 1$, 
while $K=50$ is optimal for $\lambda = 10^{2}$. 
This result indicates that we need $K=30$ clusters to represent the action distribution 
for our $N=161$ stars if we adopt the greedy optimistic clustering, 
almost independent of the strength of the penalty term (unless the penalty term is too strong). 
Thus, we choose $K=30$ as the fiducial value. 
With $K=30$, we have a freedom to choose $0 \leq \lambda \leq 1$. 
For simplicity, we choose $\lambda=0$ as the fiducial value, 
motivated by the mock analysis in \cite{OkunoHattori2022}.

\subsection{Fiducial result} \label{subsec:fiducial_result}

Here, we focus on the result obtained 
for the fiducial set of hyper parameters, 
$(K, \lambda, \sigma_J) = (30, 0, 100 \kpc\kms)$. 
We call the clusters of $r$-II stars in our fiducial result 
simply as $r$-II clusters. 
Also, following recent convention \citep{Yuan2020ApJ...891...39Y, Gudin2021ApJ...908...79G}, 
we label the $k$th $r$-II cluster as H22:DTC-$k$, where DTC stands for dynamically tagged cluster.\footnote{
We use dynamically tagged {\it clusters} rather than dynamically tagged {\it groups}, 
in order not to be confused with the merger groups. 
}

Since we are most interested in the orbital properties of the $r$-II stars and $r$-II clusters, 
we visualize their distribution in the $\vector{J}$-space. 
The top row in Fig.~\ref{fig:action_distribution} 
shows the distribution of the greedy optimistic estimate 
$\vector{J}_{i,\beta_i}$ (black cross) for $N=161$ $r$-II stars, 
along with their uncertainty set $D_i^\text{action}$ (gray dots).\footnote{
\new{
For some stars,
the greedy optimistic estimate is located near the edge of the uncertainty set. 
We point out that this phenomenon is 
not troublesome in our clustering analysis, 
as described in Appendix \ref{sec:edge}. 
}
} 
The second row in Fig.~\ref{fig:action_distribution} 
shows the distribution of centroids of the orbital action, $\langle \vector{J} \rangle_k$, for $K=30$ clusters.

Table~\ref{table:clusters} summarizes 
the properties of $r$-II clusters, 
including 
the cluster size $N_{\text{member},k}$, 
orbital action of the centroid $\langle \vector{J} \rangle_k$, 
the total internal dispersion of the orbital action $\sigma_k$, 
and the chemical properties. 
As seen in Table~\ref{table:clusters}, 
$\sigma_k$ is approximately $100 \kpc\kms$, 
which is a natural consequence of the way we choose our hyper parameter $K$. 
We note that the chemical information is not used in our clustering analysis, 
but will be used to validate our clustering results (see Section \ref{subsec:percentile}).

Tables~\ref{table:members_all_info} and \ref{table:members_additional_info}
present the detailed chemical and dynamical information of 
individual $r$-II stars analyzed in this paper. 
Due to their length, we put them in Appendix \ref{sec:list_of_stars}.

Table~\ref{table:members_all_info} 
lists the chemical and kinematical properties 
of the member stars for each cluster. 
For an easy comparison of the results in \cite{Roederer2018} with ours, 
we add a column \texttt{R18} in Table~\ref{table:members_all_info}, 
which indicates those stars that are included in groups A-H in \cite{Roederer2018}. 
We will comment on this comparison in Section \ref{subsec:consistency_with_R18}. 
\new{
Table~\ref{table:members_all_info} also 
lists the reference for the chemical data.
}
The orbital action listed in Table~\ref{table:members_all_info} 
is the greedy optimistic estimate $\vector{J}_{i,\beta_i}$
(shown on the top row of Fig.~\ref{fig:action_distribution}). 
For a reference, we also list the orbital energy $E$, 
which is computed as $E=\vector{v}_{i,\beta_i}^2/2 + \Phi(\vector{x}_{i,\beta_i})$, 
where $\Phi$ is the Milky Way's gravitational potential \citep{McMillan2017}. 
Unlike \cite{Roederer2018}, 
we do not use $E$ in our clustering, 
because $E$ is a function of the orbital action and 
therefore it provides duplicated information. 
The last column in Table~\ref{table:members_all_info} 
is the quantity \texttt{parallax\_over\_error} ($=\varpi/\sigma_\varpi$) 
taken from Gaia EDR3 
(and not corrected for the zero-point offset of the parallax). 
We list this information here to 
make it easier to understand 
which member stars benefit from the greedy optimistic clustering. 
For example, 
a star with poor parallax measurements $(\varpi/\sigma_\varpi \lesssim 10)$ 
may not be assigned to the right cluster with conventional clustering methods; 
while 
a star with good parallax measurements $(\varpi/\sigma_\varpi \gtrsim 10)$ is 
likely assigned to the right cluster independent of the clustering methods.
\new{
(See Appendix~\ref{sec:standardGMM} where we analyze $N=119$ $r$-II stars with $\varpi/\sigma_\varpi > 10$ by using standard GMM.)
}

Table~\ref{table:members_additional_info} lists 
additional information of the member stars in each cluster. 
All quantities in this table are not used in the clustering analysis, 
but they are presented to offer an intuitive understanding of 
the dynamical properties of stars. 
The Galactocentric position and velocity listed in this table 
correspond to $(\vector{x}_{i,\beta_i}, \vector{v}_{i,\beta_i})$, 
respectively. 
By using the position-velocity pair, 
we integrate the orbit for $100 \Gyr$ 
to obtain the 
pericentric radius ($r_\mathrm{peri}$), 
apocentric radius ($r_\mathrm{apo}$), 
maximum distance from the Galactic disk plane ($z_\mathrm{max}$), 
and orbital eccentricity $ecc = (r_\mathrm{apo}-r_\mathrm{peri})/(r_\mathrm{apo}+r_\mathrm{peri})$. 
Because the member stars of a given cluster 
have similar orbital actions, 
they also have similar values of 
$(r_\mathrm{peri}, r_\mathrm{apo}, z_\mathrm{max}, ecc)$. 
In contrast, 
the member stars of a given cluster 
have very different $(\vector{x}_{i,\beta_i}, \vector{v}_{i,\beta_i})$.  
This result is consistent with a scenario wherein the progenitor system 
of a cluster accreted to the Milky Way long ago, 
and the stars stripped from the progenitor system 
are completely phase-mixed today.

\subsection{Consistency with \cite{Roederer2018}} 
\label{subsec:consistency_with_R18}

Because we extend the analysis in \cite{Roederer2018}, 
we briefly check the consistency between 
their results and ours. 
As summarized in Table~\ref{table:clusters}, 
all the stars in groups 
A, D, E, G, and H in \cite{Roederer2018} 
are found in clusters 
H22:DTC-$3, 2, 15, 5$, and $15$, respectively. 
Also, 
the majority of the members in groups 
B (3 stars out of 4 stars), 
C (3 out of 4), and  
F (2 out of 3) 
are found in 
H22:DTC-$9, 4$, and $3$, respectively. 
These results are reassuring in that 
\cite{Roederer2018} and this work 
apply different clustering methods 
to data sets with different quality 
and still mostly agree with each other.

Intriguingly, 
groups A and F are included in a single cluster 
H22:DTC-3
in our analysis; 
and 
groups E and H (and another star from group F) 
are included in a single cluster 
H22:DTC-15. 
In our analysis,
the clusters H22:DTC-$3$ and $15$ are the biggest clusters, 
containing $N_{\text{member},k}=18$ stars each. 
Our result may reflect either of the two possibilities. 
The first possibility is that 
our new clustering method is superior to conventional methods.  
(Namely, our method can find a cluster 
that is seen as multiple clusters with conventional methods due to the observational uncertainty.) 
The second possibility is that 
our method is too optimistic. 
(Namely, multiple clusters with a slightly different orbital properties 
are mistakenly regarded as a single one due to the optimistic nature of our method.) 
Although we do not have a decisive conclusion on 
which of these possibilities is correct, 
it is worth mentioning that 
the cluster H22:DTC-3 
is one of the six clusters in our analysis 
that we have the highest confidence based on the tight distribution in [Fe/H] and [Eu/H] 
(see Section \ref{subsubsec:tier1}). 
\new{
According to our additional analysis using high-quality data alone (see Appendix \ref{sec:standardGMM}), 
10 stars among 18 stars in H22:DTC-3 are more likely to be associated with each other than the remaining 8 stars. 
}
Also, 
the cluster H22:DTC-15 
has a tight distribution in [Fe/H] and [Eu/H] 
\new{
if we remove one metal-rich outlier star}  
(see Appendix \ref{subsec:tier4}).

\begin{deluxetable*}{rr ccr cccc l }
\tablecaption{Properties of the clusters 
\label{table:clusters}}
\rotate 
\tablewidth{0pt}
\tabletypesize{\scriptsize}
\tablehead{
\colhead{Cluster name} &
\colhead{$N_{\mathrm{member},k}$} &
\colhead{$(J_r, J_z, J_\phi)$} &
\colhead{$(\sigma_{Jr}, \sigma_{Jz}, \sigma_{J\phi})$} &
\colhead{$\sigma_{k}$} &
\colhead{$\langle \mathrm{[Fe/H]} \rangle$} &
\colhead{$\sigma_\mathrm{[Fe/H]}$ ($q_\mathrm{[Fe/H]}$)} &
\colhead{$\langle \mathrm{[Eu/H]} \rangle$} &
\colhead{$\sigma_\mathrm{[Eu/H]}$ ($q_\mathrm{[Eu/H]}$)} &
\colhead{Comment$^{\mathrm{(a)}}$} \\
\colhead{(H22:DTC-$k$)} &
\colhead{} &
\colhead{$\kpc \kms$} &
\colhead{$\kpc \kms$} &
\colhead{$\kpc \kms$} &
\colhead{dex} &
\colhead{dex (percentile)} &
\colhead{dex} &
\colhead{dex (percentile)} 
}
\startdata
H22:DTC-{\Iiiii}  \phantom{14} & {9} & {$(129, 265, 1209)$} & {$(107, 140, 112)$} & {121} & {$-2.78$} & {0.22$\;$ (0.56)} & {$-1.64$} & {0.32$\;$ (5.06)} & {Tier-1 -- New}\\
H22:DTC-{\Oviiii} \phantom{09} & {9} & {$(942, 52, 102)$} & {$(113, 49, 104)$} & {93} & {$-1.65$} & {0.25$\;$ (1.08)} & {$-0.62$} & {0.22$\;$ (0.78)} & {Tier-1 -- D$^{3/3}$(R18)}\\
H22:DTC-{\IIo}    \phantom{20} & {18} & {$(464, 118, -711)$} & {$(123, 113, 117)$} & {118} & {$-2.37$} & {0.37$\;$ (1.86)} & {$-1.45$} & {0.35$\;$ (1.20)} & {Tier-1 -- A$^{4/4}$,F$^{2/3}$(R18), DTG38(Y20)}\\
H22:DTC-{\Ov}     \phantom{05} & {12} & {$(115, 195, -889)$} & {$(63, 56, 67)$} & {63} & {$-2.42$} & {0.33$\;$ (2.76)} & {$-1.48$} & {0.33$\;$ (2.78)} & {Tier-1 -- C$^{3/4}$(R18)}\\
H22:DTC-{\Ovi}    \phantom{06} & {5} & {$(954, 354, 773)$} & {$(70, 114, 15)$} & {78} & {$-2.62$} & {0.21$\;$ (4.92)} & {$-1.60$} & {0.20$\;$ (4.30)} & {Tier-1 -- G$^{2/2}$(R18)}\\
H22:DTC-{\IIii}   \phantom{22} & {2} & {$(67, 67, -2504)$} & {$(35, 65, 7)$} & {43} & {$-2.55$} & {0.05$\;$ (10.11)} & {$-1.36$} & {0.36$\;$ (63.38)} & {Tier-3 -- New}\\
H22:DTC-{\Oiiii}  \phantom{04} & {2} & {$(971, 206, -2749)$} & {$(57, 30, 86)$} & {62} & {$-2.83$} & {0.05$\;$ (10.83)} & {$-1.82$} & {0.11$\;$ (21.37)} & {Tier-2 -- New}\\
H22:DTC-{\Iv}     \phantom{15} & {2} & {$(3519, 3390, 163)$} & {$(8, 71, 110)$} & {76} & {$-1.66$} & {0.06$\;$ (11.65)} & {$-0.39$} & {0.23$\;$ (43.04)} & {Tier-3 -- New}\\
H22:DTC-{\Oi}     \phantom{01} & {6} & {$(112, 873, 829)$} & {$(60, 193, 79)$} & {125} & {$-2.87$} & {0.31$\;$ (12.58)} & {$-1.65$} & {0.33$\;$ (13.92)} & {Tier-1 -- B$^{3/4}$(R18), DTG10(Y20)}\\
H22:DTC-{\Oiii}   \phantom{03} & {4} & {$(878, 1190, 1208)$} & {$(51, 110, 39)$} & {73} & {$-2.27$} & {0.26$\;$ (15.28)} & {$-1.34$} & {0.26$\;$ (16.82)} & {Tier-2 -- New}\\
H22:DTC-{\Iviii}  \phantom{18} & {2} & {$(936, 388, 1757)$} & {$(101, 32, 0)$} & {61} & {$-1.39$} & {0.08$\;$ (17.42)} & {$-0.58$} & {0.03$\;$ (7.01)} & {Tier-2 -- New}\\
H22:DTC-{\Iviiii} \phantom{19} & {2} & {$(256, 1282, -1180)$} & {$(13, 54, 180)$} & {108} & {$-2.48$} & {0.09$\;$ (18.15)} & {$-1.40$} & {0.07$\;$ (14.35)} & {Tier-2 -- New}\\
H22:DTC-{\Oviii}  \phantom{08} & {6} & {$(450, 807, 47)$} & {$(77, 103, 106)$} & {96} & {$-2.51$} & {0.35$\;$ (18.28)} & {$-1.43$} & {0.26$\;$ (6.44)} & {Tier-2 -- New}\\
H22:DTC-{\IIviiii} \phantom{29} & {2} & {$(969, 254, -1940)$} & {$(47, 111, 167)$} & {119} & {$-2.90$} & {0.11$\;$ (22.26)} & {$-1.43$} & {0.86$\;$ (97.43)} & {Tier-3 -- New}\\
\hline 
H22:DTC-{\Io}     \phantom{10} & {18} & {$(518, 153, -177)$} & {$(105, 109, 105)$} & {106} & {$-2.43$} & {0.50$\;$ (30.62)} & {$-1.51$} & {0.49$\;$ (30.46)} & {Tier-4 -- E$^{3/3}$,F$^{1/3}$,H$^{2/2}$(R18), DTG38(Y20)}\\
H22:DTC-{\IIi}    \phantom{21} & {13} & {$(340, 214, 521)$} & {$(131, 120, 64)$} & {109} & {$-2.21$} & {0.50$\;$ (34.18)} & {$-1.25$} & {0.54$\;$ (52.98)} & {B$^{1/4}$(R18)}\\
H22:DTC-{\Oo}     \phantom{00} & {4} & {$(129, 1145, -419)$} & {$(153, 65, 88)$} & {108} & {$-1.94$} & {0.42$\;$ (43.80)} & {$-0.60$} & {0.39$\;$ (38.32)} & {Tier-4 -- New}\\
H22:DTC-{\Ovii}   \phantom{07} & {3} & {$(370, 243, -1889)$} & {$(32, 4, 38)$} & {29} & {$-2.64$} & {0.41$\;$ (51.06)} & {$-1.42$} & {0.41$\;$ (51.54)} & {--}\\
H22:DTC-{\Ivi}    \phantom{16} & {7} & {$(107, 292, -287)$} & {$(88, 130, 86)$} & {103} & {$-2.12$} & {0.53$\;$ (54.58)} & {$-1.11$} & {0.56$\;$ (67.74)} & {--}\\
H22:DTC-{\IIvi}   \phantom{26} & {4} & {$(1308, 342, -1017)$} & {$(18, 98, 49)$} & {64} & {$-2.20$} & {0.49$\;$ (58.04)} & {$-1.32$} & {0.51$\;$ (63.82)} & {--}\\
H22:DTC-{\Iiii}   \phantom{13} & {2} & {$(2163, 215, -1052)$} & {$(12, 11, 39)$} & {24} & {$-2.36$} & {0.43$\;$ (70.65)} & {$-1.17$} & {0.01$\;$ (2.44)} & {Tier-3 -- New}\\
H22:DTC-{\IIviii} \phantom{28} & {2} & {$(1497, 1253, 363)$} & {$(11, 13, 38)$} & {24} & {$-2.78$} & {0.55$\;$ (81.60)} & {$-1.86$} & {0.74$\;$ (93.86)} & {--}\\
H22:DTC-{\Ii}     \phantom{11} & {3} & {$(167, 595, -1688)$} & {$(9, 53, 95)$} & {63} & {$-2.48$} & {0.66$\;$ (85.46)} & {$-1.41$} & {0.57$\;$ (78.02)} & {--}\\
H22:DTC-{\IIiii}  \phantom{23} & {14} & {$(105, 80, -1344)$} & {$(97, 87, 116)$} & {101} & {$-2.04$} & {0.68$\;$ (92.02)} & {$-1.19$} & {0.57$\;$ (64.82)} & {Tier-4 -- C$^{1/4}$(R18). }\\
H22:DTC-{\IIiiii} \phantom{24} & {2} & {$(5062, 2249, -367)$} & {$(10, 115, 233)$} & {150} & {$-1.90$} & {0.72$\;$ (92.49)} & {$-1.07$} & {0.75$\;$ (93.99)} & {--}\\
H22:DTC-{\Ivii}   \phantom{17} & {4} & {$(1668, 165, -210)$} & {$(99, 24, 146)$} & {103} & {$-1.85$} & {0.83$\;$ (97.70)} & {$-0.93$} & {0.84$\;$ (97.96)} & {--}\\
\hline
H22:DTC-{\Oii}    \phantom{02} & {1} & {$(1485, 1831, 579)$} & {(--,$\;$ --,$\;$ --)} & {--} & {$-1.82$} & {--$\;$ (--)} & {$-1.12$} & {--$\;$ (--)} & {}\\
H22:DTC-{\Iii}    \phantom{12} & {1} & {$(727, 1284, -585)$} & {(--,$\;$ --,$\;$ --)} & {--} & {$-1.88$} & {--$\;$ (--)} & {$-1.10$} & {--$\;$ (--)} & {}\\
H22:DTC-{\IIv}    \phantom{25} & {1} & {$(6120, 342, -2857)$} & {(--,$\;$ --,$\;$ --)} & {--} & {$-3.71$} & {--$\;$ (--)} & {$-2.71$} & {--$\;$ (--)} & {Most metal-poor}\\
H22:DTC-{\IIvii}  \phantom{27} & {1} & {$(8014, 274, -3751)$} & {(--,$\;$ --,$\;$ --)} & {--} & {$-3.41$} & {--$\;$ (--)} & {$-2.52$} & {--$\;$ (--)} & {2nd most metal-poor}\\
\hline 
\enddata
\tablecomments{
(a) R18 denotes \cite{Roederer2018} and Y20 denotes \cite{Yuan2020ApJ...891...39Y}. 
The group C in R18 contains 4 member stars, 
among which 3 stars are included in our cluster H22:DTC-$\Ov$. 
This is why we put a comment `C$^{3/4}$(R18)' for our cluster H22:DTC-$\Ov$. 
}
\end{deluxetable*}


\section{Discussion} \label{sec:discussion}

\subsection{Chemical homogeneity of the clusters} \label{subsec:percentile}

Up to this point, 
we only use the orbital action of the $r$-II stars, 
and we do not use the chemical properties of these stars. 
(The chemical information is used to construct the sample, 
but the chemical information is not used for the clustering analysis.) 
Because our clustering analysis 
is based on a few assumptions, 
we intentionally reserve the chemical information 
so that we can check the validity of our clustering result.

The sibling stars born in the same dwarf galaxy 
are expected to have similar chemical abundances, 
such as [Fe/H] or [Eu/H]. 
Indeed, the seven stars in the Reticulum~II with enhanced [Eu/Fe] abundance 
show a dispersion in [Fe/H] of $\sigma_\mathrm{[Fe/H]}=0.35$ 
\citep{Ji2016Natur.531..610J}. 
If, within a given cluster we found a dispersion that
satisfies $\sigma_\mathrm{[Fe/H]} \leq 0.35$, 
there is a reasonably high probability that the cluster is genuine.\footnote{
There is no guarantee that a genuine cluster has a tight [Fe/H] distribution (or other chemical abundance distribution). 
However, if a dynamically identified cluster 
happens to have a tight distribution in [Fe/H], 
we are more confident with our clustering result. 
This is the motivation to use chemical abundance information to {\it validate} our results. 
Even if a given cluster identified in our analysis 
has a broad distribution in [Fe/H], 
it does not necessarily mean that the cluster is illusory; 
rather, it signals that we are {\it less confident} about that cluster. 
} 
\footnote{
\new{
Simulated Eu-rich dwarf galaxies also have similar chemical properties. 
According to Fig. 15(d) of \cite{Hirai2022MNRAS.517.4856H}, low-mass dwarf galaxies named Halo-11, 12, 13, 14, and 16 in their simulation have two or more $r$-II stars. 
These systems respectively exhibit $\sigma_\mathrm{[Fe/H]} = 0.19, 0.38, 0.55, 0.39$, and $0.15$ and $\sigma_\mathrm{[Eu/Fe]}= 0.19, 0.09, 0.53, 0.26$, and $0.18$.}
}
Among 26 clusters with $N_{\mathrm{member},k} \geq 2$, 
13 clusters satisfy $\sigma_\mathrm{[Fe/H]} \leq 0.35$, 
having metallicity dispersion equivalent to or smaller than that of Reticulum~II. 
This result indicates that 
many of the field $r$-II stars 
may have originated from disrupted dwarf galaxies.

Following the procedure in \cite{Roederer2018}, 
we evaluated the statistical significance of the chemical homogeneity 
of each cluster as follows. 
First, for each cluster H22:DTC-$k$ with $N_{\mathrm{member},k} \geq 2$, 
we compute the dispersions in [Fe/H] and [Eu/H], 
namely $\sigma_\mathrm{[Fe/H]}$ and $\sigma_\mathrm{[Eu/H]}$, respectively. 
We then randomly draw $N_{\mathrm{member},k}$ stars from the 161 sample stars 
and compute sample standard deviation in [Fe/H] and [Eu/H]. 
We repeat this process 5000 times to derive the probability distribution of 
the randomly drawn dispersions, 
$\sigma_\mathrm{[Fe/H]}^\mathrm{random}$ and 
$\sigma_\mathrm{[Eu/H]}^\mathrm{random}$. 
Finally, we compute the percentile rank of the $\sigma_\mathrm{[Fe/H]}$ and $\sigma_\mathrm{[Eu/H]}$, 
which we denote $q_\mathrm{[Fe/H]}$ and $q_\mathrm{[Eu/H]}$.\footnote{
\new{
In principle, we could perform the same analysis by using [Eu/Fe] and derive $\sigma_\mathrm{[Eu/Fe]}$ and $q_\mathrm{[Eu/Fe]}$. 
However, it turns out that these quantities are not meaningful. 
This is because the distribution of [Eu/Fe] in our sample is highly skewed 
and a large fraction of our sample stars has [Eu/Fe] very close to the lower limit (0.7). 
Due to this skewed distribution, 
a randomly chosen stars have high probability of having small dispersion in [Eu/Fe]. 
For example, randomly chosen pair of stars have 75\% probability of 
having $\sigma_\mathrm{[Eu/Fe]}<0.2$. 
}
}

In Table~\ref{table:clusters}, 
the clusters are shown in an ascending order of $q_\mathrm{[Fe/H]}$. 
Among the 26 clusters with $N_{\mathrm{member},k} \geq 2$, 
14 clusters have $q_\mathrm{[Fe/H]}$ 
below a threshold of $q_\mathrm{[Fe/H]}<25\%$. 
(These 14 clusters have $\sigma_\mathrm{[Fe/H]} \leq 0.35$ 
except for the cluster H22:DTC-3
for which $\sigma_\mathrm{[Fe/H]} = 0.37$.) 
Therefore, about half of the clusters have tight distribution in [Fe/H]. 
A similar trend is also seen for $\sigma_\mathrm{[Eu/H]}$. 

\new{
To investigate the significance of the tight distribution in [Fe/H] 
for our $r$-II clusters, 
we did an additional analysis by using carefully constructed sample of $N=161$ non-$r$-II stars in the literature. 
We find that our $r$-II clusters have tighter distribution in [Fe/H] 
than the clusters of non-$r$-II stars found in this additional analysis 
(see Appendix~\ref{sec:normal_stars} for detail). 
}

Our result indicates that, 
in agreement with \cite{Roederer2018} and \cite{Gudin2021ApJ...908...79G},
$r$-enhanced stars with similar orbits tend to have similar chemistry. 
This result is consistent with a scenario 
wherein $r$-II stars were born in $r$-enhanced dwarf galaxies, similar to the UFD Reticulum~II,
that were later tidally disrupted when they merged with the Milky Way.

\subsection{Individual clusters} \label{subsec:5class}

Among 30 clusters, 26 clusters have $N_{\mathrm{member},k}\geq2$ and 
four clusters have $N_{\mathrm{member},k}=1$. 
Based on the distribution of the chemical abundances in each cluster, 
we categorize 18 clusters 
that have $N_{\mathrm{member},k}\geq2$ 
and show a tight chemical abundance distribution into Tiers-1, 2, 3, and 4. 
This classification reflects our confidence, 
such that we have the highest confidence on Tier-1 clusters. 
In addition, there are four clusters with one member star ($N_{\mathrm{member},k}=1$). 
We do not further discuss the remaining eight clusters 
which have $N_{\mathrm{member},k}\geq2$ and broad distributions in 
[Fe/H] and [Eu/H].

\subsubsection{Tier-1: Six best clusters} 
\label{subsubsec:tier1}

Among 26 clusters with $N_{\mathrm{member},k}\geq2$, 
six clusters 
H22:DTC-$1, 2, 3, 4, 5$, and $9$ 
have tight distributions in both [Fe/H] and [Eu/H]. 
Specifically, 
both $q_\mathrm{[Fe/H]}$ and $q_\mathrm{[Eu/H]}$ 
are below $15\%$. 
(Both $q_\mathrm{[Fe/H]}$ and $q_\mathrm{[Eu/H]}$ 
are below $\sim 5\%$ for five clusters, 
H22:DTC-$1, 2, 3, 4$, and $5$.)
Thus, the probability that all of these six clusters reflect a chance alignment of [Fe/H] values
is extremely low. 
We label these clusters as Tier-1 clusters. 
Based on the chemical properties, 
we are most confident with our clustering results 
for Tier-1 clusters.

Fig.~\ref{fig:TierI} shows the distribution of 
$\vector{J}$, [Fe/H], and [Eu/H] for Tier-1 clusters. 
We see that the member stars 
are distributed roughly within $200 \kpc \kms$ from the cluster centroid 
in the $\vector{J}$-space. 
We visually confirm the tight distribution of [Fe/H] and [Eu/H], 
which characterizes Tier-1 clusters.

To obtain additional insights 
into the progenitor systems of Tier-1 clusters, 
Fig.~\ref{fig:TierI_alpha} shows the distribution 
of [Fe/H] and [$\alpha$/Fe] of their member stars. 
The values of [Mg/Fe] and [Ca/Fe] 
are curated from high-resolution spectroscopic measurements 
in the literature. 
Although some member stars do not have reliable measurements
and are not shown, 
we can grasp the overall trend 
of the abundance of $\alpha$-elements from Fig.~\ref{fig:TierI_alpha}. 
Importantly, 
all Tier-1 clusters show a reasonably tight distribution
in [Fe/H]-[Mg/Fe] and [Fe/H]-[Ca/Fe] planes. 
The fact that Tier-1 clusters have tight distributions 
not only in (Fe, Eu) but also in (Mg, Ca) abundances 
is an supporting evidence that these clusters are remnants of dwarf galaxies that merged with the Milky Way. 
Furthermore, 
we see a hint of decreasing trend of [$\alpha$/Fe] as a function of [Fe/H] 
for clusters H22:DTC-$1$ and $4$, 
which is reminiscent of the trends seen in classical dwarf galaxies, 
such as Sculptor and Ursa Minor
(e.g., 
\citealt{Tolstoy2009ARA&A..47..371T}; \citealt{Kirby2011alpha}),
and UFDs, such as Hercules and Reticulum~II 
\new{\citep{Vargas2013, Ji2022arXiv220703499J}.} 
If this trend is real, 
it may provide a clue that the progenitor systems of these clusters 
experienced a quiescent star formation activity.

The distribution of position $\vector{x}$ and velocity $\vector{v}$ 
of the member stars 
provides additional insights into the progenitor systems of Tier-1 clusters. 
As seen in the top row in Fig.~\ref{fig:position_velocity_distribution}, 
the member stars in each Tier-1 cluster 
show a wide spatial distribution. 
This spatial distribution suggests that 
the member stars have different orbital phases, 
although they have similar orbits. 
The bottom row in Fig.~\ref{fig:position_velocity_distribution} 
further supports this view. 
We see that, for clusters H22:DTC-$1,2,3$, and $4$, 
there are almost equal numbers of stars with 
$v_R>0$ (approaching apocenter), 
$v_R<0$ (approaching pericenter), 
$v_z>0$ (moving upward), and 
$v_z<0$ (moving downward). 
For clusters H22:DTC-$5$ and $9$, 
there are almost equal numbers of stars with 
$v_R>0$ and $v_R<0$, 
while stars with $v_z<0$ dominate. 
(The member stars in each cluster 
have similar values of $v_\phi$ 
because they have similar values of $J_\phi = R v_\phi$.) 
This result means that \new{most} Tier-1 clusters 
are apparently completely phase-mixed, 
suggesting that the disruption of the progenitor systems 
happened long ago. 
Therefore, Tier-1 clusters are likely 
the remnants of completely disrupted dwarf galaxies 
that merged with the ancient Milky Way. 
\new{
Of course, it is premature to conclude decisively that they 
really are disrupted dwarf galaxies, 
and we discuss the prospects 
to confirm these $r$-II clusters in Section \ref{sec:future_prospects}. 
}

In the following, we summarize 
other basic properties of Tier-1 clusters: 

\uline{\bf{Tier-1 cluster H22:DTC-1}.}

This cluster is a newly discovered cluster 
characterized by retrograde, round orbits 
with $\langle J_\phi \rangle = 1209 \kpc \kms$. 
This cluster is metal-poor, 
with $\langle \mathrm{[Fe/H]} \rangle = -2.78$. 
This cluster is rather large ($N_{\mathrm{member},k}=9$), 
but its metallicity dispersion is only $\sigma_\mathrm{[Fe/H]}=0.22$. 
In terms of $q_\mathrm{[Fe/H]}$, 
this cluster has the tightest distribution of [Fe/H] among 30 clusters. 
The clusters 
H22:DTC-$1$ and $9$
have a similar orbit and chemistry, 
which will be discussed in Section \ref{subsec:new_merger_group}. 
Intriguingly, 
two member stars of the cluster H22:DTC-$1$ 
({2MASS J09544277+5246414} and {BPS CS 22896--154}) 
were also analyzed by 
\cite{Roederer2018}; 
however they are regarded as $r$-II stars 
that are not associated to any kinematic groups in \cite{Roederer2018}.

\uline{\bf{Tier-1 cluster H22:DTC-2}.}

This cluster is 
characterized by highly eccentric orbits 
with $r_\mathrm{peri}< 1 \kpc$ and $r_\mathrm{apo}> 10 \kpc$. 
This cluster is a metal-rich cluster, 
with $\langle \mathrm{[Fe/H]} \rangle = -1.65$. 
In terms of $q_\mathrm{[Eu/H]}$, 
this cluster has the tightest distribution in Eu abundance. 
This cluster corresponds to the group D in \cite{Roederer2018}, 
containing all three stars in the group D. 
One of the member stars, {HD 222925}, 
is the most well-studied $r$-II star 
in terms of its chemistry \citep{Roederer2022}.

\uline{\bf{Tier-1 cluster H22:DTC-3}.}

This cluster is 
characterized by prograde, eccentric orbits 
with $\langle J_\phi \rangle = -711 \kpc \kms$. 
This cluster is one of the largest clusters in our analysis, 
with $N_{\mathrm{member},k}=18$. 
The face value of the dispersion in [Fe/H] is 
relatively large, $\sigma_\mathrm{[Fe/H]}=0.37$, 
but its low percentile $q_\mathrm{[Fe/H]} = 1.86\%$ 
suggests that this cluster has a tight distribution in [Fe/H] 
(given its large $N_{\mathrm{member},k}$). 
This cluster includes 
4 out of 4 stars from the group A 
and 
2 out of 3 stars from the group F 
in \cite{Roederer2018}. 
This cluster also includes a $r$-II star ({2MASS J22562536$-$0719562}) 
that is associated with the DTG38 in \cite{Yuan2020ApJ...891...39Y}.\footnote{
The DTG38 in \cite{Yuan2020ApJ...891...39Y} contains not only 
{2MASS J22562536$-$0719562} but also another $r$-II star 
{2MASS J00405260$-$5122491}, 
which is included in the cluster H22:DTC-15. 
Although \cite{Yuan2020ApJ...891...39Y} 
claims these two $r$-II stars are associated with a single group (DTG38), 
our analysis associate them to two clusters (clusters H22:DTC-3 and 15) 
with similar orbital properties. 
} 

\uline{\bf{Tier-1 cluster H22:DTC-4}.}

This cluster is a dynamically cold cluster 
with its internal action dispersion of $\sigma_k = 63 \kpc \kms$. 
The member stars show prograde, mildly eccentric orbits 
with $\langle J_\phi \rangle = -889 \kpc \kms$, 
which corresponds to a guiding center radius of $R \sim 4 \kpc$. 
The very metal-poor nature of this group, 
$\langle \mathrm{[Fe/H]} \rangle = -2.42$, 
is in contrast to the majority of the disk stars at $R \sim 4 \kpc$. 
This cluster corresponds to the group C in \cite{Roederer2018}, 
containing 3 out of 4 stars in the group C. 
Our method finds 9 additional members of this group. 

\uline{\bf{Tier-1 cluster H22:DTC-5}.}

This cluster is 
characterized by retrograde, eccentric orbits 
with $\langle J_\phi \rangle = 773 \kpc \kms$. 
This cluster has the smallest $\sigma_\mathrm{[Fe/H]}$ and $\sigma_\mathrm{[Eu/H]}$ among Tier-1 clusters. 
This cluster corresponds to the group G in \cite{Roederer2018}, 
containing 2 out of 2 stars in the group G. 
Our method finds 3 additional member stars of this group. 
Two of the new members ({SMSS J183647.89$-$274333.1} and {HE~0300$-$0751}) 
have poor measurement of the parallax 
($\varpi/\sigma_\varpi<4$; see Table~\ref{table:members_all_info}), 
which highlights the advantage of using our clustering method.

\uline{\bf{Tier-1 cluster H22:DTC-9}.}

This cluster is 
characterized by retrograde, round orbits 
with $\langle J_\phi \rangle = 829 \kpc \kms$. 
This cluster is one of the lowest metallicity clusters, 
with $\langle \mathrm{[Fe/H]} \rangle = -2.87$. 
It corresponds to the group B in \cite{Roederer2016AJ....151...82R}, 
containing 3 out of 4 stars in the group B. 
This cluster also corresponds to a dynamically tagged group (DTG),
DTG10, in \cite{Yuan2020ApJ...891...39Y}.\footnote{
The DTG10 in \cite{Yuan2020ApJ...891...39Y} includes two $r$-II stars 
({BPS CS 31082$-$001} and {SDSS J235718.91$-$005247.8}), 
both of which are included in the group B in \cite{Roederer2018}. 
} 

\subsubsection{Tiers-2, 3 and 4: Other 12 clusters} 
\label{subsubsec:tier234}

Apart from Tier-1 clusters, 
we choose 12 clusters with reasonably small 
$q_\mathrm{[Fe/H]}$ and/or $q_\mathrm{[Eu/H]}$ 
and categorize them as Tiers-2, 3, and 4 clusters 
(see comment in Table~\ref{table:clusters}). 
All the clusters in Tiers-2 and 3 are newly discovered. 
The details are described in Appendix \ref{sec:tier234}.

\subsubsection{One-member clusters} \label{subsubsec:single_member_cluster}

Four clusters have only one member star 
(H22:DTC-$27, 28, 29$, and $30$). 
The star in the cluster 
H22:DTC-$29$ 
({SMSS J024858.41$-$684306.4}) has [Fe/H]$=-3.71$; 
while the star in the cluster 
H22:DTC-$30$ 
({SMSS J063447.15$-$622355.0}) has [Fe/H]$=-3.41$. 
These two stars are the two most metal-poor stars in our catalog.

Although our clustering analysis assign 
these extremely metal-poor $r$-II stars to 
separate single-member clusters, 
they have similar orbital properties. 
For example, as seen in Table~\ref{table:members_additional_info}, 
both stars have prograde orbits with high eccentricity ($ecc \simeq 0.88$)
and have similar pericentric and apocentric radii 
($r_\mathrm{peri} \simeq 8 \kpc$ and $r_\mathrm{apo} > 90 \kpc$). 
These similarities indicate that 
their origins might be related to each other, 
but we do not further discuss their connection in this paper.

In our analysis, 
the extremely metal-poor stars 
{SMSS J024858.41$-$684306.4} and {SMSS J063447.15$-$622355.0} 
(which are the only member of clusters 
H22:DTC-$29$ and $30$, respectively)
do not have sibling stars in our catalog. 
This result can be understood from chemical and dynamical points of view.
If each of these stars was formed in a dwarf galaxy, 
the total stellar mass of its progenitor dwarf galaxy 
may have been small, 
according to the mass-metallicity relationship of the dwarf galaxies 
\citep{Kirby2013ApJ...779..102K,Naidu2022arXiv220409057N}. 
Therefore, it is reasonable that these extremely metal-poor stars 
have a rather small number of sibling stars ($\lesssim 10^3$ stars) in the Galactic halo. 
Also, 
the fact that these two stars 
have highly eccentric orbit (with $r_\mathrm{apo}>90 \kpc$) 
means that most of their sibling stars are in the outer part of the Galactic halo. 
Because our catalog is restricted to bright $r$-II stars, 
and most of the $r$-II stars in our catalog are within $10 \kpc$ from the Sun, 
the apparent lack of sibling stars may be understood as the spatial selection bias of the sample stars 
(cf. \citealt{Hattori2011MNRAS.418.2481H}).

\begin{figure*}
\centering
\includegraphics[width=1.1in]
{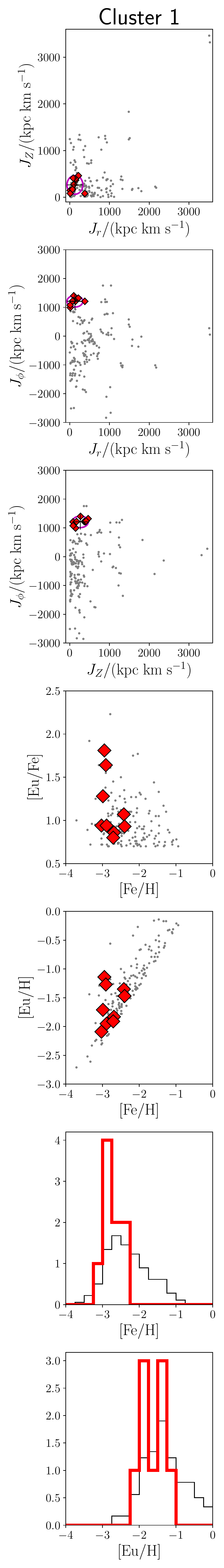}
\includegraphics[width=1.1in]
{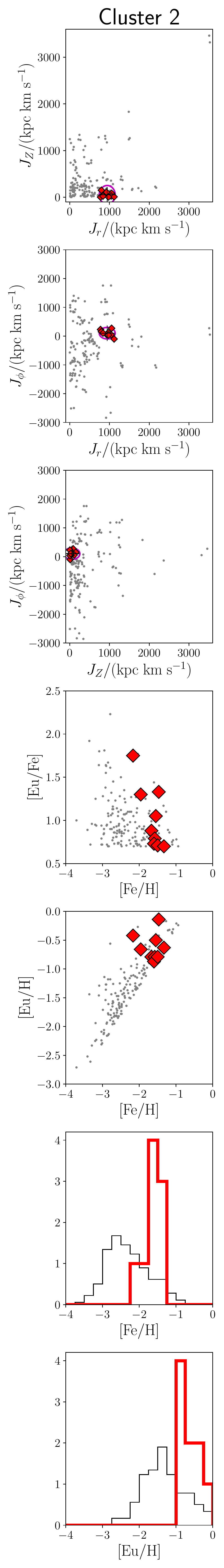}
\includegraphics[width=1.1in]
{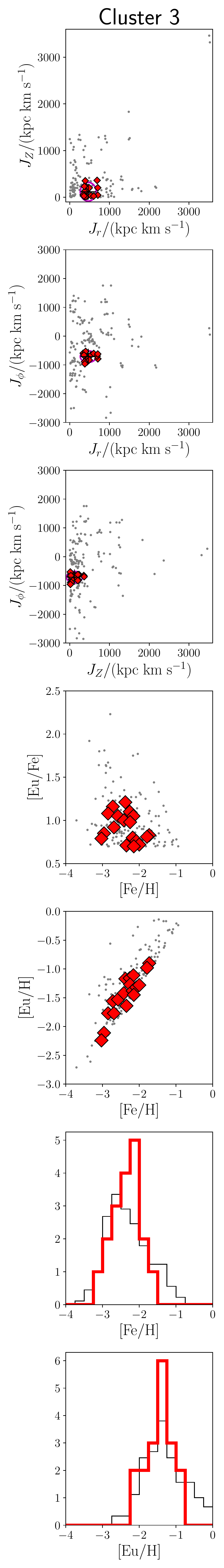}
\includegraphics[width=1.1in]
{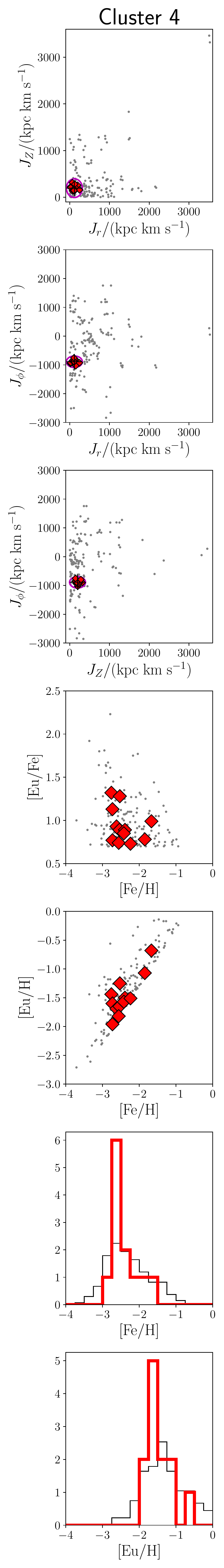}
\includegraphics[width=1.1in]
{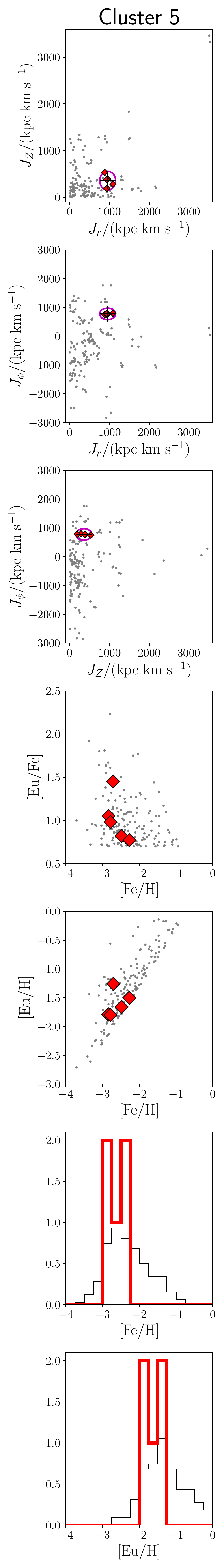}
\includegraphics[width=1.1in]
{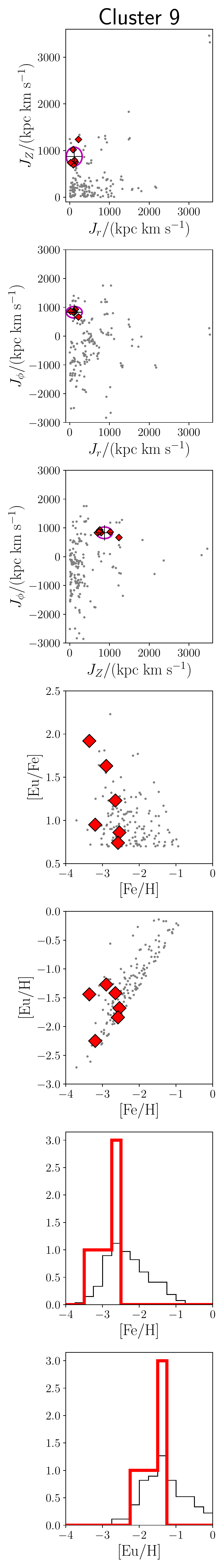}
\caption{
The properties of stars in Tier-1 clusters, 
for which we have the highest confidence. 
From left to right column, 
we show clusters 
H22:DTC-$1,2,3,4,5$, and $9$. 
In each panel, 
the 
red \new{diamonds} correspond to the member stars, 
while small gray dots correspond to the other stars. 
(The first three rows):
The distribution of the orbital action. 
The black cross is the location of the centroid 
$(\langle J_r \rangle, \langle J_z \rangle, \langle J_\phi \rangle)$. 
The magenta circle with a radius of $200 \kpc \kms$ is shown 
to guide the eye. 
(Fourth and fifth rows): 
The distribution of stars in the chemical abundance space. 
We note that the chemical abundance information is not used 
in the clustering analysis, 
and is used for validation. 
(Sixth and seventh rows):
The distribution of [Fe/H] and [Eu/H] in the cluster are shown by the red histogram. 
The black histogram shows the distribution of [Fe/H] and [Eu/H] for the entire $r$-II sample, 
which is normalized so that the areas under the black and red histograms are the same. 
}
\label{fig:TierI}
\end{figure*}

\begin{figure*}
\centering
\includegraphics[width=6.0in]{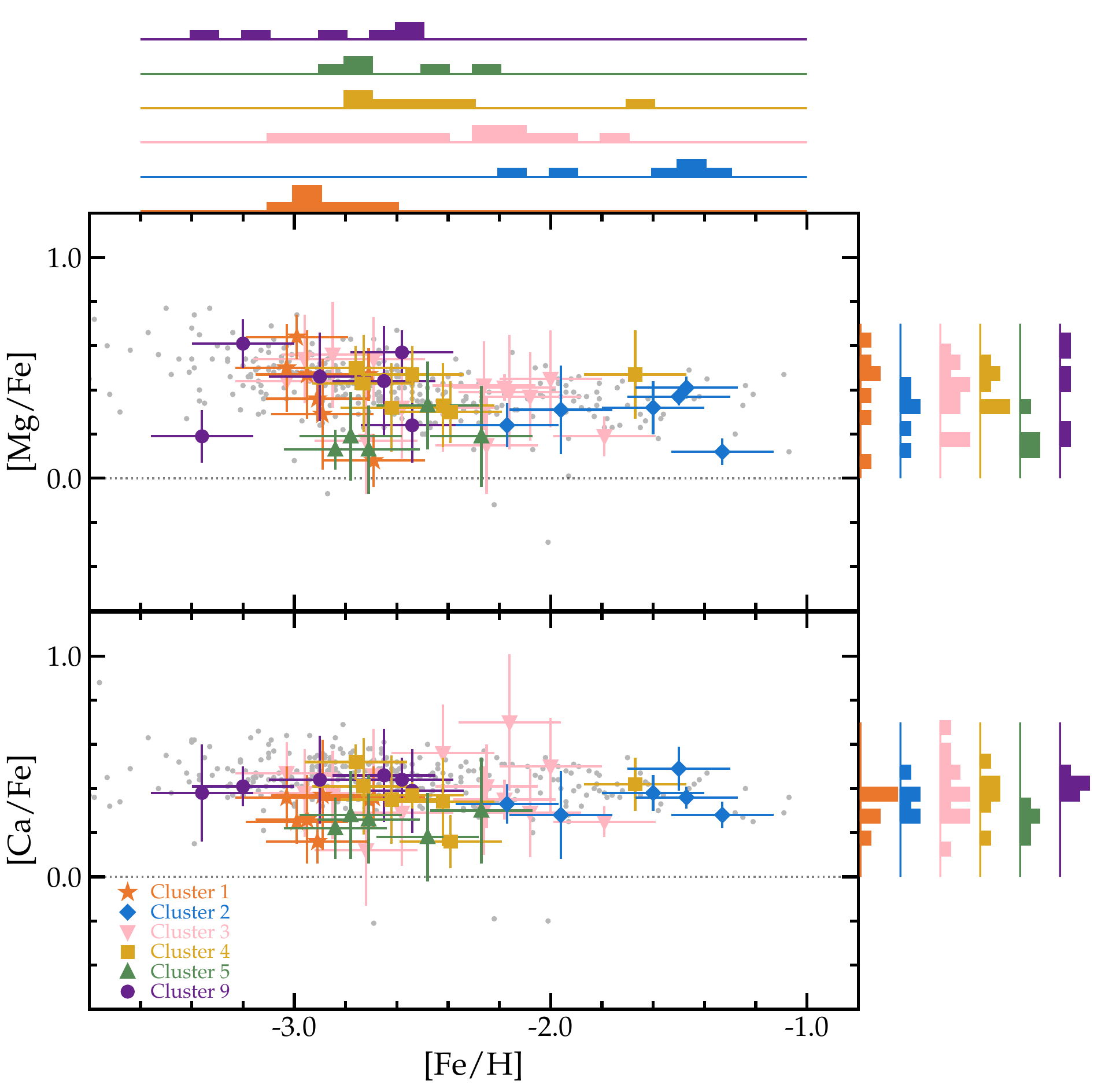} 
\caption{
The distributions of stars in the [Fe/H]-[Mg/Fe] space (top)
and [Fe/H]-[Ca/Fe] space (bottom) 
for Tier-1 clusters 
(H22:DTC-$1, 2, 3, 4, 5$, and $9$). 
These distributions are reminiscent of 
those of dwarf galaxies orbiting around the Milky Way. 
Histograms of [Fe/H], [Mg/Fe], and [Ca/Fe] are shown to the top or right of each panel.
The abundance ratios for individual stars
in the clusters identified in this work
are adopted from
\citet{Westin2000},
\citet{Hill2002},
\citet{Cayrel2004},
\citet{Honda2004b},
\citet{Barklem2005},
\citet{Ivans2006},
\citet{Francois2007},
\citet{Frebel2007},
\citet{Aoki2010},
\citet{Mashonkina2010},
\citet{Hollek2011},
\citet{Hansen2012},
\citet{Ishigaki2012,Ishigaki2013},
\citet{Johnson2013},
\citet{Roederer2014c,Roederer2018},
\citet{Howes2015},
\citet{Jacobson2015},
\citet{Navarrete2015},
\citet{Placco2017},
\citet{Cain2018},
\citet{Hansen2018},
\citet{Holmbeck2018},
\citet{Sakari2018},
\citet{Bandyopadhyay2020},
\citet{Ezzeddine2020},
\citet{Hanke2020},
\citet{Rasmussen2020},
and
\citet{Zepeda2022}.
The small gray points mark field stars
from \citet{Roederer2014c}.
The dotted lines mark the Solar ratios.
}
\label{fig:TierI_alpha}
\end{figure*}

\begin{figure*}
\centering
\includegraphics[width=6.3in]
{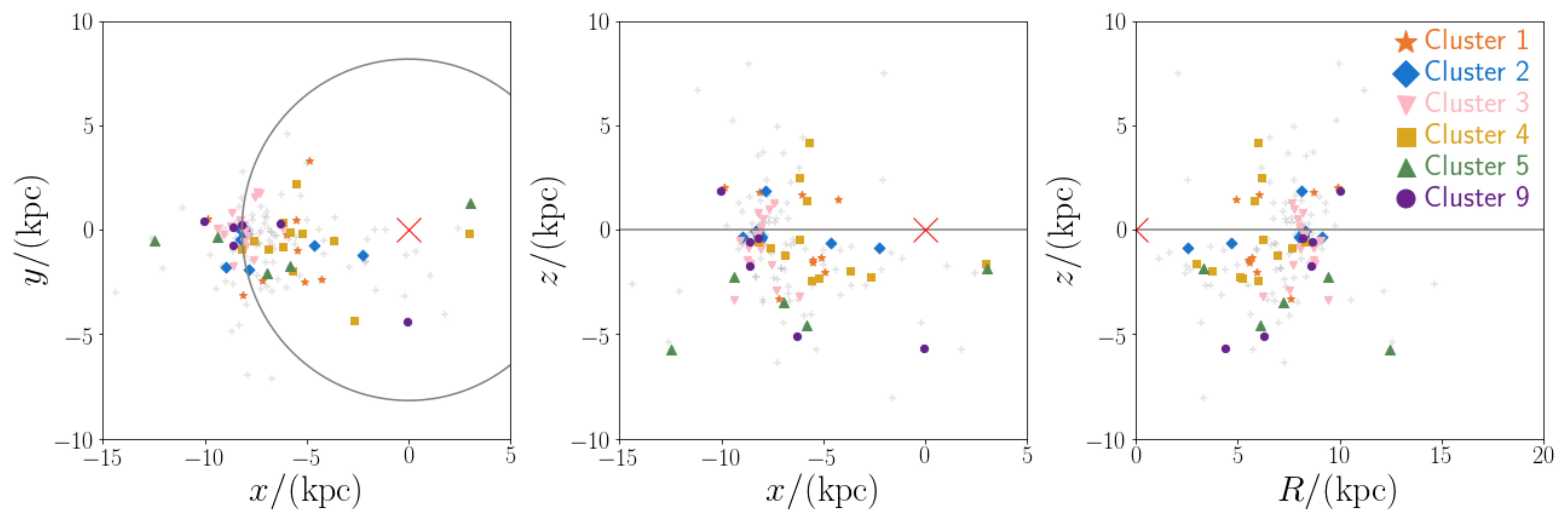}\\
\includegraphics[width=6.3in]
{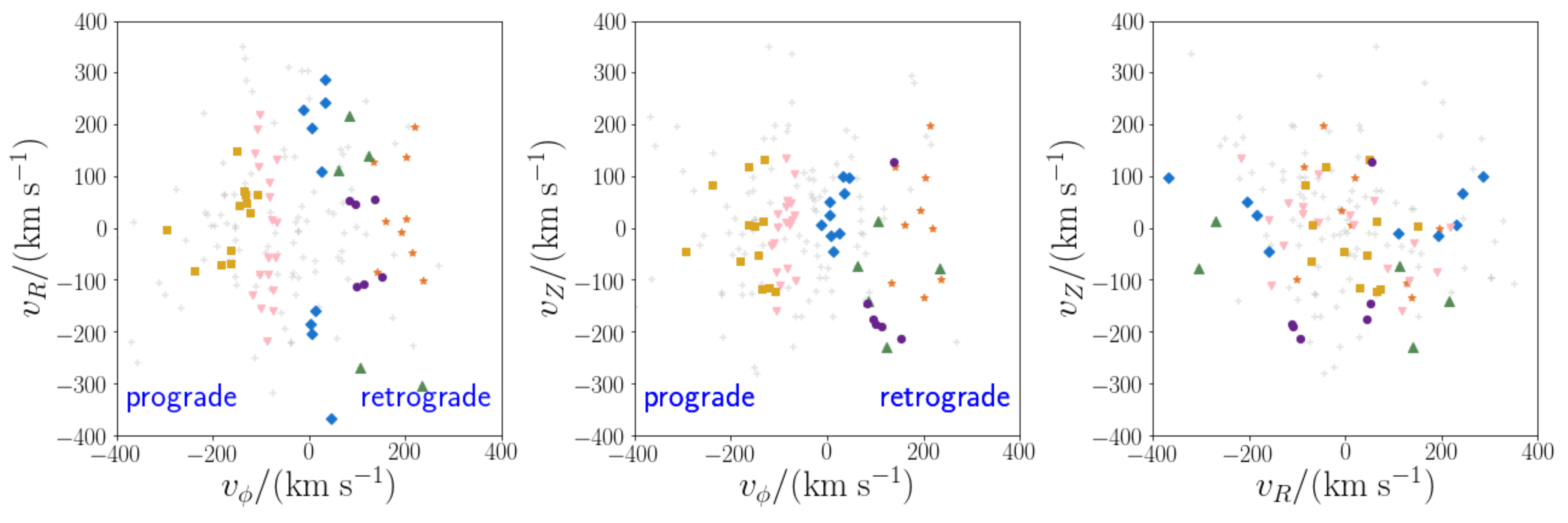}
\caption{
Position and velocity of the member stars of Tier-1 clusters 
(H22:DTC-$1,2,3,4,5$, and $9$; colored symbols) 
and other $r$-II stars (light-gray cross). 
On the top panels, 
we use the Galactocentric Cartesian coordinate $(x,y,z)$, 
such that the Sun is located at $(x,y,z)=(-8.178, 0, 0) \kpc$. 
As a reference, 
we show the Solar circle (gray circle on top left panel) 
and the location of the Galactic center (red $\times$ on top panels). 
On the bottom panels, 
we use the velocity in the Galactocentric cylindrical coordinate $(v_R, v_\phi, v_z)$, 
such that prograde stars have $v_\phi<0$.
}
\label{fig:position_velocity_distribution}
\end{figure*}


\begin{figure*}
\centering
\includegraphics[width=6.3in]
{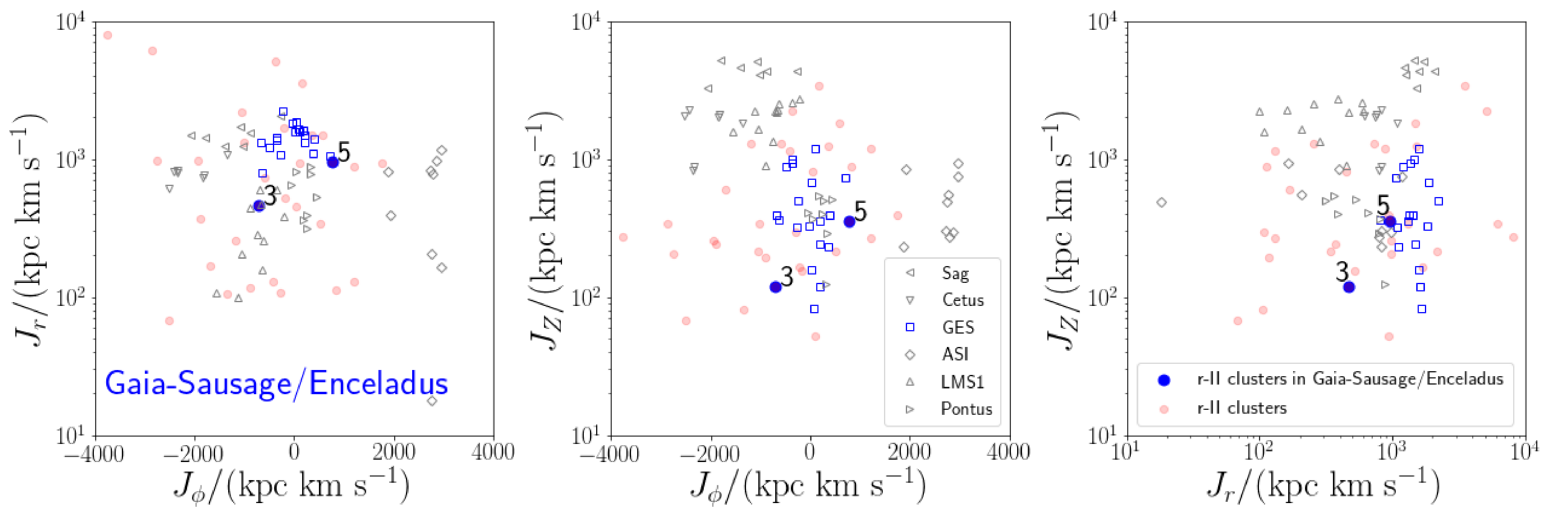}\\
\includegraphics[width=6.3in]
{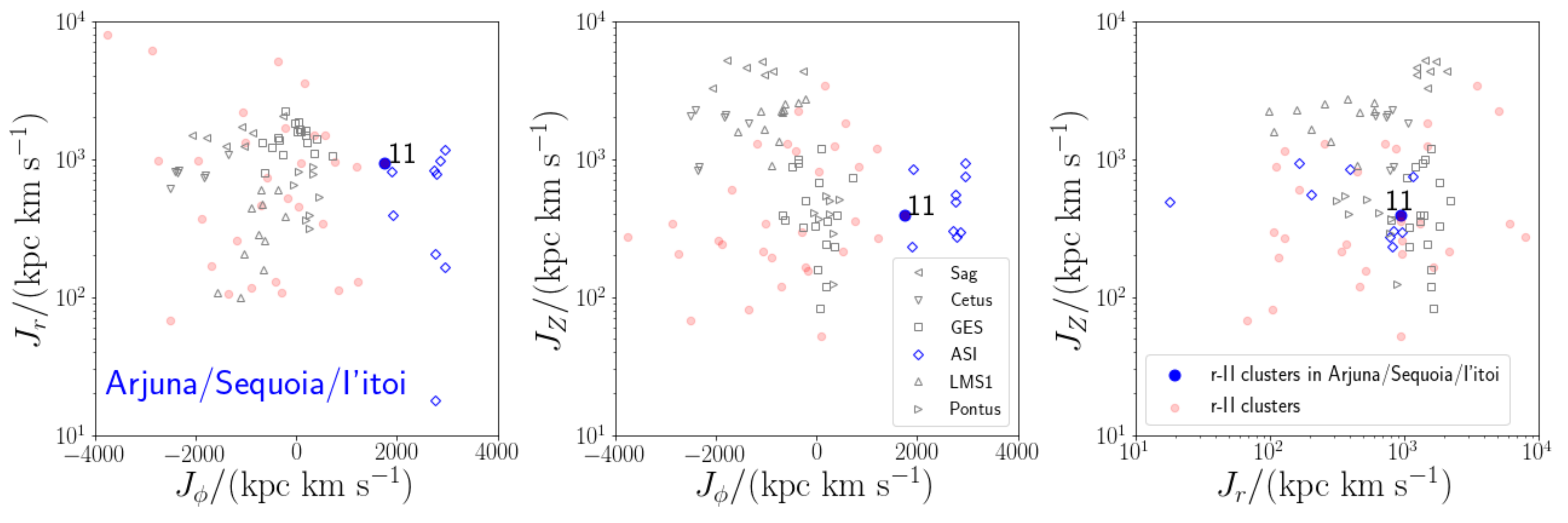}\\
\includegraphics[width=6.3in]
{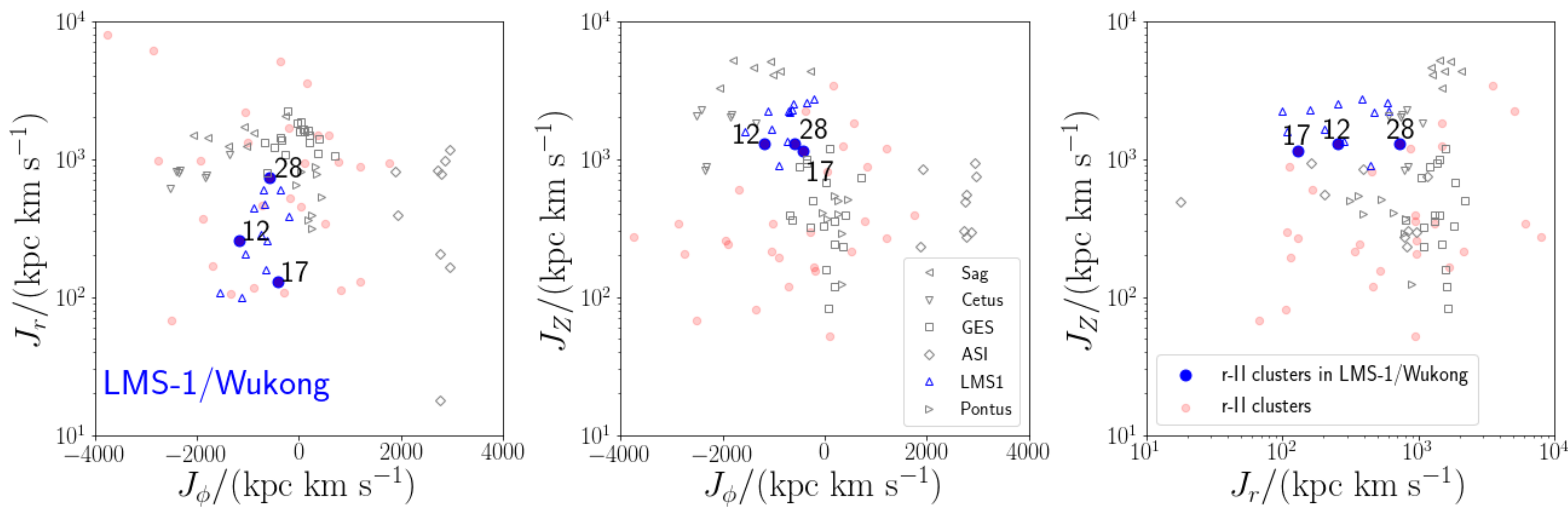}\\
\includegraphics[width=6.3in]
{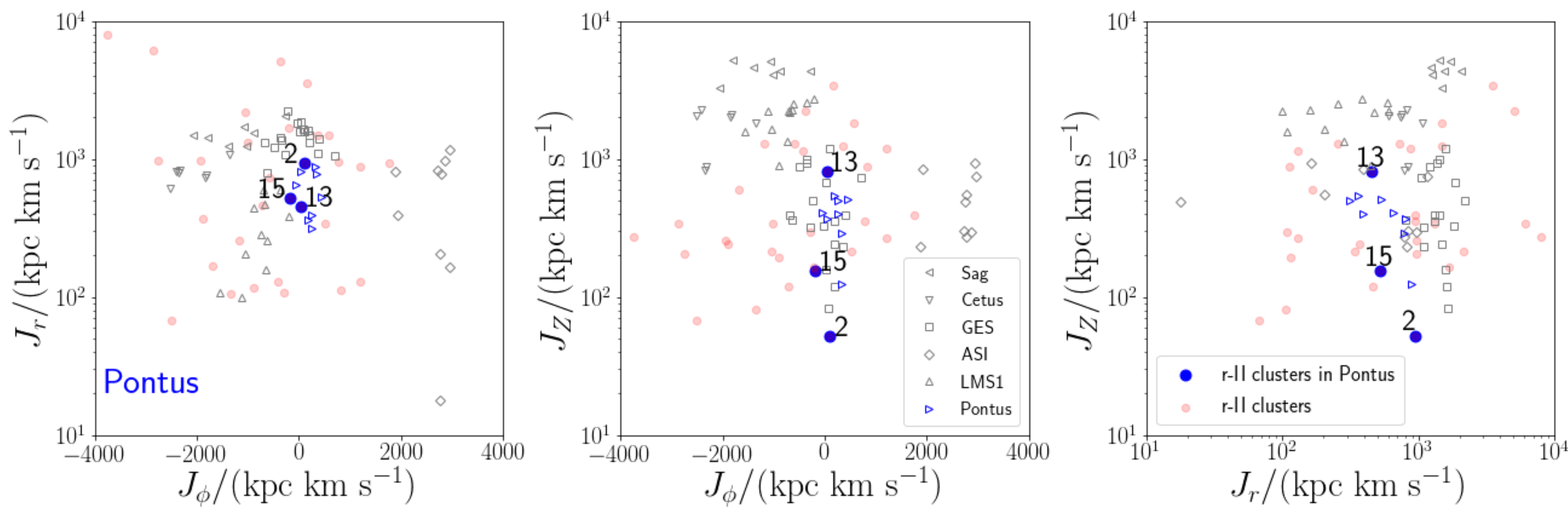}
\caption{
Associations of our clusters and 
the known big merger groups in \cite{Malhan2022ApJ...926..107M}. 
In all panels, 
the filled symbols represent the $r$-II clusters in our analysis, 
and the open symbols represent 
the stellar streams and globular clusters in the merger groups. 
In each row, 
we highlight one of the merger groups with blue symbols. 
The rest of the merger groups are shown by gray symbols. 
The clusters associated with the highlighted merger group 
are shown by blue filled circles. 
(The number corresponds to $k$ of the $r$-II cluster H22:DTC-$k$.)
The rest of the clusters are shown by light-red filled circles. 
We do not find clusters associated with the merger groups 
Sagittarius and Cetus. 
}
\label{fig:malhan_groups}
\end{figure*}

\subsection{Connection to the past merger events}
\label{subsec:Malhan_groups}

\cite{Malhan2022ApJ...926..107M} analyzed the kinematics 
of the dynamical tracers in the Milky Way
(i.e., stellar streams, globular clusters, and dwarf galaxies) 
and found that 
these tracers show a clumpy distribution 
in the orbital action and energy space 
(see also similar analysis by \citealt{Bonaca2021ApJ...909L..26B}). 
They identified six groups of dynamical tracers, namely, 
(i) the Gaia-Sausage/Enceladus merger group; 
(ii) the Arjuna/Sequoia/I’itoi merger group;
(iii) the LMS-1/Wukong merger group;
(iv) the Pontus merger group; 
(v) the Cetus merger group; 
and 
(vi) the Sagittarius merger group. 
They claimed that these groups correspond to 
the past merger events in the Milky Way.

To investigate if any of our clusters 
are associated with these merger groups, 
we check the action distributions 
of the clusters and the merger groups in \cite{Malhan2022ApJ...926..107M}. 
First, 
we adopt the literature values of 
the point-estimate of the orbital actions 
for the stellar streams from \cite{Malhan2022ApJ...926..107M} 
and 
for the globular clusters from \cite{Vasiliev2021MNRAS.505.5978V} 
(see bottom row in Fig.~\ref{fig:action_distribution}). 
We note that they computed the orbital actions with a slightly different assumptions on the position and velocity of the Sun,
but this difference has a minor effect on the values of actions 
(with a typical difference of $\sim 100 \kpc \kms$). 
We neglect this minor effect, because our conclusion is not affected. 
Secondly, 
we compute the Euclidian distance from our $r$-II clusters to 
the stellar streams and globular clusters in the action space. 
Thirdly, 
for each $r$-II cluster, 
we find the nearest neighbor stellar stream and globular cluster. 
If the distance to the nearest neighbor object (stream or globular cluster) 
satisfies $|| \vector{J}_{r\mathrm{IIcluster}} - \vector{J}_\mathrm{nearest} || < 600 \kpc \kms$, 
and if the nearest neighbor object is a member of a certain merger group (among the above-mentioned six merger groups (i)-(vi)), 
we interpret that the $r$-II cluster is also associated with the same merger group.

As a result, 
we find eight $r$-II clusters that are associated 
with four merger groups (as summarized below);
and there are no $r$-II clusters 
that are associated with the Cetus and Sagittarius merger groups.

{\bf{(i) Gaia-Sausage/Enceladus merger group}}

Two $r$-II clusters, 
H22:DTC-$3$ and $5$, 
are dynamically associated with the Gaia-Sausage/Enceladus merger group 
(see top row in Fig.~\ref{fig:malhan_groups}).
The mean metallicity of these $r$-II clusters are 
$\langle$[Fe/H]$\rangle = -2.37$ and $-2.62$, respectively. 
These $r$-II clusters are more metal-poor than
the majority of the stellar streams and globular clusters associated with the Gaia-Sausage/Enceladus merger group 
($-2.4 \leq$ [Fe/H] $\leq -1.1$) \citep{Malhan2022ApJ...926..107M}. 
\begin{itemize}
\item 
For the $r$-II cluster 
H22:DTC-$3$, 
the nearest neighbor object is the NGC~7089 stream 
([Fe/H]$=-1.46\pm0.06$; \citealt{Horta2020MNRAS.493.3363H}), 
which is the member of the Gaia-Sausage/Enceladus merger group. 
As seen in Fig.~\ref{fig:TierI}, this $r$-II cluster shows 
a wide distribution in [Fe/H] (covering $-3.03 \leq$[Fe/H]$\leq -1.73$), 
which is roughly consistent with the lower-metallicity part of 
the Gaia-Sausage/Enceladus merger group. 

\item 
For the $r$-II cluster 
H22:DTC-$5$, 
the nearest neighbor objects are 
the C-7 stream (metallicity is unknown) and 
NGC~2298 ([Fe/H]$= -1.76 \pm 0.05$; \citealt{Baeza2022}),
both of which are the members of the this merger group. 
Given that this $r$-II cluster has 
$\langle$[Fe/H]$\rangle = -2.62$, 
it may be the most metal-poor member of this merger group. 
\end{itemize}

{\bf{(ii) Arjuna/Sequoia/I’itoi merger group}}

One $r$-II cluster, 
H22:DTC-$11$, 
is dynamically associated with the Arjuna/Sequoia/I’itoi merger group
(see second row in Fig.~\ref{fig:malhan_groups}). 
The mean metallicity of this $r$-II cluster is 
$\langle$[Fe/H]$\rangle=-1.39$. 
This metallicity is slightly higher than 
the metallicities of the stellar streams and globular clusters 
associated with the Arjuna/Sequoia/I’itoi merger group 
($-2.24 \leq$ [Fe/H] $\leq-1.56$) \citep{Malhan2022ApJ...926..107M}. 

\begin{itemize}
\item 
The nearest neighbor to the $r$-II cluster 
H22:DTC-$11$ 
is the Phlegethon stream 
([Fe/H]$=-1.96\pm0.05$; \citealt{Martin2022arXiv220101310M}), 
which is a member of this merger group \citep{Malhan2022ApJ...926..107M}. 
Given that this $r$-II cluster 
has a mean metallicity $\langle$[Fe/H]$\rangle = -1.39$, 
it may be the most metal-rich member of this merger group. 
\end{itemize}

{\bf{(iii) LMS-1/Wukong merger group}}

Three $r$-II clusters, 
H22:DTC-$12, 17$, and $28$, 
are dynamically associated with the LMS-1/Wukong merger group
(see third row in Fig.~\ref{fig:malhan_groups}). 
The mean metallicities of these $r$-II clusters are 
$\langle$[Fe/H]$\rangle=-2.48, -1.94$, and $-1.88$, respectively. 
These metallicities are consistent with the metallicities of 
the stellar streams and globular clusters associated with the LMS-1/Wukong merger group ($-3.38 \leq$[Fe/H]$\leq-1.35$) \citep{Malhan2022ApJ...926..107M}. 

\begin{itemize}
\item 
For the $r$-II cluster 
H22:DTC-$12$,
the nearest neighbor object is 
the Pal 5 ([Fe/H]$=-1.35\pm0.06$; \citealt{Ishigaki2016ApJ...823..157I}), 
which is a member of this merger group \citep{Malhan2022ApJ...926..107M}. 

\item 
The $r$-II cluster 
H22:DTC-$17$
is also associated with this merger group;  
but our reasoning mainly comes from external knowledge 
that this cluster includes an $r$-II star  
\citep{Hansen2021ApJ...915..103H} 
associated with the Indus stream 
([Fe/H]$=-1.96\pm0.33$; \citealt{Li2022ApJ...928...30L}) and 
that the Indus stream is a member of the LMS-1/Wukong group. 

\item 
For the $r$-II cluster 
H22:DTC-$28$
which is a one-member cluster, the nearest neighbor objects are Pal 5 and NGC~5272 (M3;
[Fe/H]$=-1.45\pm0.03$; 
\citealt{Sneden2004}),
both of which are members of this merger group. 
\end{itemize}

{\bf{(iv) Pontus merger group}}

Three $r$-II clusters, 
H22:DTC-$2, 13$, and $15$, 
are dynamically associated with the Pontus merger group
(see bottom row in Fig.~\ref{fig:malhan_groups}). 
The mean metallicities of these $r$-II clusters are 
$\langle$[Fe/H]$\rangle=-1.65,-2.51$, and $-2.43$, respectively. 
It is intriguing that two of these $r$-II clusters are more metal-poor than 
the stellar streams and globular clusters associated with the Pontus merger group 
($-2.3 \leq$[Fe/H]$\leq-1.3$) \citep{Malhan2022ApJ...926..107M}.

\begin{itemize}
\item 
For the $r$-II cluster 
H22:DTC-$2$, 
the nearest neighbor object is the
anomalous globular cluster NGC~5286 ([Fe/H]$=-1.72\pm0.11$; \citealt{Marino2015}),
which is a member of the Pontus merger group \citep{Malhan2022ApJ...926..107M}. 
\item 
For the $r$-II cluster 
H22:DTC-$13$, 
the nearest neighbor objects are 
NGC~6341 (M92; [Fe/H]= $-2.34 \pm 0.04$;
\citealt{Sneden2000})
and NGC~7099 (M30; [Fe/H]$=-2.34\pm0.07$; 
\citealt{Carretta2009feh}),
both of which are members of this merger group. 
\item
For the $r$-II cluster 
H22:DTC-$15$, 
the nearest neighbor objects are M92 
and NGC~4833 ([Fe/H] = $-2.19 \pm 0.02$;
\citealt{Roederer2015}),
both of which are members of this merger group. 
\end{itemize}

\subsection{A very-metal-poor merger group candidate}
\label{subsec:new_merger_group}

Two Tier-1 clusters, 
H22:DTC-$1$ and $9$, 
are very metal-poor with $\langle\text{[Fe/H]}\rangle \simeq -2.8$. 
As seen in Section \ref{subsubsec:tier1},
these clusters have tight distributions in chemical abundances, 
making them reliable candidates \new{for} disrupted dwarf galaxies. 
Interestingly, these two clusters have very similar dynamical and chemical properties.

The orbital actions of these $r$-II clusters are characterized by 
\eq{
&( J_r, J_z, J_\phi) = \nonumber \\
& 
\begin{cases}
(129\pm107, 265\pm140,1209\pm112) \;\; \text{(H22:DTC-$1$)} \\
(112\pm 60, 873\pm193, 829\pm 79) \;\;\;\;\;\;\; \text{(H22:DTC-$9$)}. 
\end{cases}
}
The distance between these $r$-II clusters in the $\vector{J}$-space 
is only $717 \kpc \kms$, 
which is much smaller than the typical distance between two points 
randomly drawn in the action space. 
For example, if we randomly draw two $r$-II clusters from the 30 clusters in Table~\ref{table:clusters}, 
we have only 4.4\% chance of getting a distance smaller than $717 \kpc \kms$. 
Therefore, it is likely that these clusters are dynamically associated with each other. 
Intriguingly, 
as seen in the third row in Fig.~\ref{fig:action_distribution}, 
these two $r$-II clusters are away from 
any of the big merger groups in \cite{Malhan2022ApJ...926..107M}, 
such as the Gaia-Sausage/Enceladus merger group or Pontus merger group 
(see also Section \ref{subsec:Malhan_groups}).

The chemistry of these clusters is characterized by 
\eq{
&(\mathrm{[Fe/H]}, \mathrm{[Eu/H]} ) = \nonumber \\
& 
\begin{cases}
(-2.78\pm0.22, -1.64\pm0.32) \;\; \text{(H22:DTC-$1$)} \\
(-2.87\pm0.31, -1.65\pm0.33) \;\; \text{(H22:DTC-$9$)}. 
\end{cases} \label{eq:mean_FeH_EuH}
}
The similarity of the chemical properties 
indicates that these clusters may have experienced a similar star formation history.

Based on the chemo-dynamical similarity 
and the fact that they are separated from other merger groups, 
we postulate that these $r$-II clusters 
may be the remnants of two very-metal-poor dwarf galaxies that merged with the Milky Way together,
corresponding to yet another merger event.

In the following, 
we  estimate the stellar mass of the progenitor dwarf galaxies 
of the $r$-II clusters 
H22:DTC-$1$ and $9$, 
which we refer to as 
$M_{*}(\text{H22:DTC-1})$ and $M_{*}(\text{H22:DTC-9})$, 
respectively. 
According to the mass-mean metallicity relationship of the disrupted dwarf galaxies
(\citealt{Naidu2022arXiv220409057N}; see also \citealt{Kirby2013ApJ...779..102K}), 
\eq{
\langle \text{[Fe/H]} \rangle = 
(-2.11^{+0.11}_{-0.12}) + (0.36^{+0.12}_{-0.04}) 
\log_{10} \left( \frac{M_{*}}{10^6 M_\odot} \right),
}
disrupted dwarf galaxies with 
$\langle \text{[Fe/H]} \rangle \simeq -2.8$ 
typically have a stellar mass $M_{*} \sim 10^4 M_\odot$. 
Thus, given the mean metallicity in equation (\ref{eq:mean_FeH_EuH}), 
we have 
$M_{*}(\text{H22:DTC-1}) \sim 
 M_{*}(\text{H22:DTC-9}) \sim 10^4 M_\odot$. 

Our estimate of $M_{*}$ 
is supported by the distribution of stars in the 
color-magnitude diagram for each $r$-II cluster. 
Based on the Padova stellar evolution model (\citealt{Bressan2012MNRAS.427..127B}; ver 3.6), 
a stellar population with an initial total mass of $10^4 M_\odot$ 
and metallicity of $\text{[Fe/H]} = -2.8$ 
is expected to have, on average, $4.4$ stars with $M_\mathrm{G}<0$ 
(brigter than the majority of the horizontal branch stars)
at the age of 10 Gyr. 
The expected number (4.4) is, at face value, 
consistent with the observation:
we have 7 and 3 stars with $M_\mathrm{G}<0$ 
in $r$-II clusters 
H22:DTC-$1$ and $9$, 
respectively. 
Of course, we admit that 
this argument needs to be treated with care, 
because we do not know the completeness of the member stars with $M_\mathrm{G}<0$ 
and the contamination in our $r$-II clusters. 
However, 
the distribution of stars in the 
color-magnitude diagram is at least consistent with the view that 
the progenitor dwarf galaxies 
of our $r$-II clusters 
H22:DTC-$1$ and $9$ 
had $M_{*}\sim 10^4 M_\odot$ before they were disrupted.

\subsection{\new{Prospects for confirming our $r$-II clusters}}
\label{sec:future_prospects}

\new{
The aim of this paper is to find candidates for $r$-II clusters which are likely remnants of disrupted dwarf galaxies. Due to the optimistic nature of our clustering method, at this moment it is premature to conclusively determine whether any of these $r$-II clusters are real. Here we discuss two prospects to confirm or refute the reality of these $r$-II clusters. 
}

\new{
Because dwarf galaxies have their own chemical enrichment history, dwarf galaxies show different trends in ([Fe/H], [X/Fe])-space \citep{Tolstoy2009ARA&A..47..371T}. By obtaining spectroscopic measurements of various elemental abundances, we may be able to find real $r$-II clusters that are remnants of dwarf galaxies. 
}

\new{
An issue of our greedy optimistic clustering method is that stars with poor parallax (or distance) measurements can contaminate real clusters. This kind of contamination effect can be reduced by improving the distance estimates of the $r$-II stars, by using parallax from future data releases of {\it Gaia} or using better photometric distances.  
}

\subsection{Caveats in our analysis}

\subsubsection{\new{Assumptions on the gravitational potential}}

In this paper, 
we performed a clustering analysis of $r$-II stars in the orbital action space. 
In computing the orbital action $\vector{J}(\vector{x}, \vector{v})$ 
from the observed position-velocity data 
$(\vector{x}, \vector{v})$, 
we assumed a gravitational potential model of the Milky Way
$\Phi(\vector{x})$
and the Solar position and velocity with respect to the Galactic center. 
Therefore, 
our clustering analysis is necessarily affected 
by any bias on these assumptions. 
For example, seven stars travel beyond $r>30 \kpc$ 
(see $r_\mathrm{apo}$ in Table~\ref{table:members_additional_info}), 
where the perturbation from the Large Magellanic Cloud may be important 
\citep{Besla2010,Erkal2019,Erkal2021MNRAS.506.2677E,GaravitoCamargo2019,Koposov2019,Conroy2021Natur.592..534C,Petersen2021NatAs...5..251P,Shipp2021ApJ...923..149S}. 
Also, some groups of stars that pass near the bulge region with prograde motion might be affected by the rotating potential of the Galactic bar 
\citep{Hattori2016,PriceWhelan2016MNRAS.455.1079P}. 
Although these complexities are not included in our study, 
we believe they do not seriously affect our results, 
especially the result on Tier-1 clusters, 
because we validated our results 
with chemical abundance information 
that are independent from the dynamical analysis of this paper. 

\subsubsection{\new{Assumptions on the distribution of the orbital action of $r$-II stars}} \label{sec:caveat_action_distribution}

\new{
We assume that the distribution of the $r$-II stars in the $\vector{J}$-space 
is described by a mixture of isotropic Gaussian distributions with identical dispersion $\sigma_J^2$, as expressed in equation (\ref{eq:GMM}). 
This is a natural assumption 
if all the $r$-II stars in our catalog 
originate from small dwarf galaxies such as UFDs, 
but in reality it is uncertain if this assumption is valid. 
For example, if some fraction of $r$-II stars 
originate from disrupted large dwarf galaxy 
(such as Gaia-Sausage/Enceladus), 
we may expect diffuse and smooth background of $r$-II stars not associated with small clumps. 
Indeed, 
there are some indications that 
large dwarf galaxies do include $r$-enhanced stars 
(\citealt{Matsuno2021A&A...650A.110M} for Gaia-Sausage/Enceladus;  \citealt{Reichert2021ApJ...912..157R} for Fornax dwarf galaxy; see also \citealt{Xing_et_al__2019}, who discuss the origin of an $r$-II star 
that exhibits low [Mg/Fe]). 
Because the fraction of $r$-II stars in 
large dwarf galaxies are expected to be low \citep{Hirai2022MNRAS.517.4856H}, 
we believe it is justifiable to apply our method 
to $r$-II stars.  
However, since we did not analyze the effect of 
diffuse and smooth background of $r$-II stars 
in our mock-data analysis \citep{OkunoHattori2022}, 
it is unclear how such a background population 
may affect our results \citep{Brauer2022arXiv220607057B}. 
This issue may be a scope for future studies. 
}

\subsubsection{\new{Very-metal-poor $r$-II stars in our sample}} \label{sec:caveat_FeH}

\new{
Our analysis is motivated by the discovery of an Eu-rich UFD, Reticulum II. 
Given that known members of surviving UFDs typically have $\mathrm{[Fe/H]}<-2$, 
it may be interesting to see how our results are affected 
if we pre-select very-metal-poor $r$-II stars with [Fe/H]$<-2$ 
before performing the clustering analysis. 
We did an additional analysis in this direction, 
as described in Appendix \ref{sec:VMP}. 
We found that the clustering results are similar to our fiducial result. 
With this simple [Fe/H] selection, 
one may fail to discover some interesting clusters, 
such as H22:DTC-2 (which corresponds to group B in \citealt{Roederer2018}), 
whose mean metallicity is  
$\langle\mathrm{[Fe/H]}\rangle = -1.65$. 
Thus, we assert that using the entire sample of $r$-II stars available 
is more informative than using only very-metal-poor $r$-II stars. 
}

\section{Conclusion} \label{sec:conclusion}

In this paper, 
we extended the work in \cite{Roederer2018} 
and performed a clustering analysis of $N=161$ $r$-II stars 
([Eu/Fe]$\geq0.7$ and [Ba/Fe]$<0$)
in the orbital action ($\vector{J}$) space. 
Our data set is the largest catalog of $r$-II stars, 
which includes 
$r$-II stars discovered before the end of 2020 
(see Tables~\ref{table:members_all_info} and \ref{table:members_additional_info}). 
For all the sample stars, 
we have astrometric data from Gaia EDR3, 
and therefore our catalog supersedes 
the catalog in \cite{Roederer2018} 
in which Gaia DR2 data were used. 
To our updated catalog, 
we applied a newly-developed {\it greedy optimistic clustering method} \citep{OkunoHattori2022}, 
which allows us to analyze 
not only stars with good observational data 
but also stars with poor data. 
As a result, we were able to analyze 
all the $r$-II stars in our catalog, 
without discarding stars with large observational uncertainty.

The summary of this paper is as follows. 
\begin{itemize}

\item
By using the Gaussian Mixture Model equipped with 
the greedy optimistic clustering algorithm, 
we performed the clustering analysis in the orbital action space. 
The orbital action distribution of $N=161$ $r$-II stars 
can be described by 
$K=30$ clusters with intrinsic internal dispersion of \textcolor{red}{$\sigma_J$}$=100 \kpc\kms$ (see Table~\ref{table:clusters} and Fig.~\ref{fig:action_distribution}).

\item 
The groups A-H discovered in \cite{Roederer2018} are 
recovered in our analysis (see Table~\ref{table:clusters}). 
Specifically, groups B, C, D, and G are identified as separate clusters in our analysis. 
Groups A and F are identified as a single (big) cluster in our analysis. 
Groups E and H (and a star in group F) are identified as a single (big) cluster in our analysis (see Section \ref{subsec:consistency_with_R18}). 
For all groups in \cite{Roederer2018}, 
we found additional member stars 
(see Table~\ref{table:members_all_info}).

\item
Among 26 clusters with $N_{\mathrm{member},k} \geq 2$ member stars, 
13 clusters have metallicity dispersion of 
$\sigma_\mathrm{[Fe/H]}<0.35$, 
which is equivalent to or smaller than the dispersion 
in the $r$-enhanced UFD, Reticulum II. 
This result indicates that many of the filed $r$-II stars may have originated from disrupted dwarf galaxies
(see Section \ref{subsec:percentile}).

\item 
We validated our clustering result 
by using the chemical abundance data, 
which we did not use in the clustering analysis. 
Based on the tightness of the distribution of [Fe/H] and [Eu/H], 
we categorized our clusters into five categories:
Tier-1 clusters (for which we have the highest confidence); 
Tiers-2, 3, and 4 clusters (with decreasing confidence); 
and single-member clusters 
(see Sections \ref{subsec:percentile} and \ref{subsec:5class}).

\item 
We found six $r$-II clusters 
(H22:DTC-$1,2,3,4,5$, and $9$; Tier-1 clusters) 
with tight distributions in [Fe/H], [Eu/H], [Mg/Fe], and [Ca/Fe] 
(see Figs.~\ref{fig:TierI} and \ref{fig:TierI_alpha}). 
Because the chemical information is not used 
in the clustering analysis, 
the chemical homogeneity of these clusters suggests 
that these six clusters are 
\new{likely}
to be genuine clusters. 
Given that the member stars of Tier-1 clusters are 
apparently completely phase-mixed 
(see Fig.~\ref{fig:position_velocity_distribution}), 
we interpret Tier-1 clusters 
as the remnants of completely disrupted dwarf galaxies 
that merged with the ancient Milky Way (see Section \ref{subsubsec:tier1}). 
\new{However, more data are desired to confirm this scenario (see Section \ref{sec:future_prospects}).}

\item
The cluster 
H22:DTC-$1$ 
is a newly discovered cluster with 
$N_{\mathrm{member},k}=9$ member stars. 
This cluster has the tightest distribution in [Fe/H] 
in terms of the quantities $q_\text{[Fe/H]}$ introduced in Section \ref{subsec:percentile}. 
Two of the member stars 
are among 35 $r$-II stars 
analyzed in \cite{Roederer2018}  
and regarded by that study as $r$-II stars not associated with any groups.
The fact that these two stars (as well as the other 7 member stars) 
are successfully considered as a single cluster 
highlights the advantage of using the greedy optimistic clustering method 
(see Section \ref{subsubsec:tier1}).

\item 
Apart from a Tier-1 cluster 
H22:DTC-$1$, 
we identified many new $r$-II clusters 
that have not been identified in previous studies. 
All the nine clusters in Tiers-2 and 3 are newly discovered (see Appendix \ref{sec:tier234}).

\item 
We found four $r$-II clusters 
(H22:DTC-$27, 28, 29$, and $30$) 
with a single member star. 
Two of these clusters, 
H22:DTC-$29$ and $30$,
are the most metal-poor $r$-II clusters 
characterized by a highly eccentric orbit 
with $ecc \simeq 0.88$ and $r_\mathrm{apo}>90 \kpc$ 
(see Section \ref{subsubsec:single_member_cluster}).

\item 
In the hierarchical galaxy formation paradigm, 
some small stellar systems such as dwarf galaxies or globular clusters 
merge together as a group. 
In accordance with this scenario, 
it has been claimed that 
some stellar streams, globular clusters, and dwarf galaxies 
in the Milky Way 
are clustered in phase space 
\citep{Bonaca2021ApJ...909L..26B, Malhan2022ApJ...926..107M}. 
Recently, \cite{Malhan2022ApJ...926..107M} reported 
that there are six big merger groups. 
We checked the 30 $r$-II clusters obtained in this study 
and found that eight $r$-II clusters are associated with four of the merger groups: 
Gaia-Sausage/Enceladus, Arjuna/Sequoia/I’itoi, 
LMS-1/Wukong, and Pontus 
(see Section \ref{subsec:Malhan_groups} and Fig.~\ref{fig:malhan_groups}).

\item
Two Tier-1 clusters 
H22:DTC-$1$ and $9$ 
are very metal poor ([Fe/H]$\simeq-2.8$), 
which indicates that their progenitor systems 
were low-mass UFD-like systems with stellar mass of $M_*\sim 10^4 M_\odot$ 
according to the mass-metallicity relationship 
\citep{Kirby2013ApJ...779..102K,Naidu2022arXiv220409057N}. 
Intriguingly, these two clusters have similar chemistry and orbits. 
They may be the remnants of two 
$r$-enhanced dwarf galaxies that merged with the Milky Way as a group. 
If these clusters are associated, 
they may constitute a new merger group 
which is separate from previously known merger groups 
found in \cite{Bonaca2021ApJ...909L..26B} and \cite{Malhan2022ApJ...926..107M} 
(see Section \ref{subsec:new_merger_group}).

\end{itemize}

\acknowledgments

K.H. thanks useful suggestions from Hideitsu Hino on the split-and-merge EM algorithm, and Yukito Iba for comments on mock data analyses. 
K.H. thanks Kazuhei Kikuchi for encouraging this project. 
K.H. is supported by JSPS KAKENHI Grant Numbers JP21K13965 and JP21H00053.
A.O. is supported by JSPS KAKENHI Grant Number JP21K17718, and JST CREST Number JPMJCR21N3. 
I.U.R.\ is supported by the U.S.\ National Science Foundation grants PHY~14-30152 (Physics Frontier Center/JINA-CEE), AST~1815403/1815767, and AST~2205847, as well as the
NASA Astrophysics Data Analysis Program grant 80NSSC21K0627. 
This work has made use of data from the European Space Agency (ESA) mission
{\it Gaia} (\url{https://www.cosmos.esa.int/gaia}), processed by the {\it Gaia}
Data Processing and Analysis Consortium (DPAC,
\url{https://www.cosmos.esa.int/web/gaia/dpac/consortium}). Funding for the DPAC
has been provided by national institutions, in particular the institutions
participating in the {\it Gaia} Multilateral Agreement.

\facility{Gaia}

\software{
AGAMA \citep{Vasiliev2019_AGAMA},\;
matplotlib \citep{Hunter2007},
numpy \citep{vanderWalt2011},
scipy \citep{Jones2001}}

\bibliographystyle{aasjournal}
\bibliography{mybibtexfile}

\appendix

\section{A demonstration of the greedy optimistic clustering}
\label{sec:idea}

To perform a clustering analysis for a nosiy data set, 
we introduce the greedy optimistic clustering method \citep{OkunoHattori2022}. 
In the greedy optimistic clustering, 
we simultaneously estimate both 
(i) the centroids of the clusters and 
(ii) the {\it true} orbital action $\vector{J}_i^\text{true}$. 
By contrast, 
in the conventional clustering methods, 
we are mainly interested in estimating the centroids of the clusters by using the point estimate of the orbital action $\widehat{\vector{J}_i}$. 
Here we explain how these clustering methods work differently. 
As a demonstration, 
we focus on a clustering of three $r$-II stars 
under an assumption that these stars are members of a single cluster. 
Fig.~\ref{fig:point_estimate_and_greedy_optimistic} 
shows the distribution of $\vector{J}$ for stars A, B, and C. 
Stars A, B, and C respectively have 
$\varpi/\sigma_\varpi = 16.63, 11.18$, and $26.83$; 
and  
$\Delta J / (\kpc \kms)= 126, 353$, and $138$. 
The names of these stars are described in the caption of Fig.~\ref{fig:point_estimate_and_greedy_optimistic}. 
These stars are taken from the cluster 
H22:DTC-$1$
that we find in our main analysis of this paper. 
Apart from this fact, the content of this Appendix is independent from the main analysis of this paper.

\subsection{Conventional clustering methods}

On the left-hand panels of Fig.~\ref{fig:point_estimate_and_greedy_optimistic}, 
the point-estimate $\widehat{\vector{J}}_i$ ($i \in \{A,B,C\}$) 
of each star (big black symbol) is 
derived from the point-estimate of the observables. 
Under an assumption that stars A, B, and C are associated with one cluster, 
conventional clustering methods find the centroid 
in the `middle' of these three point estimates. 
(The definition of the `middle' depends on the clustering methods.)

The small gray symbols on the left-hand panels of Fig.~\ref{fig:point_estimate_and_greedy_optimistic} 
represent the uncertainty set of $\vector{J}$ 
for these stars. 
We see that the distribution of the uncertainty set for each star is highly elongated, 
despite the relatively small parallax uncertainty  ($\varpi/\sigma_\varpi>10$). 
The elongated shape arises from the fact 
that the parallax uncertainty dominates the uncertainty in $\vector{J}$, 
as explained in Section \ref{sec:why}. 
For example, 
if we adopt a small value of $\varpi$ for star B, 
both $J_r$ and $J_z$ become large; and vice versa.

As we can see from the uncertainty set, 
the point estimate is just one of many possibilities. 
Even when the parallax uncertainty is small, 
the point estimate may be very different from the true orbital action. 
Because conventional clustering methods use the point estimate 
$\widehat{\vector{J}}_i$, 
conventional clustering methods may fail to work 
when the observational uncertainty is not negligible.

\subsection{Greedy optimistic clustering method} 

\begin{figure*}
\centering
\includegraphics[width=6.3in]{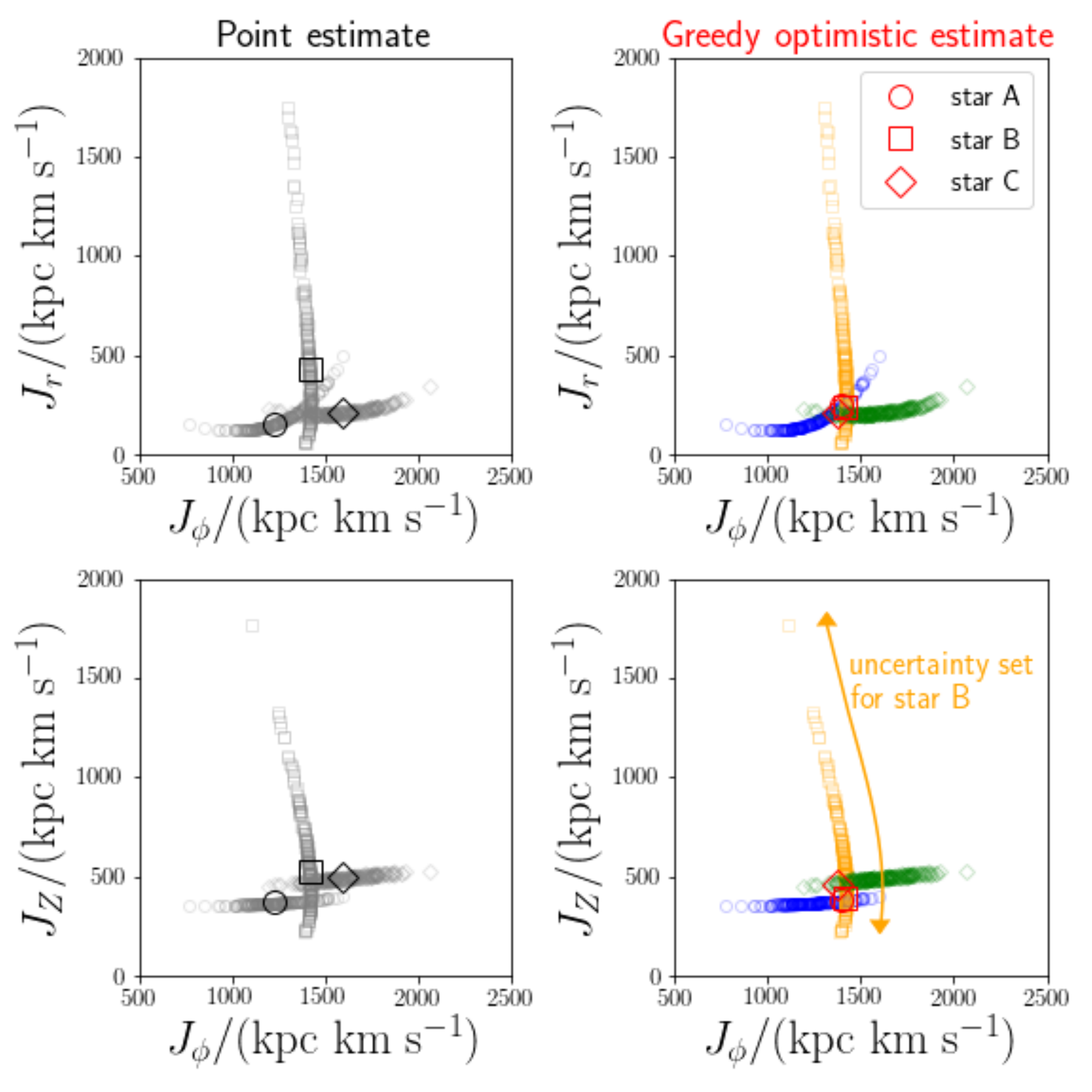}
\caption{
The distribution of the orbital action for
stars A (HE~1523$-$0901), 
B (RAVE J203843.2$-$002333), and 
C (2MASS J09544277+5246414), 
which are taken from the cluster 
H22:DTC-1 
in our main analysis. 
(Left panels)
The big black symbols are the point estimate $\widehat{\vector{J}}_{i}$ ($i \in \{A,B,C\}$). 
The small gray symbols represent the uncertainty set for each star, which are shown for a reference. 
(Right panels)
The big red symbols are the greedy optimistic estimate $\vector{J}_{i,\beta_i}$. 
The small colored symbols represent the uncertainty set for each star.
The uncertainty set for star B is marked 
to illustrate its banana-shaped distribution. 
We see that the distribution of 
$(\vector{J}_{A,\beta_A}, \vector{J}_{B,\beta_B}, \vector{J}_{C,\beta_C})$ 
is more condensed than that of 
$(\widehat{\vector{J}}_A, \widehat{\vector{J}}_B, \widehat{\vector{J}}_C)$. 
}
\label{fig:point_estimate_and_greedy_optimistic}
\end{figure*}

In the greedy optimistic clustering, we estimate the {\it true} orbital action $\vector{J}_i^\text{true}$ under some assumptions 
and then perform a clustering analysis. 
Because the estimation of $\vector{J}_i^\text{true}$ is essential in this clustering method, 
we demonstrate how we estimate $\vector{J}_i^\text{true}$ 
by using the example case 
in Fig.\ref{fig:point_estimate_and_greedy_optimistic}.

On the right-hand panels of Fig.~\ref{fig:point_estimate_and_greedy_optimistic}, the uncertainty sets of $\vector{J}$ are shown with small colored symbols. 
In the greedy optimistic clustering, 
we assume that the true orbital action 
$\vector{J}_i^\text{true}$ of star $i$ 
is very close to one of the $M$ instances (realizations) 
$\{ \vector{J}_{i,j} \}$
in the uncertainty set of that star. 
In other words,
we assume that one of the $M^3$ combinations of 
$\{ (\vector{J}_{A, j_A}, \vector{J}_{B, j_B}, \vector{J}_{C, j_C}) \}$ 
is very close to the true orbital actions of these stars 
$(\vector{J}_{A}^\text{true}, \vector{J}_{B}^\text{true}, \vector{J}_{C}^\text{true})$. 
In addition, we {\it optimistically} assume that the 
true orbital actions 
$(\vector{J}_{A}^\text{true}, \vector{J}_{B}^\text{true}, \vector{J}_{C}^\text{true})$ 
are highly clustered in the $\vector{J}$-space. 
Under these assumptions, 
we perform a {\it greedy} search for the 
best combination. 
As a demonstration, 
among $M^3$ combinations of $\{ (\vector{J}_{A, j_A}, \vector{J}_{B, j_B}, \vector{J}_{C, j_C}) \}$, 
we find the best combination 
$(\vector{J}_{A, \beta_A}, \vector{J}_{B, \beta_B}, \vector{J}_{C, \beta_C})$ that achieves the minimum internal dispersion in $\vector{J}$ 
with a brute-force approach. 
The orbital action estimated through this procedure 
is the {\it greedy optimistic estimate} of the orbital action.
The greedy optimistic estimates 
are shown by 
the big red symbols on the right-hand panels of Fig.~\ref{fig:point_estimate_and_greedy_optimistic}. 
By comparing 
$\vector{J}_{i, \beta_i}$ 
(right-hand panels)
with 
$\widehat{\vector{J}}_{i}$ 
(left-hand panels), 
we can visually confirm that 
the greedy optimistic estimates are more condensed. 
Indeed, 
the dispersion of $\{ \vector{J}_{i, \beta_i} \}$ 
is only $41 \kpc \kms$, 
while 
the dispersion of $\{ \widehat{\vector{J}}_{i} \}$ 
is as large as $203 \kpc \kms$.

The most important assumption in the greedy optimistic clustering 
is the assumption that the true orbital actions are highly clustered in $\vector{J}$-space. 
Because there is no guarantee that 
this optimistic assumption is valid, 
it is important to validate our clustering results 
with an independent set of data. 
This is why, 
in the main analysis of this paper, 
we check the chemical information for each cluster.


\section{A comment on the computational cost of GOEM algorithm}

In this paper, we use GOEM algorithm 
to find the best solution of the greedy optimistic GMM. 
Here we comment on the computational cost of this algorithm.

If we had an infinite amount of computing resources, we could use the following steps to find the best solution. 
(1) We try $M^N$ combinations of $\{ \beta_i \}$. 
(2) We apply the conventional GMM to each of the $M^N$ configurations of the data points. For each configuration, we find the best parameters (i.e., the centrids and weights) that maximize the object function in equation (\ref{eq:lnL_optimisticEM}).  
(3) Find the best configurations and parameters that maximize the object function in equation (\ref{eq:lnL_optimisticEM}). 
Obviously, such a brute-force strategy is computationally expensive, 
because we would need to perform GMM fitting for $M^N \sim 10^{322}$ times for our case with $(M,N)=(101,161)$. 
(In Appendix \ref{sec:idea}, we use a brute-force strategy because it is easier to understand.) 
Fortunately, 
in the mock data analysis in \cite{OkunoHattori2022}, 
we typically need $\sim20$ iterations of GO, E, and M steps 
to reach a good solution. 
This computational cost is obviously much smaller than a brute-force approach.

\section{Clusters in Tiers 2, 3, and 4}
\label{sec:tier234}

In the main body of this paper, 
we showed the distribution of orbital action and chemical abundances 
of Tier-1 clusters, for which we have the highest confidence in our results. 
Here we show the distribution of stars 
for other clusters that are worth mentioning.

\subsection{Tier-2: Five interesting clusters}
\label{subsec:tier2}

\begin{figure*}
\centering
\includegraphics[width=1.1in]
{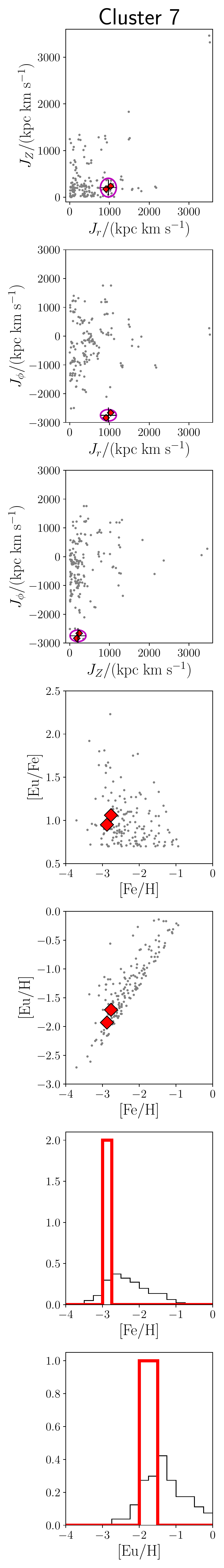}
\includegraphics[width=1.1in]
{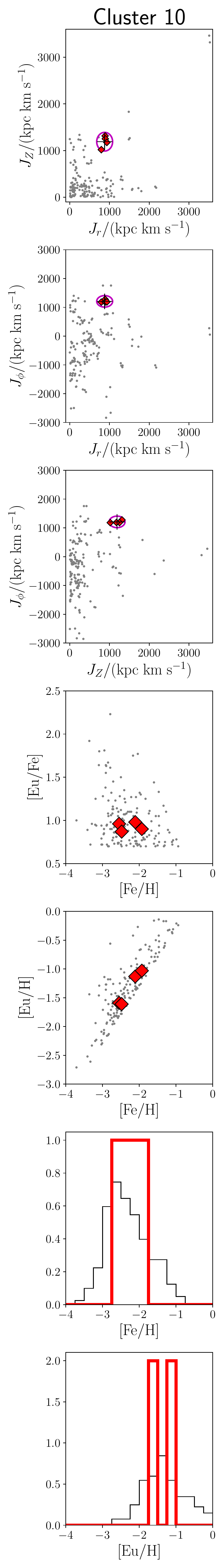}
\includegraphics[width=1.1in]
{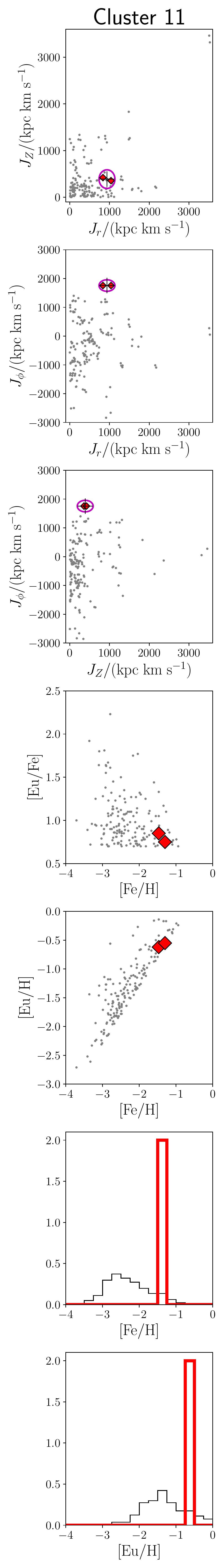}
\includegraphics[width=1.1in]
{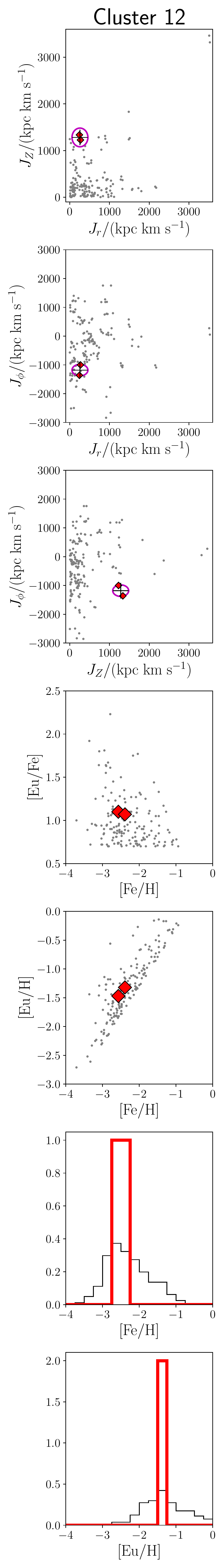}
\includegraphics[width=1.1in]
{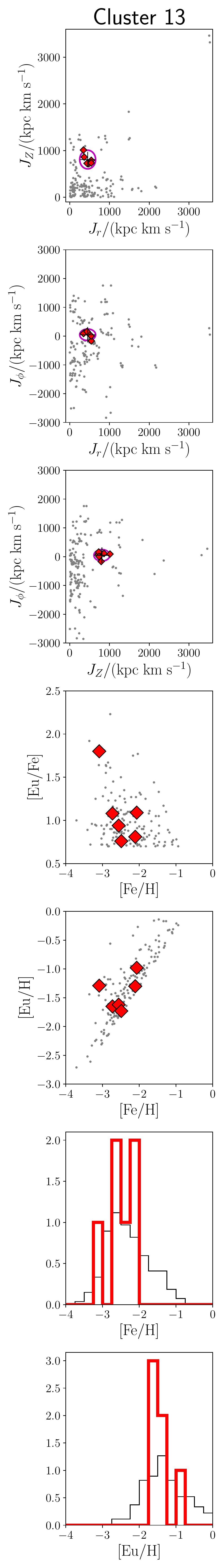}
\caption{
The same as Fig.~\ref{fig:TierI}, 
but for Tier-2 clusters 
(H22:DTC-$7, 10, 11, 12$, and $13$). 
}
\label{fig:TierII}
\end{figure*}

Apart from Tier-1 clusters, 
there are five clusters that satisfy both 
$q_\mathrm{[Fe/H]} < 25 \%$ and 
$q_\mathrm{[Eu/H]} < 25 \%$. 
We label these clusters as Tier-2 clusters, 
which include clusters with 
H22:DTC-$7, 10, 11, 12$, and $13$. 
All of these clusters are newly discovered. 
Fig.~\ref{fig:TierII} shows that both [Fe/H] or [Eu/H] 
have a moderately tight distributions (see Fig.~\ref{fig:TierII}).

\uline{\bf{Tier-2 cluster H22:DTC-7}.}

This cluster is one of the lowest metallicity and highly prograde cluster 
with 
$\langle \mathrm{[Fe/H]} \rangle = -2.83$ 
and 
$\langle \mathrm{[Eu/H]} \rangle = -1.82$. 
It contains only $N_{\mathrm{member},k}=2$ member stars, 
but their [Fe/H] and [Eu/H] are very close to each other.

\uline{\bf{Tier-2 cluster H22:DTC-10}.}

This cluster has $N_{\mathrm{member},k}=4$. 
Apart from the most metal-poor member, BPS CS 29491-069, 
the other three stars have poor parallax measurement 
with $\varpi / \sigma_\varpi < 5$ (see Table~\ref{table:members_all_info}). 
This is a clear example that our optimistic clustering 
can find a candidate group even if the observational error is large.

\uline{\bf{Tier-2 cluster H22:DTC-11}.}

This cluster has $N_{\mathrm{member},k}=2$. 
Its mean metallicty is the highest among the 30 clusters, 
with $\langle \mathrm{[Fe/H]} \rangle = -1.39$. 

\uline{\bf{Tier-2 cluster H22:DTC-12}.}

This cluster has $N_{\mathrm{member},k}=2$ 
with a shell-like orbit charactrized by small $J_r$ and large $J_z$.

\uline{\bf{Tier-2 cluster H22:DTC-13}.}

This cluster has $N_{\mathrm{member},k}=6$. 
It has the smallest $||\langle J_\phi \rangle|| = 47 \kpc \kms$, 
which corresponds to highly radial orbits. 

\subsection{Tier-3: Four promising clusters}
\label{subsec:tier3}

\begin{figure*}
\centering
\includegraphics[width=1.1in]
{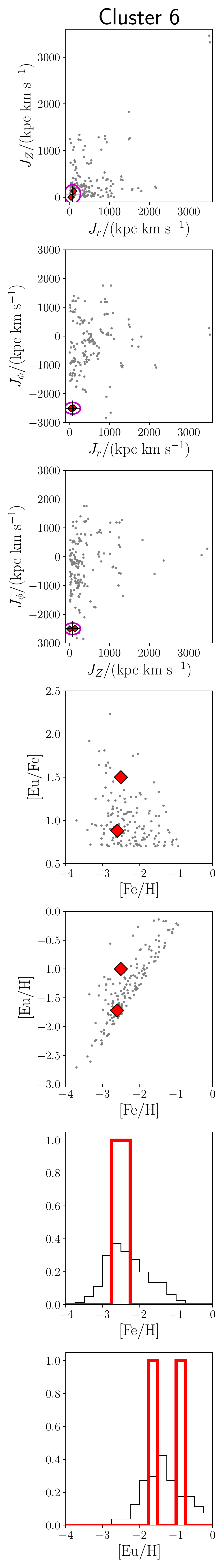}
\includegraphics[width=1.1in]
{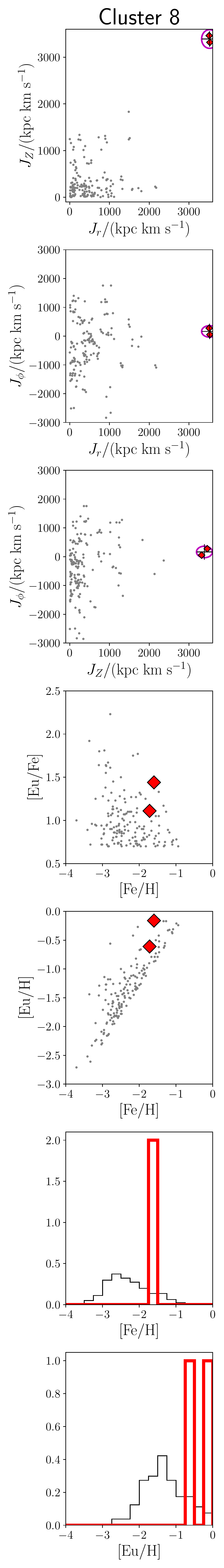}
\includegraphics[width=1.1in]
{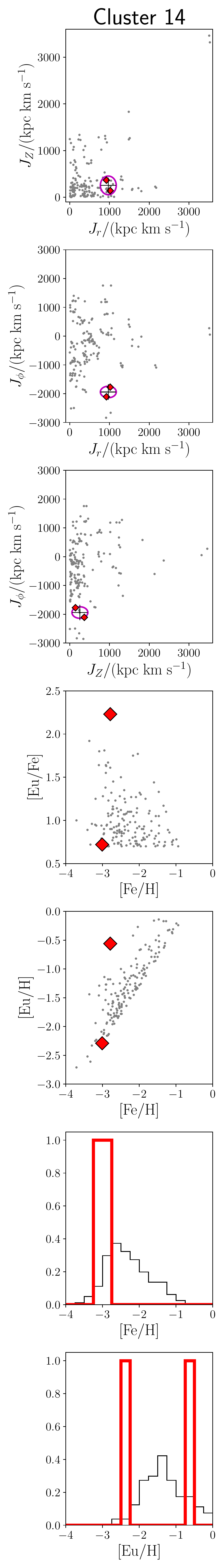}
\includegraphics[width=1.1in]
{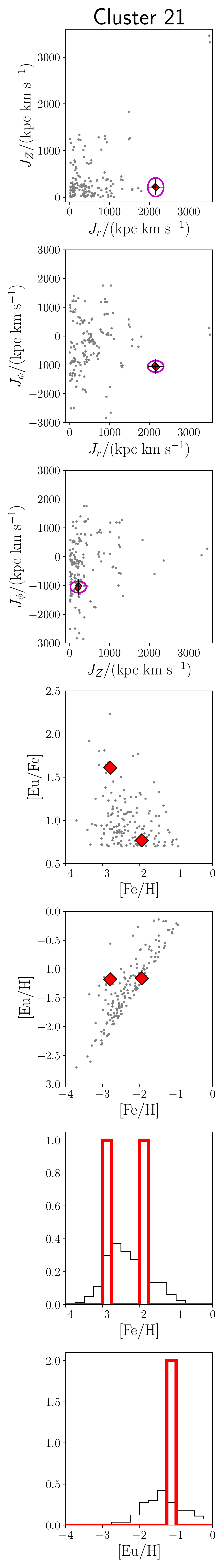}
\caption{
The same as Fig.~\ref{fig:TierI}, 
but for Tier-3 clusters 
(H22:DTC-$6, 8, 14$, and $21$). 
}
\label{fig:TierIII}
\end{figure*}

Apart from Tiers-1 and 2 clusters, 
there are four clusters that satisfy either 
$q_\mathrm{[Fe/H]} < 25 \%$ or 
$q_\mathrm{[Eu/H]} < 25 \%$. 
We label these clusters as Tier-3 clusters, 
which include clusters with 
H22:DTC-$6, 8, 14$, and $21$. 
All of the Tier-3 clusters are newly found, 
and have $N_{\mathrm{member},k}=2$ member stars. 
The distribution of stars in the action and chemistry space is shown in Fig.~\ref{fig:TierIII}. 

\uline{\bf{Tier-3 cluster H22:DTC-6}.}

This group is characterized by a prograde, nearly circular orbit with 
$\langle J_\phi \rangle = -2504 \kpc \kms$, 
corresponding to a guiding center radius of $R \sim 10 \kpc$.

\uline{\bf{Tier-3 cluster H22:DTC-8}.}

This group is characterized by a highly radial orbit 
with 
$||\langle J_\phi \rangle|| = 163 \kpc \kms$.

\uline{\bf{Tier-3 cluster H22:DTC-14}.}

One of the member stars of this cluster is {2MASS J15213995$-$3538094}, 
which has the most enhanced value of [Eu/Fe]$=+2.23$ in our catalog \citep{Cain2020ApJ...898...40C}. 

\uline{\bf{Tier-3 cluster H22:DTC-21}.}

This cluster is characterized by a relatively large dispersion in [Fe/H], 
$\sigma_\mathrm{[Fe/H]}=0.43$; 
but has a very small dispersion in [Eu/H],  $\sigma_\mathrm{[Eu/H]}=0.01$.

\subsection{Tier-4: Three possibly promising clusters}
\label{subsec:tier4}

\begin{figure}
\centering
\includegraphics[width=1.10in]
{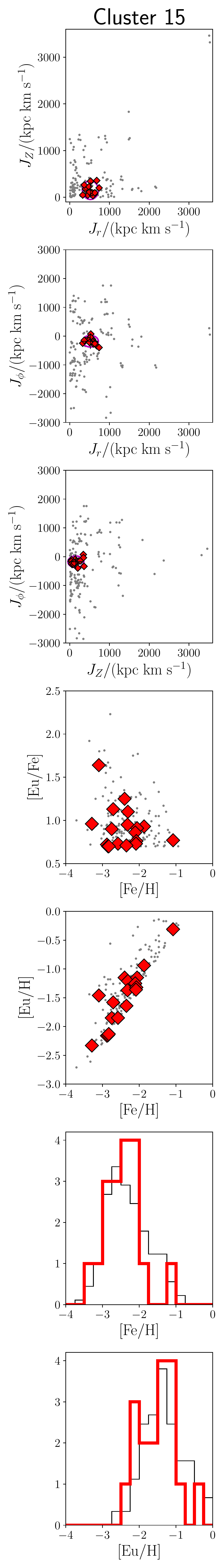}
\hspace{-2mm}
\includegraphics[width=1.10in]
{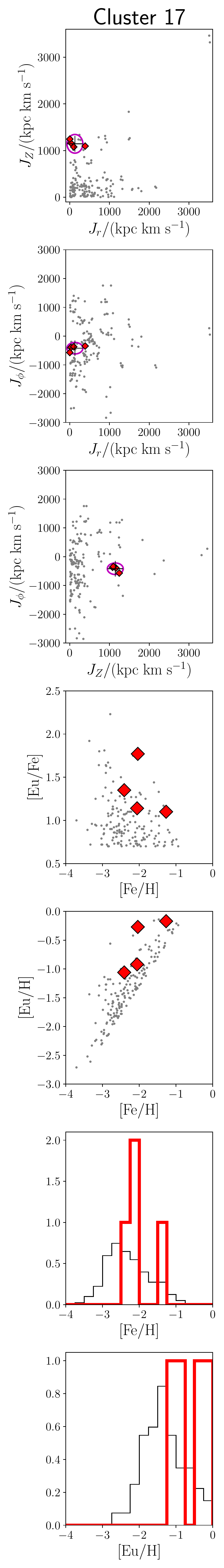}
\hspace{-2mm}
\includegraphics[width=1.10in]
{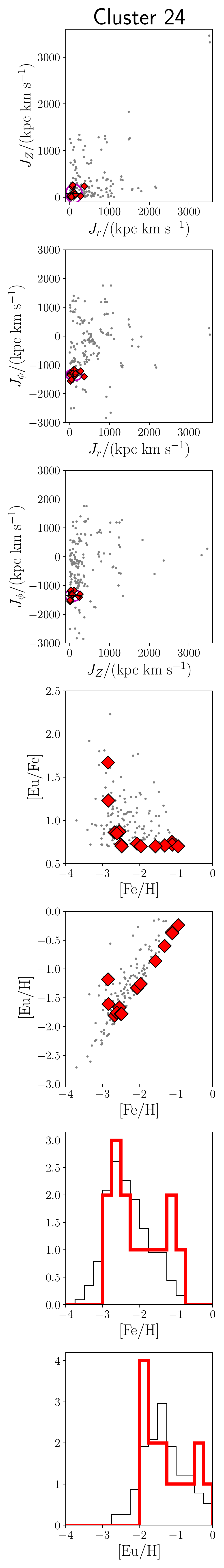}
\caption{
The same as Fig.~\ref{fig:TierI}, 
but for Tier-4 clusters 
(H22:DTC-$15, 17$, and $24$).
}
\label{fig:TierIV}
\end{figure}

There are three additional clusters 
(H22:DTC-$15, 17$, and $24$) 
that are worth attention. 
We label them as Tier-4.  
The distribution of stars in the action and chemistry space is shown in Fig.~\ref{fig:TierIV}. 

\uline{\bf{Tier-4 cluster H22:DTC-15}.}

This group has $N_{\mathrm{member},k}=18$ member stars. 
Its relatively large dispersion in [Fe/H] 
is mostly due to a single outlier, {G210-33}, 
which has [Fe/H]$=-1.08$. 
If we manually exclude this star, 
the standard deviations in [Fe/H] and [Eu/H] are 
$0.40$ and $0.41$, respectively. 
The corresponding percentile values are 
$q_\mathrm{[Fe/H]}=4.88$ and $q_\mathrm{[Eu/H]}=7.90$, 
respectively. 
Therefore, without one outlier star, 
this cluster could be classified as a Tier-1 cluster. 
Even after removing {G210-33}, 
this cluster contains 
all the stars in groups E and H 
and a star in group F in \cite{Roederer2018}. 

\uline{\bf{Tier-4 cluster H22:DTC-17}.}

This is a cluster of size $N_{\mathrm{member},k}=4$. 
This cluster includes {Gaia DR2 6412626111276193920} 
(also known as {Indus\_13})
which is a member of the Indus stream \citep{Hansen2021ApJ...915..103H}. 
The remaining three $r$-II stars in this group 
have very different orbital phases, 
suggesting that they are unlikely to be the members of the Indus stream. 
These $r$-II stars might have originated from different dwarf galaxies 
that accreted to the Milky Way together.

\uline{\bf{Tier-4 cluster H22:DTC-24}.}

This group has $N_{\mathrm{member},k}=14$ member stars. 
\textcolor{red}{\textbf{As}} seen in Fig.~\ref{fig:TierIV} 
(\textcolor{red}{\textbf{right}} column), 
these 14 stars show a L-shaped distribution in [Fe/H]-[Eu/Fe] diagram. 
The 7 stars above [Fe/H]$=-2.1$ have [Eu/Fe]$\simeq 0.7$, 
which is the lower boundary of our sample selection. 
If we manually select stars below [Fe/H]$=-2.1$, 
the standard deviations in [Fe/H] and [Eu/H] are 
$0.14$ and $0.20$, respectively. 
The corresponding percentile values are 
$q_\mathrm{[Fe/H]}=0.16$ and $q_\mathrm{[Eu/H]}=1.18$, 
respectively. 
Therefore, with this (arbitrary) manual selection of 7 stars, 
this cluster could be classified as a Tier-1 cluster.

\section{Member stars of all the clusters}
\label{sec:list_of_stars}

Table~\ref{table:members_all_info} 
lists the member stars of all the $K=30$ clusters 
and the basic chemical and dynamical properties of the member stars. 
Table~\ref{table:members_additional_info}  
lists additional kinematical and orbital information of the $r$-II stars. 
See Section \ref{subsec:fiducial_result} for the description of these tables.

\section{A comment on the `edge effect' of greedy optimistic solution} \label{sec:edge}

\new{
An alert reader may notice that,
in the top row in Fig.~\ref{fig:action_distribution}, 
the greedy optimistic estimates of some stars are located near the edge of their uncertainty sets. 
To confirm the validity of our result, we conduct an additional test on our fiducial solution. In this test, we use $K=30$ centroids 
$\langle \vector{J} \rangle_k$ 
($k=1,\cdots,K$) obtained from the fiducial analysis and the as-observed orbital action of $N=161$ stars $\vector{J}_{i} \equiv \vector{J}_{i,50}$ ($i=1,\cdots,N$). 
For each star $i$, we find the nearest centroid $k^\prime$ such that $|| \vector{J}_{i} - \langle \vector{J} \rangle_{k^\prime} ||$ is minimized. 
For each star $i$, we compare $k^\prime$ and the value of $k$ in our fiducial analysis. 
As a result, we find 
that (i) 137 stars among 161 stars (85 percent) satisfy $k^\prime=k$;
and that (ii) 53 stars among 59 stars in Tier-1 clusters (90 percent) satisfy $k^\prime=k$.  
These results indicate that, in most cases, the choice of the instance of the uncertainty set is not critical in assigning the cluster. 
As demonstrated in Appendix \ref{sec:idea}, our algorithm tries to shrink the cluster in the action space; and depending on the configuration of the uncertainty set and the centroid, the edge of the uncertainty set is chosen in our method. 
}

\clearpage

\section{Clustering analysis using standard GMM and high-quality data}
\label{sec:standardGMM}

\new{
The originality of this paper is that we use a new clustering algorithm, 
namely the greedy optimistic GMM, 
in finding candidates for clusters in the $\vector{J}$-space. 
Although our approach has an advantage in that stars with low-quality kinematic data can be used, it has a disadvantage in that stars with low-quality data might contaminate clusters. 
In this regard, the resultant clusters in our fiducial analysis might be contaminated by stars that are not supposed to be true member stars. 
In the main text of this paper, we carefully analyze the chemical information of the member stars to conclude that Tier-1 clusters 
are more plausible candidates for disrupted dwarf galaxies than other $r$-II clusters. 
}

\new{
To further examine the validity of our fiducial result, 
here we perform an additional clustering analysis using 
$N=119$ $r$-II stars with high-quality astrometric data  
defined by \texttt{parallax\_over\_error} $>10$. 
In this analysis, 
we fix $\sigma_J = 100 \kpc \kms$ and $K=30$ as in our fiducial analysis. 
Also, we use the standard GMM instead of the greedy optimistic GMM. 
}

\new{
In the fiducial result, 
we have $K=30$ clusters consisting of $N=161$ stars. 
Among these 30 clusters, 
we have 
(i) 11 clusters in which all the member stars have high-quality kinematic data 
(\texttt{parallax\_over\_error} $>10$); 
(ii) 2 clusters in which all the member stars have low-quality kinematic data 
(\texttt{parallax\_over\_error} $<10$); and 
(iii) 17 clusters which include both stars with high-quality data and stars with low-quality data. 
From the fiducial result, we discard 2 clusters in item (ii). 
Also, we discard member stars with \texttt{parallax\_over\_error} $<10$ 
from 17 clusters in item (iii). 
As a result, we end up with $K=28$ clusters consisting of 
$N=119$ stars with \texttt{parallax\_over\_error} $>10$. 
}

\new{
To investigate the similarity between the two sets of clusters mentioned above, 
namely 
$30$ clusters described in the second paragraph 
(i.e., an additional clustering result using stars with high-quality data) 
and 
$28$ clusters described in the third paragraph 
(i.e., a subset of the fiducial analysis), 
we compute four similarity indices. 
We find that 
the purity is $0.883$, 
Normalized Mutual Information is $0.888$, 
Rand index is $0.968$, 
and 
F-measure is $0.846$. 
These indices are close to unity, 
which means that these two sets of clusters are similar to each other. }

\new{
We also investigate how the member stars of Tier-1 clusters in the fiducial result are classified in our additional analysis. 
As a result, most member stars in each Tier-1 cluster are successfully identified as a single cluster. 
This result indicates that, as long as we use high-quality data only, 
Tier-1 clusters can be identifiable independent of the adopted clustering method (i.e., the standard GMM or greedy optimistic GMM), 
supporting the plausibility of Tier-1 clusters. 
The details of individual Tier-1 clusters are summarized below.
}

\new{
(H22:DTC-1)
Among 9 member stars, all the stars satisfy \texttt{parallax\_over\_error} $>10$. 
Among them, 7 stars (except for {J14592981-3852558} and {BPS CS 22896-154}) are found in the same cluster. 
This cluster (which is a subset of H22:DTC-1) is chemically homogeneous, with $\sigma_\mathrm{[Fe/H]} = 0.20$ and $\sigma_\mathrm{[Eu/H]}= 0.35$. 
These quantities correspond to $q_\mathrm{[Fe/H]} = 0.9$ and $q_\mathrm{[Eu/H]}= 16.8$, respectively.
}

\new{
(H22:DTC-2)
Among 9 member stars, 8 stars satisfy \texttt{parallax\_over\_error} $>10$ (except for {SMSS J175046.30-425506.9}, which happens to be the only very-metal-poor member star). 
These 8 stars are found in the same cluster. 
This cluster (which is a subset of H22:DTC-2 and a superset of group D in \citealt{Roederer2018}) 
is chemically homogeneous, with $\sigma_\mathrm{[Fe/H]} = 0.17$ and $\sigma_\mathrm{[Eu/H]}=0.22$. 
These quantities correspond to $q_\mathrm{[Fe/H]} = 0.2$ and $q_\mathrm{[Eu/H]}=1.1$, respectively.
}

\new{
(H22:DTC-3)
Among 18 member stars, all stars satisfy \texttt{parallax\_over\_error} $>10$. 
These stars are separated into three clusters. 
The biggest cluster contains 10 member stars 
({2MASS J00512646-1053170}, 
{HE 0430-4901}, 
{2MASS J22562536-0719562}, 
{J03422816-6500355}, 
{BPS CS 22958-052}, 
{SDSS J004305.27+194859.20}, 
{J01425445-0904162}, 
{HD 221170}, 
{J12044314-2911051}, and 
{2MASS J15271353-2336177}). 
This cluster (which is a subset of H22:DTC-3) is chemically homogeneous, with $\sigma_\mathrm{[Fe/H]} = 0.24$ and $\sigma_\mathrm{[Eu/H]} = 0.21$. 
These quantities correspond to $q_\mathrm{[Fe/H]} = 1.1$ and $q_\mathrm{[Eu/H]}=1.4$, respectively. 
}

\new{
(H22:DTC-4)
Among 12 member stars, 8 stars satisfy \texttt{parallax\_over\_error} $>10$. 
These 8 stars are found in the same cluster. 
This cluster (which is a subset of H22:DTC-4) 
is chemically homogeneous, with $\sigma_\mathrm{[Fe/H]} = 0.27$ and $\sigma_\mathrm{[Eu/H]}=0.18$. 
These quantities correspond to $q_\mathrm{[Fe/H]} = 2.5$ and $q_\mathrm{[Eu/H]}=0.5$, respectively.
}

\new{
(H22:DTC-5)
Among 5 member stars, 3 stars satisfy \texttt{parallax\_over\_error} $>10$. 
Among these 3 stars, 2 stars ({2MASS J02462013-1518419} and {BPS CS 22953-003}) are found in a single cluster. 
This cluster (which is a subset of H22:DTC-5) has $\sigma_\mathrm{[Fe/H]} = 0.065$ and $\sigma_\mathrm{[Eu/H]}= 0.27$. 
These quantities correspond to $q_\mathrm{[Fe/H]} = 13.4$ and $q_\mathrm{[Eu/H]}= 51.6$, respectively.
}

\new{
(H22:DTC-9)
Among 6 member stars, 4 stars satisfy \texttt{parallax\_over\_error} $>10$. 
Among these 4 stars, 3 stars ({SDSS J235718.91-005247.8}, {BPS CS 31082-001} and {SMSS J051008.62-372019.8}) are found in a single cluster. 
This cluster (which is a subset of H22:DTC-9) has $\sigma_\mathrm{[Fe/H]} = 0.19$ and $\sigma_\mathrm{[Eu/H]}= 0.43$. 
These quantities correspond to $q_\mathrm{[Fe/H]} = 15.7$ and $q_\mathrm{[Eu/H]}= 59.8$, respectively.
}

\newpage

\section{Clustering analysis using stars with $\mathrm{[Fe/H]} < -2$}
\label{sec:VMP}


\new{
In our $r$-II star catalog, we have $N=118$ very-metal-poor stars with $\mathrm{[Fe/H]} < -2$, similar to most stars in UFDs known to date (including Reticulum II). 
Motivated by the observed chemical properties of UFDs, 
we perform an additional clustering analysis using these $N=118$ $r$-II stars. 
We fix $\lambda=0$ and $\sigma_J = 100 \kpc \kms$ as in our fiducial analysis and set $K=25$. 
}


\new{
In the fiducial result, 
we have $K=30$ clusters consisting of $N=161$ stars. 
Among these 30 clusters, 
we have 
(i) 4 clusters in which all the member stars are metal-rich 
($\mathrm{[Fe/H]} \geq -2$); 
(ii) 12 clusters in which all the member stars are very metal-poor 
($\mathrm{[Fe/H]} < -2$); and 
(iii) 14 clusters which include both very-metal-poor stars 
and metal-rich stars. 
From the fiducial result, we discard 4 metal-rich clusters 
in item (i). 
Also, we discard metal-rich member stars from 14 clusters in item (iii). 
As a result, we end up with $K=26$ clusters consisting of 
$N=118$ very-metal-poor stars. 
}

\new{
To investigate the similarity between the two sets of clusters mentioned above, 
namely 
$25$ clusters described in the first paragraph 
(i.e., an additional clustering result using very-metal-poor stars only) 
and 
$26$ clusters described in the second paragraph 
(i.e., a subset of the fiducial analysis), 
we compute four similarity indices. 
We find that 
the purity is $0.839$, 
Normalized Mutual Information is $0.868$, 
Rand index is $0.966$, 
and 
F-measure is $0.815$. 
These indices are close to unity, 
which means that these two sets of clusters are similar to each other. }

\new{
We also investigate how the member stars of Tier-1 clusters in the fiducial result  
(except for H22:DTC-2 because only one out of its nine member stars is very metal-poor) are classified in our additional analysis. 
H22:DTC-1 is divided into two separate clusters.
All the very-metal-poor member stars in H22:DTC-3 are found in a single cluster. 
H22:DTC-4 is divided into two separate clusters.
All the very-metal-poor member stars in H22:DTC-5 are found in a single cluster. 
Five out of six member stars in H22:DTC-9 are found in a single cluster. 
These results indicate that the pre-selection of the sample by [Fe/H] does not drastically affect the clustering results of Tier-1 clusters, 
while some Tier-1 clusters (H22:DTC-3 and H22:DTC-9) are less sensitive to the pre-selection than others. 
}

\section{Analysis of non-$r$-II stars} 
\label{sec:normal_stars}

\new{
In the fiducial analysis, we find that 
many dynamically identified $r$-II clusters 
have relatively tight distribution in [Fe/H]. 
Here we investigate whether or not our result is statistically significant 
by performing the same analysis for non-$r$-II stars. 
}

\new{
For this test, 
we carefully construct a sample of $N=161$ non-$r$-II stars such that its [Fe/H] histogram (with a bin size of 0.25 dex) is almost identical to that of our $r$-II sample. For this sample, [Fe/H] is determined from high-resolution spectroscopy, either from Subaru observation \citep{Li2022ApJ...931..147L} or Gaia-ESO survey DR5.0 
(\citealt{Gilmore2012Msngr.147...25G}; \url{https://www.gaia-eso.eu/data-products/public-data-releases}). 
}

\new{
We analyze this sample in the same manner as our fiducial analysis with $K=30$. 
We note that we do not have [Eu/H] or [Eu/Fe] measurements for most of our non-$r$-II stars partly because Eu abundance is difficult to measure for most non-$r$-II stars. 
Thus, we compare the tightness in the [Fe/H] distribution for clusters of $r$-II stars and those of non-$r$-II stars. 
As a result, we find  
\begin{itemize}
\item 
that non-$r$-II sample has 3 clusters with $\sigma_\mathrm{[Fe/H]} < 0.35$ (while our $r$-II sample has 14 clusters); 
\item  
that non-$r$-II sample has 0 clusters with $q_\mathrm{[Fe/H]} < 5$  (while our $r$-II sample has 5 clusters); and  
\item 
that non-$r$-II sample has 4 clusters with $q_\mathrm{[Fe/H]} < 15$  (while our $r$-II sample has 9 clusters). 
\end{itemize}
We see that our $r$-II clusters have tighter [Fe/H] distributions than the corresponding clusters of non-$r$-II stars. 
In other words, $r$-II stars with similar orbits tend to have smaller [Fe/H] dispersion than normal stars with similar orbits do, 
supporting the main result in this paper. 
}

\clearpage

\startlongtable
\begin{deluxetable*}{l l c lccc }
\tablecaption{Basic properties of the member stars of the clusters 
\label{table:members_all_info}}
\tablewidth{0pt}
\tabletypesize{\scriptsize}
\tablehead{
\colhead{$k$} &
\colhead{R18} &
\colhead{Name} &
\colhead{([Fe/H],[Eu/H],[Eu/Fe]), Reference} &
\colhead{$(J_r,J_z,J_\phi)$} &
\colhead{$E$} &
\colhead{$\varpi/\sigma_\varpi$} \\
\colhead{} &
\colhead{} &
\colhead{} &
\colhead{dex} &
\colhead{$\kpc \kms$} &
\colhead{$10^5 \; \mathrm{km^2\;s^{-2}}$} &
\colhead{} 
}
\startdata
{1} & {} & {HE 1523-0901} & {$(-2.95, -1.14, 1.81)$, \cite{Frebel2007}} & {$(148, 366, +1219)$} & {$-1.558$} & {$16.626$}\\
{1} & {} & {RAVE J203843.2-002333} & {$(-2.91, -1.27, 1.64)$, \cite{Placco2017}} & {$(99, 269, +1400)$} & {$-1.559$} & {$11.179$}\\
{1} & {} & {2MASS J09544277+5246414} & {$(-2.99, -1.71, 1.28)$, \cite{Holmbeck2018}} & {$(216, 462, +1319)$} & {$-1.482$} & {$26.830$}\\
{1} & {} & {2MASS J17225742-7123000} & {$(-2.42, -1.35, 1.07)$, \cite{Hansen2018}} & {$(143, 396, +1212)$} & {$-1.553$} & {$18.841$}\\
{1} & {} & {2MASS J20050670-3057445} & {$(-3.03, -2.09, 0.94)$, \cite{Cain_et_al__2018}} & {$(378, 80, +1208)$} & {$-1.565$} & {$20.871$}\\
{1} & {} & {HE 1044-2509} & {$(-2.89, -1.95, 0.94)$, \cite{Barklem2005}} & {$(87, 410, +1237)$} & {$-1.564$} & {$11.977$}\\
{1} & {} & {J14592981-3852558} & {$(-2.40, -1.47, 0.93)$, \cite{Holmbeck2020}} & {$(10, 146, +993)$} & {$-1.790$} & {$13.810$}\\
{1} & {} & {BPS CS 22896-154} & {$(-2.69, -1.83, 0.86)$, \cite{Francois_et_al__2007}} & {$(10, 83, +1076)$} & {$-1.793$} & {$17.928$}\\
{1} & {} & {2MASS J01555066-6400155} & {$(-2.71, -1.91, 0.80)$, \cite{Ezzeddine2020}} & {$(70, 171, +1218)$} & {$-1.662$} & {$17.600$}\\
\hline
{2} & {} & {SMSS J175046.30-425506.9} & {$(-2.17, -0.42, 1.75)$, \cite{Jacobson2015}} & {$(966, 55, +117)$} & {$-1.606$} & {$2.798$}\\
{2} & {} & {HD 222925} & {$(-1.47, -0.14, 1.33)$, \cite{Roederer_et_al__2018}} & {$(1052, 84, +273)$} & {$-1.520$} & {$189.056$}\\
{2} & {} & {RAVE J071142.5-343237} & {$(-1.96, -0.66, 1.30)$, \cite{Sakari2018}} & {$(774, 3, +228)$} & {$-1.709$} & {$43.802$}\\
{2} & {} & {2MASS J18024226-4404426} & {$(-1.55, -0.50, 1.05)$, \cite{Hansen2018}} & {$(796, 150, +154)$} & {$-1.642$} & {$16.274$}\\
{2} & {} & {J11404944-1615396} & {$(-1.67, -0.79, 0.88)$, \cite{Holmbeck2020}} & {$(925, 109, +57)$} & {$-1.620$} & {$16.332$}\\
{2} & {} & {J07352232-4425010} & {$(-1.58, -0.79, 0.79)$, \cite{Holmbeck2020}} & {$(1041, 27, +46)$} & {$-1.604$} & {$184.629$}\\
{2} & {} & {HD 20} & {$(-1.60, -0.87, 0.73)$, \cite{Hanke2020}} & {$(1115, 9, -99)$} & {$-1.572$} & {$92.637$}\\
{2} & {} & {2MASS J01530024-3417360} & {$(-1.50, -0.79, 0.71)$, \cite{Hansen2018}} & {$(972, 5, +37)$} & {$-1.657$} & {$231.958$}\\
{2} & {} & {HD 3567} & {$(-1.33, -0.63, 0.70)$, \cite{Hansen2012}} & {$(835, 24, +105)$} & {$-1.692$} & {$412.636$}\\
\hline
{3} & {} & {2MASS J00512646-1053170} & {$(-2.38, -1.17, 1.21)$, \cite{Ezzeddine2020}} & {$(420, 219, -810)$} & {$-1.600$} & {$189.026$}\\
{3} & {} & {HE 0430-4901} & {$(-2.72, -1.56, 1.16)$, \cite{Barklem2005}} & {$(427, 121, -719)$} & {$-1.669$} & {$27.471$}\\
{3} & {} & {2MASS J22562536-0719562} & {$(-2.26, -1.16, 1.10)$, \cite{Sakari2018}} & {$(371, 121, -636)$} & {$-1.726$} & {$11.873$}\\
{3} & {} & {SDSS J173025.57+414334.7} & {$(-2.85, -1.77, 1.08)$, \cite{Bandyopadhyay2020}} & {$(706, 45, -788)$} & {$-1.557$} & {$39.280$}\\
{3} & {} & {HE 2224+0143} & {$(-2.58, -1.53, 1.05)$, \cite{Barklem2005}} & {$(338, 51, -592)$} & {$-1.805$} & {$16.152$}\\
{3} & {} & {J03422816-6500355} & {$(-2.16, -1.11, 1.05)$, \cite{Holmbeck2020}} & {$(365, 14, -688)$} & {$-1.778$} & {$98.567$}\\
{3} & {} & {BPS CS 22958-052} & {$(-2.42, -1.42, 1.00)$, \cite{Roederer_et_al__2014}} & {$(435, 21, -895)$} & {$-1.660$} & {$68.063$}\\
{3} & {} & {SDSS J004305.27+194859.20} & {$(-2.25, -1.27, 0.98)$, \cite{Bandyopadhyay2020}} & {$(534, 36, -652)$} & {$-1.679$} & {$35.415$}\\
{3} & {} & {BPS CS 22875-029} & {$(-2.69, -1.77, 0.92)$, \cite{Roederer_et_al__2014}} & {$(389, 352, -655)$} & {$-1.618$} & {$11.905$}\\
{3} & {} & {HD 115444} & {$(-2.96, -2.11, 0.85)$, \cite{Westin2000}} & {$(601, 25, -603)$} & {$-1.667$} & {$67.006$}\\
{3} & {} & {J01425445-0904162} & {$(-1.73, -0.90, 0.83)$, \cite{Holmbeck2020}} & {$(372, 37, -891)$} & {$-1.684$} & {$47.161$}\\
{3} & {} & {G206-23} & {$(-1.79, -0.98, 0.81)$, \cite{Ishigaki2013}} & {$(688, 361, -695)$} & {$-1.476$} & {$305.366$}\\
{3} & {} & {HD 221170} & {$(-2.18, -1.38, 0.80)$, \cite{Ivans2006}} & {$(379, 16, -962)$} & {$-1.669$} & {$79.320$}\\
{3} & {} & {HE 0420+0123a} & {$(-3.03, -2.24, 0.79)$, \cite{Hollek2011}} & {$(367, 232, -622)$} & {$-1.680$} & {$41.097$}\\
{3} & {} & {RAVE J015656.3-140211} & {$(-2.08, -1.32, 0.76)$, \cite{Sakari2018}} & {$(704, 203, -621)$} & {$-1.543$} & {$14.705$}\\
{3} & {} & {BPS CS 22877-015} & {$(-2.00, -1.28, 0.72)$, \cite{Roederer_et_al__2014}} & {$(399, 12, -531)$} & {$-1.819$} & {$103.131$}\\
{3} & {} & {J12044314-2911051} & {$(-2.35, -1.64, 0.71)$, \cite{Holmbeck2020}} & {$(489, 202, -829)$} & {$-1.574$} & {$31.947$}\\
{3} & {} & {2MASS J15271353-2336177} & {$(-2.15, -1.45, 0.70)$, \cite{Hansen2018}} & {$(357, 28, -659)$} & {$-1.778$} & {$46.190$}\\
\hline
{4} & {} & {2MASS J21064294-6828266} & {$(-2.76, -1.44, 1.32)$, \cite{Hansen2018}} & {$(62, 314, -901)$} & {$-1.716$} & {$40.082$}\\
{4} & {} & {J05383296-5904280} & {$(-2.53, -1.25, 1.28)$, \cite{Holmbeck2020}} & {$(198, 257, -890)$} & {$-1.670$} & {$80.688$}\\
{4} & {} & {BPS CS 22945-017} & {$(-2.73, -1.60, 1.13)$, \cite{Roederer_et_al__2014}} & {$(150, 200, -1029)$} & {$-1.673$} & {$35.846$}\\
{4} & {} & {2MASS J18174532-3353235} & {$(-1.67, -0.68, 0.99)$, \cite{Johnson2013}} & {$(126, 146, -871)$} & {$-1.769$} & {$3.870$}\\
{4} & {} & {RAVE J133748.9-082617} & {$(-2.62, -1.69, 0.93)$, \cite{Sakari2018}} & {$(125, 281, -795)$} & {$-1.731$} & {$5.305$}\\
{4} & {} & {2MASS J15582962-1224344} & {$(-2.54, -1.65, 0.89)$, \cite{Hansen2018}} & {$(18, 222, -951)$} & {$-1.763$} & {$22.353$}\\
{4} & {} & {2MASS J17163340-7009028} & {$(-2.39, -1.50, 0.89)$, \cite{Ezzeddine2020}} & {$(23, 207, -919)$} & {$-1.780$} & {$10.026$}\\
{4} & {} & {BPS CS 30306-132} & {$(-2.42, -1.57, 0.85)$, \cite{Honda2004b}} & {$(244, 157, -926)$} & {$-1.678$} & {$21.118$}\\
{4} & {} & {J20435776-4408037} & {$(-1.85, -1.07, 0.78)$, \cite{Holmbeck2020}} & {$(65, 138, -852)$} & {$-1.818$} & {$12.669$}\\
{4} & {} & {2MASS J21095804-0945400} & {$(-2.73, -1.96, 0.77)$, \cite{Hansen2018}} & {$(95, 143, -864)$} & {$-1.792$} & {$9.632$}\\
{4} & {} & {2MASS J19215077-4452545} & {$(-2.56, -1.82, 0.74)$, \cite{Hansen2018}} & {$(106, 176, -887)$} & {$-1.758$} & {$4.410$}\\
{4} & {} & {2MASS J17435113-5359333} & {$(-2.24, -1.51, 0.73)$, \cite{Hansen2018}} & {$(105, 144, -754)$} & {$-1.829$} & {$25.874$}\\
\hline
{5} & {} & {2MASS J02462013-1518419} & {$(-2.71, -1.26, 1.45)$, \cite{Hansen2018}} & {$(1087, 281, +796)$} & {$-1.358$} & {$21.971$}\\
{5} & {} & {BPS CS 22953-003} & {$(-2.84, -1.79, 1.05)$, \cite{Francois_et_al__2007}} & {$(951, 388, +768)$} & {$-1.383$} & {$18.803$}\\
{5} & {} & {HE 2327-5642} & {$(-2.78, -1.80, 0.98)$, \cite{Mashonkina2010}} & {$(877, 531, +749)$} & {$-1.380$} & {$12.640$}\\
{5} & {} & {SMSS J183647.89-274333.1} & {$(-2.48, -1.66, 0.82)$, \cite{Howes2015}} & {$(927, 380, +773)$} & {$-1.390$} & {$3.254$}\\
{5} & {} & {HE 0300-0751} & {$(-2.27, -1.50, 0.77)$, \cite{Barklem2005}} & {$(930, 190, +777)$} & {$-1.435$} & {$0.990$}\\
\hline
{6} & {} & {2MASS J05241392-0336543} & {$(-2.50, -1.00, 1.50)$, \cite{Ezzeddine2020}} & {$(103, 133, -2497)$} & {$-1.352$} & {$4.395$}\\
{6} & {} & {2MASS J07150266-0154092} & {$(-2.60, -1.72, 0.88)$, \cite{Ezzeddine2020}} & {$(31, 2, -2512)$} & {$-1.403$} & {$30.915$}\\
\hline
{7} & {} & {SMSS J062609.83-590503.2} & {$(-2.77, -1.71, 1.06)$, \cite{Jacobson2015}} & {$(914, 176, -2835)$} & {$-1.133$} & {$13.632$}\\
{7} & {} & {HE 2244-1503} & {$(-2.88, -1.93, 0.95)$, \cite{Barklem2005}} & {$(1029, 237, -2662)$} & {$-1.126$} & {$6.843$}\\
\hline
{8} & {} & {J07202253-3358518} & {$(-1.60, -0.16, 1.44)$, \cite{Holmbeck2020}} & {$(3527, 3318, +52)$} & {$-0.813$} & {$69.734$}\\
{8} & {} & {Gaia DR2 2233912206910720000} & {$(-1.72, -0.61, 1.11)$, \cite{Hawkins_and_Wyse__2018}} & {$(3510, 3462, +274)$} & {$-0.796$} & {$25.346$}\\
\hline
{9} & {} & {SDSS J235718.91-005247.8} & {$(-3.36, -1.44, 1.92)$, \cite{Aoki2010}} & {$(139, 755, +927)$} & {$-1.525$} & {$43.974$}\\
{9} & {} & {BPS CS 31082-001} & {$(-2.90, -1.27, 1.63)$, \cite{Hill2002}} & {$(85, 1019, +855)$} & {$-1.496$} & {$17.304$}\\
{9} & {} & {SDSS J092157.27+503404.7} & {$(-2.65, -1.42, 1.23)$, \cite{Bandyopadhyay2020}} & {$(118, 791, +834)$} & {$-1.549$} & {$27.046$}\\
{9} & {} & {SMSS J051008.62-372019.8} & {$(-3.20, -2.25, 0.95)$, \cite{Jacobson2015}} & {$(90, 695, +829)$} & {$-1.588$} & {$83.448$}\\
{9} & {} & {BPS CS 22888-047} & {$(-2.54, -1.68, 0.86)$, \cite{Roederer_et_al__2014}} & {$(21, 739, +862)$} & {$-1.595$} & {$4.769$}\\
{9} & {} & {SMSS J195931.70-643529.3} & {$(-2.58, -1.84, 0.74)$, \cite{Jacobson2015}} & {$(219, 1238, +665)$} & {$-1.444$} & {$8.568$}\\
\hline
{10} & {} & {HE 2301-4024} & {$(-2.11, -1.13, 0.98)$, \cite{Barklem2005}} & {$(794, 1019, +1181)$} & {$-1.249$} & {$2.984$}\\
{10} & {} & {BPS CS 29491-069} & {$(-2.55, -1.59, 0.96)$, \cite{Hayek_et_al__2009}} & {$(889, 1314, +1276)$} & {$-1.180$} & {$22.426$}\\
{10} & {} & {2MASS J19324858-5908019} & {$(-1.93, -1.03, 0.90)$, \cite{Hansen2018}} & {$(895, 1252, +1184)$} & {$-1.196$} & {$4.210$}\\
{10} & {} & {HE 1131+0141} & {$(-2.48, -1.61, 0.87)$, \cite{Barklem2005}} & {$(934, 1175, +1188)$} & {$-1.199$} & {$2.078$}\\
\hline
{11} & {} & {G15-13} & {$(-1.47, -0.62, 0.85)$, \cite{Ishigaki2013}} & {$(1038, 355, +1757)$} & {$-1.217$} & {$639.873$}\\
{11} & {} & {G115-58} & {$(-1.30, -0.55, 0.75)$, \cite{Ishigaki2013}} & {$(835, 420, +1756)$} & {$-1.248$} & {$116.967$}\\
\hline
{12} & {} & {2MASS J22182082-3827554} & {$(-2.57, -1.47, 1.10)$, \cite{Ezzeddine2020}} & {$(269, 1227, -1000)$} & {$-1.367$} & {$24.310$}\\
{12} & {} & {2MASS J03270229+0132322} & {$(-2.39, -1.32, 1.07)$, \cite{Ezzeddine2020}} & {$(243, 1336, -1361)$} & {$-1.294$} & {$64.065$}\\
\hline
{13} & {} & {2MASS J14534137+0040467} & {$(-3.09, -1.29, 1.80)$, \cite{Ezzeddine2020}} & {$(541, 791, -166)$} & {$-1.552$} & {$4.740$}\\
{13} & {} & {2MASS J13052137-1137220} & {$(-2.07, -0.98, 1.09)$, \cite{Ezzeddine2020}} & {$(478, 719, +84)$} & {$-1.634$} & {$6.495$}\\
{13} & {} & {HE 1127-1143} & {$(-2.73, -1.65, 1.08)$, \cite{Barklem2005}} & {$(363, 861, +110)$} & {$-1.642$} & {$1.829$}\\
{13} & {} & {SMSS J221448.33-453949.9} & {$(-2.56, -1.62, 0.94)$, \cite{Jacobson2015}} & {$(344, 1012, +90)$} & {$-1.587$} & {$0.148$}\\
{13} & {} & {2MASS J12091322-1415313} & {$(-2.11, -1.30, 0.81)$, \cite{Sakari2018}} & {$(434, 727, +160)$} & {$-1.632$} & {$16.765$}\\
{13} & {} & {RAVE J192632.8-584657} & {$(-2.49, -1.73, 0.76)$, \cite{Rasmussen2020}} & {$(538, 736, +5)$} & {$-1.614$} & {$5.545$}\\
\hline
{14} & {} & {2MASS J15213995-3538094} & {$(-2.79, -0.56, 2.23)$, \cite{Cain2020ApJ...898...40C}} & {$(922, 366, -2107)$} & {$-1.194$} & {$36.738$}\\
{14} & {} & {2MASS J01553180-4919420} & {$(-3.01, -2.29, 0.72)$, \cite{Ezzeddine2020}} & {$(1016, 142, -1773)$} & {$-1.253$} & {$58.543$}\\
\hline
{15} & {} & {BPS CS 22892-052} & {$(-3.10, -1.46, 1.64)$, \cite{Sneden_et_al__2003}} & {$(679, 358, -335)$} & {$-1.576$} & {$6.801$}\\
{15} & {} & {2MASS J21091825-1310062} & {$(-2.40, -1.15, 1.25)$, \cite{Hansen2018}} & {$(488, 136, -88)$} & {$-1.849$} & {$28.003$}\\
{15} & {} & {BPS CS 22945-058} & {$(-2.71, -1.58, 1.13)$, \cite{Roederer_et_al__2014}} & {$(624, 72, -214)$} & {$-1.753$} & {$28.693$}\\
{15} & {} & {2MASS J14543792+0830379} & {$(-2.31, -1.21, 1.10)$, \cite{Ezzeddine2020}} & {$(391, 176, -157)$} & {$-1.866$} & {$17.143$}\\
{15} & {} & {SMSS J181505.16-385514.9} & {$(-3.29, -2.33, 0.96)$, \cite{Howes2015}} & {$(452, 53, -136)$} & {$-1.916$} & {$3.207$}\\
{15} & {} & {J00041581-5815524} & {$(-2.32, -1.37, 0.95)$, \cite{Holmbeck2020}} & {$(529, 341, +74)$} & {$-1.729$} & {$14.929$}\\
{15} & {} & {2MASS J19014952-4844359} & {$(-1.87, -0.94, 0.93)$, \cite{Hansen2018}} & {$(453, 43, -125)$} & {$-1.930$} & {$19.104$}\\
{15} & {} & {BD +17 3248} & {$(-2.06, -1.15, 0.91)$, \cite{Cowan_et_al__2002}} & {$(572, 45, -110)$} & {$-1.838$} & {$95.470$}\\
{15} & {} & {BPS CS 29529-054} & {$(-2.75, -1.85, 0.90)$, \cite{Roederer_et_al__2014}} & {$(596, 59, -274)$} & {$-1.756$} & {$59.902$}\\
{15} & {} & {2MASS J00405260-5122491} & {$(-2.11, -1.25, 0.86)$, \cite{Hansen2018}} & {$(582, 35, -166)$} & {$-1.817$} & {$423.879$}\\
{15} & {} & {G210-33} & {$(-1.08, -0.31, 0.77)$, \cite{Ishigaki2013}} & {$(738, 200, -396)$} & {$-1.576$} & {$364.784$}\\
{15} & {} & {2MASS J19232518-5833410} & {$(-2.08, -1.32, 0.76)$, \cite{Hansen2018}} & {$(558, 98, -186)$} & {$-1.788$} & {$29.542$}\\
{15} & {} & {J18050641-4907579} & {$(-2.58, -1.85, 0.73)$, \cite{Holmbeck2020}} & {$(325, 48, -246)$} & {$-1.977$} & {$13.794$}\\
{15} & {} & {RAVE J000738.2-034551} & {$(-2.09, -1.36, 0.73)$, \cite{Sakari2018}} & {$(629, 95, -260)$} & {$-1.722$} & {$8.238$}\\
{15} & {} & {BPS BS 17569-049} & {$(-2.88, -2.16, 0.72)$, \cite{Francois_et_al__2007}} & {$(521, 345, -36)$} & {$-1.743$} & {$6.694$}\\
{15} & {} & {J23342332-2748003} & {$(-2.35, -1.64, 0.71)$, \cite{Holmbeck2020}} & {$(481, 226, -151)$} & {$-1.775$} & {$13.352$}\\
{15} & {} & {2MASS J13494713-7423395} & {$(-2.85, -2.15, 0.70)$, \cite{Ezzeddine2020}} & {$(363, 263, -146)$} & {$-1.845$} & {$28.769$}\\
{15} & {} & {SMSS J182601.24-332358.3} & {$(-2.83, -2.13, 0.70)$, \cite{Howes_et_al__2016}} & {$(496, 113, -218)$} & {$-1.806$} & {$7.387$}\\
\hline
{16} & {} & {HE 1219-0312} & {$(-2.92, -1.54, 1.38)$, \cite{Hayek_et_al__2009}} & {$(361, 292, +539)$} & {$-1.687$} & {$2.126$}\\
{16} & {} & {2MASS J20093393-3410273} & {$(-1.99, -0.67, 1.32)$, \cite{Hansen2018}} & {$(258, 231, +641)$} & {$-1.735$} & {$8.035$}\\
{16} & {} & {2MASS J15383085-1804242} & {$(-2.09, -0.82, 1.27)$, \cite{Sakari_et_al__2018a}} & {$(321, 150, +454)$} & {$-1.806$} & {$66.989$}\\
{16} & {} & {J07103110-7121522} & {$(-1.47, -0.42, 1.05)$, \cite{Holmbeck2020}} & {$(551, 17, +573)$} & {$-1.706$} & {$146.784$}\\
{16} & {} & {2MASS J15211026-0607566} & {$(-2.00, -1.07, 0.93)$, \cite{Sakari2018}} & {$(293, 270, +545)$} & {$-1.734$} & {$30.961$}\\
{16} & {} & {J20000364-3301351} & {$(-1.89, -0.99, 0.90)$, \cite{Holmbeck2020}} & {$(189, 198, +445)$} & {$-1.880$} & {$14.214$}\\
{16} & {} & {RAVE J194550.6-392631} & {$(-2.79, -1.89, 0.90)$, \cite{Rasmussen2020}} & {$(81, 116, +522)$} & {$-1.984$} & {$12.934$}\\
{16} & {} & {BPS BS 16543-097} & {$(-2.48, -1.63, 0.85)$, \cite{Allen_et_al__2012}} & {$(369, 93, +554)$} & {$-1.774$} & {$29.222$}\\
{16} & {} & {RAVE J093730.5-062655} & {$(-1.86, -1.01, 0.85)$, \cite{Sakari_et_al__2019}} & {$(565, 247, +608)$} & {$-1.586$} & {$18.012$}\\
{16} & {} & {BPS CS 22882-001} & {$(-2.62, -1.81, 0.81)$, \cite{Roederer_et_al__2014}} & {$(487, 415, +462)$} & {$-1.605$} & {$3.864$}\\
{16} & {} & {RAVE J130524.5-393126} & {$(-2.11, -1.35, 0.76)$, \cite{Rasmussen2020}} & {$(327, 72, +496)$} & {$-1.835$} & {$26.463$}\\
{16} & {} & {LP877-23} & {$(-1.46, -0.71, 0.75)$, \cite{Ishigaki2013}} & {$(261, 438, +410)$} & {$-1.725$} & {$385.595$}\\
{16} & {} & {HE 1430+0053} & {$(-3.03, -2.31, 0.72)$, \cite{Barklem2005}} & {$(356, 237, +530)$} & {$-1.719$} & {$16.215$}\\
\hline
{17} & {} & {Gaia DR2 6412626111276193920} & {$(-2.04, -0.27, 1.77)$, \cite{Ji_et_al__2020}} & {$(29, 1161, -393)$} & {$-1.598$} & {$0.658$}\\
{17} & {} & {2MASS J00101758-1735387} & {$(-2.41, -1.06, 1.35)$, \cite{Ezzeddine2020}} & {$(0, 1245, -569)$} & {$-1.539$} & {$74.434$}\\
{17} & {} & {2MASS J23362202-5607498} & {$(-2.06, -0.92, 1.14)$, \cite{Hansen2018}} & {$(100, 1078, -369)$} & {$-1.601$} & {$10.327$}\\
{17} & {} & {LAMOST J112456.61+453531.1} & {$(-1.27, -0.17, 1.10)$, \cite{Xing_et_al__2019}} & {$(387, 1096, -344)$} & {$-1.494$} & {$5.019$}\\
\hline
{18} & {} & {2MASS J03073894-0502491} & {$(-2.22, -0.95, 1.27)$, \cite{Ezzeddine2020}} & {$(376, 238, -1874)$} & {$-1.373$} & {$7.300$}\\
{18} & {} & {HE 0432-0923} & {$(-3.19, -1.94, 1.25)$, \cite{Barklem2005}} & {$(328, 242, -1852)$} & {$-1.388$} & {$4.483$}\\
{18} & {} & {2MASS J02165716-7547064} & {$(-2.50, -1.38, 1.12)$, \cite{Hansen2018}} & {$(407, 250, -1942)$} & {$-1.349$} & {$18.601$}\\
\hline
{19} & {} & {SMSS J183128.71-341018.4} & {$(-1.83, -0.58, 1.25)$, \cite{Howes_et_al__2016}} & {$(54, 216, -249)$} & {$-2.100$} & {$6.064$}\\
{19} & {} & {2MASS J19161821-5544454} & {$(-2.35, -1.27, 1.08)$, \cite{Hansen2018}} & {$(101, 216, -345)$} & {$-1.988$} & {$10.641$}\\
{19} & {} & {SMSS J183225.29-334938.4} & {$(-1.74, -0.66, 1.08)$, \cite{Howes_et_al__2016}} & {$(183, 311, -262)$} & {$-1.898$} & {$1.034$}\\
{19} & {} & {SMSS J175738.37-454823.5} & {$(-2.46, -1.44, 1.02)$, \cite{Jacobson2015}} & {$(38, 306, -198)$} & {$-2.079$} & {$5.578$}\\
{19} & {} & {2MASS J21224590-4641030} & {$(-2.96, -2.06, 0.90)$, \cite{Hansen2018}} & {$(139, 579, -460)$} & {$-1.725$} & {$17.135$}\\
{19} & {} & {2MASS J18294359-4924253} & {$(-1.22, -0.33, 0.89)$, \cite{Ezzeddine2020}} & {$(61, 151, -234)$} & {$-2.162$} & {$12.451$}\\
{19} & {} & {J20411424-4654315} & {$(-2.30, -1.42, 0.88)$, \cite{Holmbeck2020}} & {$(302, 225, -204)$} & {$-1.880$} & {$11.797$}\\
\hline
{20} & {} & {SMSS J155430.57-263904.8} & {$(-2.61, -1.47, 1.14)$, \cite{Jacobson2015}} & {$(1297, 444, -1027)$} & {$-1.245$} & {$6.169$}\\
{20} & {} & {2MASS J00524174-0902235} & {$(-1.46, -0.52, 0.94)$, \cite{Ezzeddine2020}} & {$(1283, 379, -940)$} & {$-1.269$} & {$29.465$}\\
{20} & {} & {HE 0240-0807} & {$(-2.68, -1.95, 0.73)$, \cite{Barklem2005}} & {$(1328, 367, -1020)$} & {$-1.260$} & {$0.323$}\\
{20} & {} & {RAVE J091858.9-231151} & {$(-2.05, -1.34, 0.71)$, \cite{Sakari2018}} & {$(1324, 179, -1080)$} & {$-1.273$} & {$15.284$}\\
\hline
{21} & {} & {2MASS J14325334-4125494} & {$(-2.79, -1.18, 1.61)$, \cite{Hansen2018}} & {$(2151, 227, -1013)$} & {$-1.116$} & {$52.981$}\\
{21} & {} & {Gaia DR2 1508756353921427328} & {$(-1.93, -1.16, 0.77)$, \cite{Hawkins_and_Wyse__2018}} & {$(2175, 204, -1091)$} & {$-1.114$} & {$23.097$}\\
\hline
{22} & {} & {2MASS J12170829+0415146} & {$(-2.22, -1.12, 1.10)$, \cite{Ezzeddine2020}} & {$(1486, 1239, +401)$} & {$-1.188$} & {$0.577$}\\
{22} & {} & {BPS CS 30315-029} & {$(-3.33, -2.61, 0.72)$, \cite{Barklem2005}} & {$(1509, 1267, +324)$} & {$-1.190$} & {$4.335$}\\
\hline
{23} & {} & {HE 1226-1149} & {$(-2.91, -1.36, 1.55)$, \cite{Cohen_et_al__2013}} & {$(156, 645, -1760)$} & {$-1.369$} & {$5.527$}\\
{23} & {} & {J22372037-4741375} & {$(-2.97, -2.13, 0.84)$, \cite{Holmbeck2020}} & {$(180, 521, -1553)$} & {$-1.428$} & {$13.029$}\\
{23} & {} & {J10401894-4106124} & {$(-1.55, -0.73, 0.82)$, \cite{Holmbeck2020}} & {$(164, 619, -1751)$} & {$-1.374$} & {$42.487$}\\
\hline
{24} & {} & {BPS CS 29497-004} & {$(-2.85, -1.18, 1.67)$, \cite{Hill_et_al__2017}} & {$(365, 239, -1406)$} & {$-1.471$} & {$12.522$}\\
{24} & {} & {BPS CS 31078-018} & {$(-2.84, -1.61, 1.23)$, \cite{Lai_et_al__2008}} & {$(275, 25, -1215)$} & {$-1.634$} & {$44.957$}\\
{24} & {} & {J22190836-2333467} & {$(-2.54, -1.67, 0.87)$, \cite{Holmbeck2020}} & {$(106, 108, -1167)$} & {$-1.690$} & {$22.914$}\\
{24} & {} & {BPS CS 22943-132} & {$(-2.67, -1.81, 0.86)$, \cite{Roederer_et_al__2014}} & {$(62, 221, -1374)$} & {$-1.598$} & {$141.707$}\\
{24} & {} & {BPS CS 22886-012} & {$(-2.61, -1.76, 0.85)$, \cite{Roederer_et_al__2014}} & {$(106, 25, -1165)$} & {$-1.738$} & {$19.888$}\\
{24} & {} & {J10191573-1924464} & {$(-1.11, -0.36, 0.75)$, \cite{Holmbeck2020}} & {$(126, 42, -1368)$} & {$-1.646$} & {$23.505$}\\
{24} & {} & {J06195001-5312114} & {$(-2.06, -1.33, 0.73)$, \cite{Holmbeck2020}} & {$(72, 258, -1299)$} & {$-1.602$} & {$149.166$}\\
{24} & {} & {J14354680-1124122} & {$(-1.10, -0.38, 0.72)$, \cite{Holmbeck2020}} & {$(16, 10, -1556)$} & {$-1.656$} & {$76.502$}\\
{24} & {} & {RAVE J183623.2-642812} & {$(-2.50, -1.78, 0.72)$, \cite{Rasmussen2020}} & {$(15, 41, -1294)$} & {$-1.730$} & {$27.378$}\\
{24} & {} & {HD 120559} & {$(-1.31, -0.60, 0.71)$, \cite{Hansen2012}} & {$(52, 15, -1493)$} & {$-1.656$} & {$136.429$}\\
{24} & {} & {2MASS J18295183-4503394} & {$(-2.48, -1.78, 0.70)$, \cite{Hansen2018}} & {$(10, 20, -1358)$} & {$-1.722$} & {$46.753$}\\
{24} & {} & {BD-10 3742 } & {$(-1.96, -1.26, 0.70)$, \cite{Hansen_C_J__et_al__2020}} & {$(81, 46, -1288)$} & {$-1.694$} & {$32.606$}\\
{24} & {} & {J06320130-2026538} & {$(-1.56, -0.86, 0.70)$, \cite{Holmbeck2020}} & {$(137, 65, -1308)$} & {$-1.647$} & {$155.373$}\\
{24} & {} & {RAVE J115941.7-382043} & {$(-0.94, -0.24, 0.70)$, \cite{Rasmussen2020}} & {$(46, 7, -1512)$} & {$-1.658$} & {$87.671$}\\
\hline
{25} & {} & {G166-37} & {$(-1.18, -0.32, 0.86)$, \cite{Ishigaki2013}} & {$(5072, 2134, -601)$} & {$-0.735$} & {$357.603$}\\
{25} & {} & {HE 2252-4225} & {$(-2.63, -1.82, 0.81)$, \cite{Mashonkina_et_al__2014}} & {$(5051, 2365, -133)$} & {$-0.747$} & {$2.050$}\\
\hline
{26} & {} & {RAVE J040618.2-030525} & {$(-1.34, -0.17, 1.17)$, \cite{Rasmussen2020}} & {$(1801, 130, -18)$} & {$-1.272$} & {$58.391$}\\
{26} & {} & {LAMOST J110901+075441} & {$(-3.17, -2.23, 0.94)$, \cite{Mardini_et_al__2020}} & {$(1727, 196, -366)$} & {$-1.289$} & {$9.180$}\\
{26} & {} & {Gaia DR2 3602288924850161792 } & {$(-1.89, -1.09, 0.80)$, \cite{Valentini_et_al__2019}} & {$(1566, 158, -339)$} & {$-1.321$} & {$27.652$}\\
{26} & {} & {CD -45 3283} & {$(-0.99, -0.21, 0.78)$, \cite{Hansen2012}} & {$(1577, 178, -119)$} & {$-1.321$} & {$153.969$}\\
\hline
{27} & {} & {Gaia DR2 4248140165233284352} & {$(-1.82, -1.12, 0.70)$, \cite{Hawkins_and_Wyse__2018}} & {$(1485, 1831, +579)$} & {$-1.102$} & {$12.291$}\\
\hline
{28} & {} & {G14-39} & {$(-1.88, -1.10, 0.78)$, \cite{Ishigaki2013}} & {$(727, 1284, -585)$} & {$-1.310$} & {$501.058$}\\
\hline
{29} & {} & {SMSS J024858.41-684306.4} & {$(-3.71, -2.71, 1.00)$, \cite{Jacobson2015}} & {$(6120, 342, -2857)$} & {$-0.662$} & {$20.514$}\\
\hline
{30} & {} & {SMSS J063447.15-622355.0} & {$(-3.41, -2.52, 0.89)$, \cite{Jacobson2015}} & {$(8014, 274, -3751)$} & {$-0.560$} & {$8.100$}\\
\hline
\enddata
\end{deluxetable*}
\startlongtable
\begin{deluxetable*}{lc cccc }
\tablecaption{Kinematics of the member stars of the clusters 
\label{table:members_additional_info}}
\tablewidth{0pt}
\tabletypesize{\scriptsize}
\tablehead{
\colhead{$k$} &
\colhead{Name} &
\colhead{$(x,y,z)$} &
\colhead{$(v_x,v_y,v_z)$} &
\colhead{$(r_\mathrm{peri},r_\mathrm{apo},z_\mathrm{max})$} &
\colhead{$e$} \\
\colhead{} &
\colhead{} &
\colhead{$\kpc$} &
\colhead{$\kms$} &
\colhead{$\kpc$} &
\colhead{} 
}
\startdata
{1} & {HE 1523-0901} & {$(-6.03, -0.22, +1.70)$} & {$(-129.96, -206.98, -133.45)$} & {$(5.12, 10.74, 6.19)$} & {$0.35$} \\
{1} & {RAVE J203843.2-002333} & {$(-4.90, +3.34, -2.05)$} & {$(-49.37, -251.97, -99.46)$} & {$(5.71, 10.28, 4.99)$} & {$0.29$} \\
{1} & {2MASS J09544277+5246414} & {$(-9.88, +0.54, +2.03)$} & {$(-134.36, -126.27, -105.30)$} & {$(5.52, 12.92, 7.89)$} & {$0.40$} \\
{1} & {2MASS J17225742-7123000} & {$(-5.08, -2.49, -1.34)$} & {$(+136.70, -171.43, +197.75)$} & {$(5.23, 10.81, 6.46)$} & {$0.35$} \\
{1} & {2MASS J20050670-3057445} & {$(-5.50, +0.50, -1.45)$} & {$(-214.54, -200.21, -0.53)$} & {$(3.32, 12.03, 2.43)$} & {$0.57$} \\
{1} & {HE 1044-2509} & {$(-8.15, -3.15, +1.80)$} & {$(+130.57, -101.43, +118.74)$} & {$(5.71, 10.06, 6.12)$} & {$0.28$} \\
{1} & {J14592981-3852558} & {$(-4.28, -2.38, +1.46)$} & {$(+82.34, -186.23, +98.55)$} & {$(4.50, 5.86, 2.39)$} & {$0.13$} \\
{1} & {BPS CS 22896-154} & {$(-5.50, -0.95, -1.58)$} & {$(+41.00, -188.75, +35.32)$} & {$(4.59, 5.87, 1.71)$} & {$0.12$} \\
{1} & {2MASS J01555066-6400155} & {$(-7.20, -2.45, -3.32)$} & {$(+39.59, -155.61, +7.25)$} & {$(4.82, 8.32, 3.35)$} & {$0.27$} \\
\hline
{2} & {SMSS J175046.30-425506.9} & {$(-2.23, -1.21, -0.84)$} & {$(+346.38, +135.11, +98.60)$} & {$(0.24, 12.48, 5.30)$} & {$0.96$} \\
{2} & {HD 222925} & {$(-7.99, -0.18, -0.34)$} & {$(-242.31, -39.80, +67.35)$} & {$(0.55, 14.65, 5.34)$} & {$0.93$} \\
{2} & {RAVE J071142.5-343237} & {$(-8.97, -1.78, -0.36)$} & {$(-101.63, -45.73, -9.88)$} & {$(0.49, 10.25, 4.42)$} & {$0.91$} \\
{2} & {2MASS J18024226-4404426} & {$(-4.64, -0.72, -0.65)$} & {$(-277.88, -76.27, +100.64)$} & {$(0.36, 11.28, 5.24)$} & {$0.94$} \\
{2} & {J11404944-1615396} & {$(-7.86, -1.91, +1.85)$} & {$(-185.77, -52.35, -15.18)$} & {$(0.29, 11.76, 5.27)$} & {$0.95$} \\
{2} & {J07352232-4425010} & {$(-8.28, -0.47, -0.08)$} & {$(+204.29, +5.85, +51.49)$} & {$(0.18, 12.53, 5.47)$} & {$0.97$} \\
{2} & {HD 20} & {$(-8.10, +0.04, -0.47)$} & {$(-229.00, +13.53, +6.35)$} & {$(0.22, 13.33, 5.55)$} & {$0.97$} \\
{2} & {2MASS J01530024-3417360} & {$(-8.21, -0.07, -0.27)$} & {$(+184.56, -2.96, +24.97)$} & {$(0.15, 11.34, 5.17)$} & {$0.97$} \\
{2} & {HD 3567} & {$(-8.19, +0.03, -0.09)$} & {$(+160.55, -13.61, -46.40)$} & {$(0.26, 10.61, 4.85)$} & {$0.95$} \\
\hline
{3} & {2MASS J00512646-1053170} & {$(-8.22, +0.06, -0.24)$} & {$(+156.24, +97.47, -109.61)$} & {$(2.54, 11.34, 5.39)$} & {$0.63$} \\
{3} & {HE 0430-4901} & {$(-8.62, -1.73, -1.63)$} & {$(-102.45, +62.92, -78.39)$} & {$(2.01, 10.07, 3.61)$} & {$0.67$} \\
{3} & {2MASS J22562536-0719562} & {$(-7.32, +1.73, -2.88)$} & {$(+106.56, +61.76, +41.20)$} & {$(1.75, 9.06, 3.24)$} & {$0.68$} \\
{3} & {SDSS J173025.57+414334.7} & {$(-7.41, +1.81, +1.26)$} & {$(-188.03, +152.28, +2.02)$} & {$(1.91, 13.08, 2.27)$} & {$0.75$} \\
{3} & {HE 2224+0143} & {$(-7.52, +1.56, -1.66)$} & {$(-0.67, +79.01, +6.80)$} & {$(1.75, 7.89, 2.67)$} & {$0.64$} \\
{3} & {J03422816-6500355} & {$(-8.04, -0.82, -0.77)$} & {$(+48.50, +90.61, +11.05)$} & {$(1.79, 8.42, 0.81)$} & {$0.65$} \\
{3} & {BPS CS 22958-052} & {$(-8.05, -0.59, -0.91)$} & {$(-151.18, +100.23, -29.59)$} & {$(2.31, 10.40, 1.21)$} & {$0.64$} \\
{3} & {SDSS J004305.27+194859.20} & {$(-8.66, +0.83, -0.88)$} & {$(+126.75, +63.23, +48.44)$} & {$(1.62, 10.31, 1.66)$} & {$0.73$} \\
{3} & {BPS CS 22875-029} & {$(-6.19, +0.07, -3.18)$} & {$(-117.12, +107.07, -160.86)$} & {$(2.52, 10.88, 7.17)$} & {$0.62$} \\
{3} & {HD 115444} & {$(-8.19, +0.15, +0.81)$} & {$(+161.39, +70.70, +14.86)$} & {$(1.41, 10.71, 1.19)$} & {$0.77$} \\
{3} & {J01425445-0904162} & {$(-8.72, +0.20, -1.42)$} & {$(+92.07, +100.09, +26.26)$} & {$(2.41, 9.76, 1.62)$} & {$0.60$} \\
{3} & {G206-23} & {$(-8.05, +0.24, +0.12)$} & {$(+219.91, +79.80, +135.48)$} & {$(2.29, 14.94, 10.11)$} & {$0.73$} \\
{3} & {HD 221170} & {$(-8.28, +0.47, -0.25)$} & {$(+135.27, +108.52, -34.37)$} & {$(2.54, 10.11, 0.94)$} & {$0.60$} \\
{3} & {HE 0420+0123a} & {$(-9.06, -0.20, -0.54)$} & {$(+54.96, +69.82, +104.20)$} & {$(2.05, 9.63, 5.17)$} & {$0.65$} \\
{3} & {RAVE J015656.3-140211} & {$(-9.40, +0.09, -3.34)$} & {$(-131.88, +67.37, -101.87)$} & {$(1.72, 13.28, 6.25)$} & {$0.77$} \\
{3} & {BPS CS 22877-015} & {$(-7.93, -0.28, +0.52)$} & {$(-12.52, +66.63, +23.92)$} & {$(1.33, 7.98, 0.69)$} & {$0.71$} \\
{3} & {J12044314-2911051} & {$(-7.63, -1.44, +1.01)$} & {$(-206.62, +69.56, -84.05)$} & {$(2.48, 12.09, 5.42)$} & {$0.66$} \\
{3} & {2MASS J15271353-2336177} & {$(-8.04, -0.04, +0.09)$} & {$(-59.40, +81.67, +52.71)$} & {$(1.73, 8.41, 1.08)$} & {$0.66$} \\
\hline
{4} & {2MASS J21064294-6828266} & {$(-6.88, -0.89, -1.18)$} & {$(-65.48, +122.55, +131.60)$} & {$(4.20, 7.41, 4.43)$} & {$0.28$} \\
{4} & {J05383296-5904280} & {$(-8.22, -0.94, -0.58)$} & {$(-75.68, +99.73, -121.73)$} & {$(3.32, 9.14, 4.80)$} & {$0.47$} \\
{4} & {BPS CS 22945-017} & {$(-7.62, -0.49, -0.89)$} & {$(-79.22, +130.01, -117.83)$} & {$(3.80, 8.82, 3.96)$} & {$0.40$} \\
{4} & {2MASS J18174532-3353235} & {$(+2.95, -0.19, -1.62)$} & {$(-21.94, -293.76, -44.82)$} & {$(3.13, 7.34, 2.93)$} & {$0.40$} \\
{4} & {RAVE J133748.9-082617} & {$(-5.69, -1.94, +4.16)$} & {$(-103.27, +104.69, +13.55)$} & {$(3.27, 7.81, 4.47)$} & {$0.41$} \\
{4} & {2MASS J15582962-1224344} & {$(-5.82, -0.08, +1.38)$} & {$(+39.50, +164.13, +118.13)$} & {$(4.54, 6.24, 3.19)$} & {$0.16$} \\
{4} & {2MASS J17163340-7009028} & {$(-2.66, -4.31, -2.24)$} & {$(-117.59, +155.10, -65.05)$} & {$(4.25, 6.17, 3.07)$} & {$0.18$} \\
{4} & {BPS CS 30306-132} & {$(-6.17, +0.33, +2.52)$} & {$(-140.90, +157.82, +3.93)$} & {$(3.03, 9.26, 3.67)$} & {$0.51$} \\
{4} & {J20435776-4408037} & {$(-5.23, -0.18, -2.31)$} & {$(+63.54, +165.16, +6.40)$} & {$(3.33, 6.30, 2.54)$} & {$0.31$} \\
{4} & {2MASS J21095804-0945400} & {$(-5.56, +2.22, -2.40)$} & {$(+12.81, +150.41, -53.20)$} & {$(3.21, 6.86, 2.73)$} & {$0.36$} \\
{4} & {2MASS J19215077-4452545} & {$(-3.69, -0.54, -1.98)$} & {$(+47.63, +247.45, +84.07)$} & {$(3.37, 7.32, 3.25)$} & {$0.37$} \\
{4} & {2MASS J17435113-5359333} & {$(-6.17, -0.81, -0.46)$} & {$(-45.91, +116.40, -115.58)$} & {$(2.79, 6.46, 2.76)$} & {$0.40$} \\
\hline
{5} & {2MASS J02462013-1518419} & {$(-9.40, -0.32, -2.24)$} & {$(-213.35, -92.13, -141.23)$} & {$(2.13, 19.64, 12.19)$} & {$0.80$} \\
{5} & {BPS CS 22953-003} & {$(-6.94, -2.08, -3.48)$} & {$(+288.98, -24.11, +13.20)$} & {$(2.21, 18.57, 11.37)$} & {$0.79$} \\
{5} & {HE 2327-5642} & {$(-5.82, -1.74, -4.54)$} & {$(-99.19, -158.43, -230.10)$} & {$(2.76, 18.21, 13.67)$} & {$0.74$} \\
{5} & {SMSS J183647.89-274333.1} & {$(+3.05, +1.27, -1.83)$} & {$(-372.02, +98.61, -77.66)$} & {$(2.35, 18.07, 11.84)$} & {$0.77$} \\
{5} & {HE 0300-0751} & {$(-12.46, -0.53, -5.76)$} & {$(-108.23, -67.08, -72.96)$} & {$(2.13, 16.62, 7.51)$} & {$0.77$} \\
\hline
{6} & {2MASS J05241392-0336543} & {$(-14.36, -3.01, -2.62)$} & {$(-43.00, +164.91, -49.98)$} & {$(9.34, 15.03, 4.26)$} & {$0.23$} \\
{6} & {2MASS J07150266-0154092} & {$(-9.73, -1.19, +0.17)$} & {$(-6.45, +257.37, -14.71)$} & {$(9.58, 12.55, 0.40)$} & {$0.13$} \\
\hline
{7} & {SMSS J062609.83-590503.2} & {$(-8.32, -4.57, -2.23)$} & {$(-30.00, +324.42, +109.52)$} & {$(8.08, 28.85, 8.82)$} & {$0.56$} \\
{7} & {HE 2244-1503} & {$(-7.23, +1.12, -2.38)$} & {$(+45.63, +361.13, -99.01)$} & {$(7.62, 29.66, 10.95)$} & {$0.59$} \\
\hline
{8} & {J07202253-3358518} & {$(-8.62, -1.02, -0.16)$} & {$(+185.82, +15.88, +403.57)$} & {$(7.40, 64.49, 64.48)$} & {$0.79$} \\
{8} & {Gaia DR2 2233912206910720000} & {$(-8.13, +2.81, +0.70)$} & {$(-109.93, +4.20, -434.35)$} & {$(8.29, 66.88, 66.53)$} & {$0.78$} \\
\hline
{9} & {SDSS J235718.91-005247.8} & {$(-8.20, +0.24, -0.41)$} & {$(+105.97, -116.30, -189.87)$} & {$(5.50, 11.26, 9.09)$} & {$0.34$} \\
{9} & {BPS CS 31082-001} & {$(-8.61, +0.13, -1.76)$} & {$(+110.91, -101.01, -185.07)$} & {$(6.64, 11.31, 9.90)$} & {$0.26$} \\
{9} & {SDSS J092157.27+503404.7} & {$(-10.01, +0.41, +1.85)$} & {$(-55.97, -81.04, -145.81)$} & {$(5.34, 10.62, 8.82)$} & {$0.33$} \\
{9} & {SMSS J051008.62-372019.8} & {$(-8.59, -0.75, -0.59)$} & {$(-36.78, -99.74, -176.05)$} & {$(5.15, 9.63, 7.78)$} & {$0.30$} \\
{9} & {BPS CS 22888-047} & {$(-6.28, +0.29, -5.08)$} & {$(-61.20, -134.63, +129.04)$} & {$(6.19, 8.57, 7.01)$} & {$0.16$} \\
{9} & {SMSS J195931.70-643529.3} & {$(-0.04, -4.38, -5.66)$} & {$(+152.99, +93.75, -213.62)$} & {$(5.97, 13.70, 12.71)$} & {$0.39$} \\
\hline
{10} & {HE 2301-4024} & {$(-6.69, -0.11, -3.09)$} & {$(+55.72, -175.75, +293.15)$} & {$(5.88, 22.78, 18.92)$} & {$0.59$} \\
{10} & {BPS CS 29491-069} & {$(-7.08, +0.27, -1.87)$} & {$(-170.60, -173.70, +279.92)$} & {$(6.93, 26.35, 22.64)$} & {$0.58$} \\
{10} & {2MASS J19324858-5908019} & {$(+1.76, -4.04, -5.75)$} & {$(+217.28, +174.23, -220.70)$} & {$(6.40, 25.64, 22.08)$} & {$0.60$} \\
{10} & {HE 1131+0141} & {$(-8.68, -4.87, +7.97)$} & {$(-154.75, -223.75, +17.18)$} & {$(6.01, 25.66, 21.71)$} & {$0.62$} \\
\hline
{11} & {G15-13} & {$(-8.10, +0.01, +0.12)$} & {$(+226.63, -217.34, +160.86)$} & {$(4.96, 25.35, 13.39)$} & {$0.67$} \\
{11} & {G115-58} & {$(-8.50, +0.04, +0.33)$} & {$(-197.67, -205.76, +176.38)$} & {$(5.52, 23.12, 12.95)$} & {$0.61$} \\
\hline
{12} & {2MASS J22182082-3827554} & {$(-6.90, +0.08, -1.89)$} & {$(+44.31, +144.55, -282.10)$} & {$(7.07, 16.07, 14.18)$} & {$0.39$} \\
{12} & {2MASS J03270229+0132322} & {$(-8.96, -0.02, -0.70)$} & {$(+23.87, +152.04, -270.56)$} & {$(8.98, 18.19, 15.54)$} & {$0.34$} \\
\hline
{13} & {2MASS J14534137+0040467} & {$(-2.04, -0.43, +7.48)$} & {$(-45.94, +71.82, +198.34)$} & {$(1.67, 12.68, 12.18)$} & {$0.77$} \\
{13} & {2MASS J13052137-1137220} & {$(-5.97, -2.78, +4.42)$} & {$(+62.41, +14.92, -171.30)$} & {$(1.51, 10.53, 10.36)$} & {$0.75$} \\
{13} & {HE 1127-1143} & {$(-7.96, -3.57, +3.74)$} & {$(+82.40, +23.00, +96.74)$} & {$(2.11, 10.09, 9.86)$} & {$0.65$} \\
{13} & {SMSS J221448.33-453949.9} & {$(-0.12, -1.20, -11.18)$} & {$(+75.39, +3.36, -12.43)$} & {$(2.67, 11.26, 11.19)$} & {$0.62$} \\
{13} & {2MASS J12091322-1415313} & {$(-7.72, -1.43, +1.65)$} & {$(+154.02, +7.74, +122.51)$} & {$(1.82, 10.49, 10.12)$} & {$0.70$} \\
{13} & {RAVE J192632.8-584657} & {$(-0.21, -3.19, -4.43)$} & {$(-14.20, -248.25, -77.29)$} & {$(1.35, 11.06, 11.05)$} & {$0.78$} \\
\hline
{14} & {2MASS J15213995-3538094} & {$(-6.72, -0.70, +0.54)$} & {$(+73.75, +321.10, +196.14)$} & {$(6.29, 25.91, 12.88)$} & {$0.61$} \\
{14} & {2MASS J01553180-4919420} & {$(-8.08, -0.67, -1.40)$} & {$(-238.57, +199.70, -125.90)$} & {$(4.72, 23.41, 7.69)$} & {$0.66$} \\
\hline
{15} & {BPS CS 22892-052} & {$(-6.26, +1.67, -3.34)$} & {$(-133.45, +89.29, -170.99)$} & {$(1.16, 12.36, 9.08)$} & {$0.83$} \\
{15} & {2MASS J21091825-1310062} & {$(-6.97, +0.89, -1.09)$} & {$(+58.32, +5.24, -76.82)$} & {$(0.24, 7.27, 4.25)$} & {$0.94$} \\
{15} & {BPS CS 22945-058} & {$(-7.51, -0.74, -1.16)$} & {$(-139.12, +14.97, +37.04)$} & {$(0.45, 9.37, 4.23)$} & {$0.91$} \\
{15} & {2MASS J14543792+0830379} & {$(-6.18, +0.22, +2.94)$} & {$(-9.22, +25.82, +53.16)$} & {$(0.49, 7.07, 3.70)$} & {$0.87$} \\
{15} & {SMSS J181505.16-385514.9} & {$(-3.62, -0.46, -0.80)$} & {$(+214.17, +64.74, +94.03)$} & {$(0.59, 6.76, 2.93)$} & {$0.84$} \\
{15} & {J00041581-5815524} & {$(-6.66, -1.54, -3.41)$} & {$(-66.28, -26.43, -122.16)$} & {$(0.35, 8.94, 7.40)$} & {$0.93$} \\
{15} & {2MASS J19014952-4844359} & {$(-5.90, -0.48, -0.90)$} & {$(+95.70, +28.98, -32.62)$} & {$(0.30, 6.82, 3.42)$} & {$0.92$} \\
{15} & {BD +17 3248} & {$(-7.64, +0.45, +0.36)$} & {$(-40.91, +16.89, +61.18)$} & {$(0.27, 8.09, 4.13)$} & {$0.94$} \\
{15} & {BPS CS 29529-054} & {$(-8.11, -0.74, -0.68)$} & {$(+91.76, +42.19, +65.32)$} & {$(0.78, 9.17, 3.69)$} & {$0.84$} \\
{15} & {2MASS J00405260-5122491} & {$(-8.14, -0.05, -0.11)$} & {$(-19.76, +20.39, -55.05)$} & {$(0.36, 8.40, 3.88)$} & {$0.92$} \\
{15} & {G210-33} & {$(-8.14, +0.21, +0.01)$} & {$(+196.53, +43.46, +106.94)$} & {$(1.07, 12.54, 6.40)$} & {$0.84$} \\
{15} & {2MASS J19232518-5833410} & {$(-6.84, -0.53, -0.71)$} & {$(+126.08, +37.06, +93.25)$} & {$(0.41, 8.86, 4.16)$} & {$0.91$} \\
{15} & {J18050641-4907579} & {$(-4.15, -1.15, -0.96)$} & {$(+143.73, +98.98, -14.45)$} & {$(0.61, 6.08, 2.74)$} & {$0.82$} \\
{15} & {RAVE J000738.2-034551} & {$(-8.31, +1.08, -2.26)$} & {$(+90.25, +19.59, +56.90)$} & {$(0.91, 9.72, 3.37)$} & {$0.83$} \\
{15} & {BPS BS 17569-049} & {$(-5.96, +4.58, -4.13)$} & {$(-8.29, +12.49, +71.41)$} & {$(0.44, 8.66, 7.55)$} & {$0.90$} \\
{15} & {J23342332-2748003} & {$(-6.94, +0.61, -4.47)$} & {$(+9.13, +21.01, +31.61)$} & {$(0.41, 8.43, 4.79)$} & {$0.91$} \\
{15} & {2MASS J13494713-7423395} & {$(-6.42, -2.34, -0.60)$} & {$(-1.62, +22.31, -122.32)$} & {$(0.62, 6.89, 5.37)$} & {$0.83$} \\
{15} & {SMSS J182601.24-332358.3} & {$(-1.66, +0.03, -1.10)$} & {$(-325.19, +137.71, +13.95)$} & {$(0.68, 8.35, 3.46)$} & {$0.85$} \\
\hline
{16} & {HE 1219-0312} & {$(-7.22, -2.85, +4.94)$} & {$(+83.95, -41.47, +13.54)$} & {$(1.89, 9.52, 5.67)$} & {$0.67$} \\
{16} & {2MASS J20093393-3410273} & {$(-3.63, +0.58, -2.64)$} & {$(+143.10, -199.81, +73.18)$} & {$(2.26, 8.37, 4.57)$} & {$0.57$} \\
{16} & {2MASS J15383085-1804242} & {$(-7.32, -0.16, +0.51)$} & {$(+58.30, -60.74, +95.84)$} & {$(1.39, 7.58, 3.76)$} & {$0.69$} \\
{16} & {J07103110-7121522} & {$(-8.05, -0.57, -0.24)$} & {$(-137.75, -80.92, +35.94)$} & {$(1.35, 9.96, 0.91)$} & {$0.76$} \\
{16} & {2MASS J15211026-0607566} & {$(-6.75, -0.10, +1.26)$} & {$(+99.28, -79.37, -137.91)$} & {$(1.96, 8.49, 5.07)$} & {$0.62$} \\
{16} & {J20000364-3301351} & {$(-4.98, +0.45, -1.69)$} & {$(-82.62, -81.94, -124.63)$} & {$(1.57, 6.26, 3.41)$} & {$0.60$} \\
{16} & {RAVE J194550.6-392631} & {$(-4.40, +0.01, -1.89)$} & {$(+12.50, -118.77, +37.76)$} & {$(1.83, 4.97, 2.01)$} & {$0.46$} \\
{16} & {BPS BS 16543-097} & {$(-7.78, +0.00, +2.38)$} & {$(-48.32, -71.31, -9.36)$} & {$(1.59, 8.51, 2.88)$} & {$0.69$} \\
{16} & {RAVE J093730.5-062655} & {$(-9.13, -1.74, +1.28)$} & {$(-136.26, -92.64, -75.55)$} & {$(1.92, 11.97, 6.48)$} & {$0.72$} \\
{16} & {BPS CS 22882-001} & {$(-7.27, -0.04, -6.36)$} & {$(-117.54, -64.38, -36.21)$} & {$(1.82, 11.28, 8.55)$} & {$0.72$} \\
{16} & {RAVE J130524.5-393126} & {$(-7.04, -1.57, +0.86)$} & {$(+59.37, -57.29, +61.05)$} & {$(1.35, 7.62, 2.91)$} & {$0.70$} \\
{16} & {LP877-23} & {$(-8.12, +0.05, -0.12)$} & {$(-27.43, -50.37, +138.34)$} & {$(1.99, 8.41, 6.71)$} & {$0.62$} \\
{16} & {HE 1430+0053} & {$(-6.58, -0.28, +2.26)$} & {$(-153.01, -87.23, -50.61)$} & {$(1.81, 8.89, 5.14)$} & {$0.66$} \\
\hline
{17} & {Gaia DR2 6412626111276193920} & {$(-1.62, -2.87, -8.05)$} & {$(-155.15, -32.15, +49.07)$} & {$(5.80, 8.73, 8.39)$} & {$0.20$} \\
{17} & {2MASS J00101758-1735387} & {$(-8.14, +0.17, -0.70)$} & {$(+12.09, +69.73, -230.17)$} & {$(7.93, 8.65, 8.16)$} & {$0.04$} \\
{17} & {2MASS J23362202-5607498} & {$(-5.33, -2.17, -5.72)$} & {$(+22.98, +78.74, -170.80)$} & {$(4.69, 9.51, 9.11)$} & {$0.34$} \\
{17} & {LAMOST J112456.61+453531.1} & {$(-11.13, +1.06, +6.69)$} & {$(+30.82, +27.96, +111.51)$} & {$(3.69, 13.41, 12.87)$} & {$0.57$} \\
\hline
{18} & {2MASS J03073894-0502491} & {$(-12.57, -0.37, -5.37)$} & {$(+105.40, +152.25, +15.28)$} & {$(6.19, 16.66, 6.76)$} & {$0.46$} \\
{18} & {HE 0432-0923} & {$(-12.04, -1.82, -2.92)$} & {$(+70.18, +164.56, +97.15)$} & {$(6.34, 15.89, 6.72)$} & {$0.43$} \\
{18} & {2MASS J02165716-7547064} & {$(-6.73, -2.94, -2.75)$} & {$(+36.32, +304.34, -32.85)$} & {$(6.41, 17.54, 7.29)$} & {$0.46$} \\
\hline
{19} & {SMSS J183128.71-341018.4} & {$(-1.14, +0.01, -1.36)$} & {$(+62.04, +218.16, -129.47)$} & {$(1.26, 3.59, 2.60)$} & {$0.48$} \\
{19} & {2MASS J19161821-5544454} & {$(-2.92, -1.78, -2.63)$} & {$(+26.46, +134.66, +4.58)$} & {$(1.50, 4.73, 3.11)$} & {$0.52$} \\
{19} & {SMSS J183225.29-334938.4} & {$(+2.41, +0.08, -2.06)$} & {$(-108.78, -112.82, +196.95)$} & {$(1.00, 6.12, 4.31)$} & {$0.72$} \\
{19} & {SMSS J175738.37-454823.5} & {$(+1.07, -2.22, -1.75)$} & {$(-123.36, +69.85, -116.03)$} & {$(1.52, 3.54, 3.04)$} & {$0.40$} \\
{19} & {2MASS J21224590-4641030} & {$(-5.51, -0.33, -2.66)$} & {$(+29.80, +85.30, +185.37)$} & {$(2.95, 7.83, 6.66)$} & {$0.45$} \\
{19} & {2MASS J18294359-4924253} & {$(-2.12, -1.56, -1.88)$} & {$(-14.13, +100.32, -22.33)$} & {$(1.03, 3.27, 2.06)$} & {$0.52$} \\
{19} & {J20411424-4654315} & {$(-4.15, -0.49, -3.13)$} & {$(-141.10, +32.51, -45.25)$} & {$(0.71, 6.48, 4.33)$} & {$0.80$} \\
\hline
{20} & {SMSS J155430.57-263904.8} & {$(-2.07, -1.52, +2.38)$} & {$(-221.60, +334.62, -153.38)$} & {$(3.00, 24.61, 15.62)$} & {$0.78$} \\
{20} & {2MASS J00524174-0902235} & {$(-8.46, +0.42, -1.54)$} & {$(+256.12, +98.35, +171.26)$} & {$(2.74, 23.42, 14.99)$} & {$0.79$} \\
{20} & {HE 0240-0807} & {$(-16.06, -0.29, -12.29)$} & {$(+88.00, +65.08, +80.21)$} & {$(2.89, 23.85, 14.66)$} & {$0.78$} \\
{20} & {RAVE J091858.9-231151} & {$(-8.98, -2.49, +0.88)$} & {$(-285.97, +40.97, +114.24)$} & {$(2.68, 23.40, 8.65)$} & {$0.79$} \\
\hline
{21} & {2MASS J14325334-4125494} & {$(-7.38, -0.61, +0.34)$} & {$(-361.21, +107.37, -108.94)$} & {$(2.21, 33.16, 17.77)$} & {$0.88$} \\
{21} & {Gaia DR2 1508756353921427328} & {$(-8.21, +1.51, +3.53)$} & {$(-255.26, +179.86, +171.52)$} & {$(2.58, 33.16, 15.27)$} & {$0.86$} \\
\hline
{22} & {2MASS J12170829+0415146} & {$(-6.70, -7.10, +16.10)$} & {$(+8.28, -51.10, +189.72)$} & {$(3.64, 27.59, 26.53)$} & {$0.77$} \\
{22} & {BPS CS 30315-029} & {$(-5.30, +1.66, -10.75)$} & {$(+185.80, -119.50, +174.85)$} & {$(3.45, 27.59, 26.80)$} & {$0.78$} \\
\hline
{23} & {HE 1226-1149} & {$(-6.94, -2.72, +3.63)$} & {$(-7.05, +251.00, +145.04)$} & {$(8.19, 15.17, 9.97)$} & {$0.30$} \\
{23} & {J22372037-4741375} & {$(-5.59, -0.66, -4.06)$} & {$(-106.23, +265.37, +72.09)$} & {$(6.64, 14.00, 8.79)$} & {$0.36$} \\
{23} & {J10401894-4106124} & {$(-8.01, -1.24, +0.36)$} & {$(-62.18, +209.15, -211.66)$} & {$(8.05, 15.03, 9.77)$} & {$0.30$} \\
\hline
{24} & {BPS CS 29497-004} & {$(-7.94, +0.22, -3.49)$} & {$(-121.16, +180.59, -111.32)$} & {$(4.70, 13.91, 6.20)$} & {$0.49$} \\
{24} & {BPS CS 31078-018} & {$(-9.14, +0.15, -0.92)$} & {$(-83.65, +134.38, -32.65)$} & {$(3.50, 10.23, 1.31)$} & {$0.49$} \\
{24} & {J22190836-2333467} & {$(-7.09, +0.64, -1.82)$} & {$(+77.42, +157.78, +67.98)$} & {$(4.20, 8.27, 2.59)$} & {$0.33$} \\
{24} & {BPS CS 22943-132} & {$(-7.81, -0.02, -0.23)$} & {$(-64.51, +175.80, -129.83)$} & {$(5.67, 9.22, 4.10)$} & {$0.24$} \\
{24} & {BPS CS 22886-012} & {$(-7.64, +0.68, -0.96)$} & {$(+9.58, +151.83, +15.09)$} & {$(3.87, 7.73, 1.03)$} & {$0.33$} \\
{24} & {J10191573-1924464} & {$(-8.61, -2.58, +1.57)$} & {$(-41.93, +146.38, -6.45)$} & {$(4.51, 9.13, 1.58)$} & {$0.34$} \\
{24} & {J06195001-5312114} & {$(-8.26, -0.54, -0.25)$} & {$(+52.52, +160.86, -127.12)$} & {$(5.43, 9.27, 4.51)$} & {$0.26$} \\
{24} & {J14354680-1124122} & {$(-7.65, -0.20, +0.56)$} & {$(-9.07, +203.15, -11.19)$} & {$(6.04, 7.68, 0.60)$} & {$0.12$} \\
{24} & {RAVE J183623.2-642812} & {$(-6.57, -0.90, -0.75)$} & {$(-30.97, +192.79, -51.91)$} & {$(5.19, 6.71, 1.22)$} & {$0.13$} \\
{24} & {HD 120559} & {$(-8.16, -0.02, +0.02)$} & {$(-25.27, +183.04, -41.38)$} & {$(5.34, 8.29, 0.78)$} & {$0.22$} \\
{24} & {2MASS J18295183-4503394} & {$(-6.47, -0.31, -0.45)$} & {$(-34.22, +208.20, +40.95)$} & {$(5.46, 6.69, 0.81)$} & {$0.10$} \\
{24} & {BD-10 3742 } & {$(-7.30, -0.67, +1.31)$} & {$(-78.18, +169.41, -13.45)$} & {$(4.51, 8.04, 1.52)$} & {$0.28$} \\
{24} & {J06320130-2026538} & {$(-8.50, -0.37, -0.10)$} & {$(-71.43, +150.84, +71.83)$} & {$(4.32, 9.18, 2.09)$} & {$0.36$} \\
{24} & {RAVE J115941.7-382043} & {$(-7.88, -0.75, +0.37)$} & {$(+16.92, +193.64, +17.56)$} & {$(5.44, 8.19, 0.50)$} & {$0.20$} \\
\hline
{25} & {G166-37} & {$(-8.11, +0.04, +0.20)$} & {$(+320.25, +72.45, +335.65)$} & {$(5.41, 79.72, 77.55)$} & {$0.87$} \\
{25} & {HE 2252-4225} & {$(+0.34, -1.05, -16.09)$} & {$(-95.53, -98.54, +349.18)$} & {$(4.94, 77.38, 77.25)$} & {$0.88$} \\
\hline
{26} & {RAVE J040618.2-030525} & {$(-8.97, -0.20, -0.61)$} & {$(-303.31, -4.82, -96.21)$} & {$(0.10, 24.05, 16.43)$} & {$0.99$} \\
{26} & {LAMOST J110901+075441} & {$(-9.42, -2.86, +5.22)$} & {$(+201.51, +100.16, -141.65)$} & {$(1.13, 23.07, 12.71)$} & {$0.91$} \\
{26} & {Gaia DR2 3602288924850161792 } & {$(-8.10, -0.84, +1.39)$} & {$(-312.86, +9.36, -27.11)$} & {$(0.76, 21.62, 14.10)$} & {$0.93$} \\
{26} & {CD -45 3283} & {$(-8.20, -0.10, -0.00)$} & {$(-304.26, +10.89, -97.82)$} & {$(0.30, 21.67, 17.24)$} & {$0.97$} \\
\hline
{27} & {Gaia DR2 4248140165233284352} & {$(-4.99, +3.17, -1.00)$} & {$(-53.60, -82.15, +408.19)$} & {$(5.89, 32.46, 31.42)$} & {$0.69$} \\
\hline
{28} & {G14-39} & {$(-8.14, -0.04, +0.11)$} & {$(-203.13, +70.86, +243.07)$} & {$(4.74, 20.27, 19.06)$} & {$0.62$} \\
\hline
{29} & {SMSS J024858.41-684306.4} & {$(-7.45, -2.19, -2.29)$} & {$(+107.90, +415.32, +215.11)$} & {$(6.28, 96.52, 42.29)$} & {$0.88$} \\
\hline
{30} & {SMSS J063447.15-622355.0} & {$(-7.93, -6.95, -3.33)$} & {$(-39.11, +439.02, +157.36)$} & {$(7.94, 129.46, 45.67)$} & {$0.88$} \\
\hline
\enddata
\end{deluxetable*}

\end{document}